\documentclass[svgnames,twocolumn]{aastex6}
\usepackage{amsmath}

\usepackage{psfrag} % for nicer fonts in figures

\usepackage{natbib}

\usepackage{graphicx}

\usepackage{txfonts}

\usepackage{bm}%Tomek for  fat greek letters!
\usepackage{soul}
\newcommand{\Eqref}[1]{Eq.\,(\ref{#1})}
\newcommand{\eqsref}[1]{Eqs.\,(\ref{#1})}
\newcommand{\figref}[1]{Fig.\,\ref{#1}}
\newcommand{\tabref}[1]{Tab.\,\ref{#1}}
\newcommand{\secref}[1]{Sec.\,\ref{#1}}
\newcommand{\bref}[1]{(\ref{#1})}

\newcommand{\timex}{ }
\newcommand{\aenus}{\textsc{Aenus}}
\newcommand{\aenusd}{\textsc{Aenus. }}

\newcommand{\tr}[1]{\textcolor{RoyalBlue}{#1}}

\newcommand{\velo}{\textsl{v}} % make velocity different from \nu in math mode

\definecolor{DarkBrown}{rgb}{.396,.263,.129}
\definecolor{LightBrown}{rgb}{.698,.463,.227}

\newcommand{\half}{\frac{1}{2}}

\newcommand{\tto}{\text{--}}

\newcommand{\deltx}{\Delta}
\newcommand{\deltt}{\Delta}

\newcommand{\vv}{\mathcal{V}}
\newcommand{\VV}{\mathcal{V}}
\newcommand{\LL}{\mathcal{L}}
\newcommand{\TT}{\mathcal{T}}

\newcommand{\nn}{\mathfrak{N}}
\newcommand{\dd}{\mathfrak{D}}
 
\newcommand{\der}{\mathrm{d}} 
\newcommand{\alf}{Alfv\'{e}n }

\newcommand{\alfc}{Alfv\'{e}n, }

\newcommand{\ff}{\bm{\mathcal{F}}}

\newcommand{\rrr}{\rightarrow }

\newcommand{\divv}{\nabla \cdot}

\newcommand{\Pm}{P_{\mathrm{m} }}
\newcommand{\Pmast}{P_{\mathrm{m} \ast}}

\newcommand{\ca}{c_{\mathrm{A}}}

\newcommand{\cs}{c_{\mathrm{s}}}
\newcommand{\cmag}{c_{\mathrm{ms}}}

\newcommand{\dda}{\dd_{\mathrm{A}}}
\newcommand{\ddms}{\dd_{\mathrm{ms}}}

\newcommand{\CFL}{C_{\mathrm{CFL}}}

\newcommand{\vek}[1]{\bf{#1}}

\newcommand{\jcop}{J. Comput. Phys.}
\newcommand{\ie}{i.e.~}

\newcommand{\eg}{e.g.~}

\newcommand{\s}{\textrm{s}}

\newcommand{\cm}{\textrm{cm}}

\newcommand{\mathsmaller}[1]{#1}

\begin{document}

%%%%%%%%%%%%%%%%%%%%%%%%%%%%%%%%%%%%%%%%%%%%%%%%%%%%%%%%%%%%%
%% Title
%%%%%%%%%%%%%%%%%%%%%%%%%%%%%%%%%%%%%%%%%%%%%%%%%%%%%%%%%%%%%
\title%[Numerical dissipation]
{
On the measurements of numerical viscosity and resistivity
in Eulerian MHD codes
}
\shorttitle{NUMERICAL VISCOSITY AND RESISTIVITY}
\shortauthors{Rembiasz et al.}

%% Use \author, \affil, plus the \and command to format author and affiliation 
%% information.  If done correctly the peer review system will be able to
%% automatically put the author and affiliation information from the manuscript
%% and save the corresponding author the trouble of entering it by hand.
%%
%% The \affil should be used to document primary affiliations and the
%% \altaffil should be used for secondary affiliations, titles, or email.

%% Authors with the same affiliation can be grouped in a single
%% %% \author and \affil call.
 \author{Tomasz Rembiasz\altaffilmark{1,2}}
 \email{tomasz.rembiasz@uv.es}
 \author{Martin Obergaulinger\altaffilmark{1}}
 \author{Pablo Cerd\'a-Dur\'an\altaffilmark{1}}
 \author{Miguel-\'Angel Aloy\altaffilmark{1}}
 \and \author{Ewald M\"uller\altaffilmark{2}}
 \affil{\altaffilmark{1} Departamento de Astronom\'{\i}a y
   Astrof\'{\i}sica,  Universidad de Valencia,   C/ Dr.~Moliner 50,
   46100 Burjassot, Spain} % ; tomasz.rembiasz@uv.es}  
 \affil{\altaffilmark{2}
Max-Planck-Institut f{\"u}r Astrophysik,   Karl-Schwarzschild-Str.~1, 85748 Garching, Germany }

\received{XX.YY.2016}
\revised{XX.YY.2016}
\accepted{XX.YY.2016}
%does not work! \cpright{AAS}{2016}

%%%%%%%%%%%%%%%%%%%%%%%%%%%%%%%%%%%%%%%%%%%%%%%%%%%%%%%%%%%%%
%% Abstract   
%%%%%%%%%%%%%%%%%%%%%%%%%%%%%%%%%%%%%%%%%%%%%%%%%%%%%%%%%%%%%
\begin{abstract}
  We propose a simple ansatz for estimating the value of the numerical
  resistivity and the numerical viscosity of any Eulerian MHD code.
  We test this ansatz with the help of simulations of the propagation
  of (magneto)sonic waves, \alf waves, and the tearing mode
  instability using the MHD code \aenusd By comparing the simulation
  results with analytical solutions of the resistive-viscous MHD
  equations and an empirical ansatz for the growth rate of tearing
  modes we measure the numerical viscosity and resistivity of \aenusd
The comparison shows that the fast-magnetosonic speed and
    wavelength are the characteristic velocity and length,
    respectively, of the aforementioned (relatively simple)
    systems. We also determine the dependance of the numerical
    viscosity and resistivity on the time integration method, the
    spatial reconstruction scheme and (to a lesser extent) the Riemann
    solver employed in the simulations. From the measured results we
    infer the numerical resolution (as a function of the spatial
    reconstruction method) required to properly resolve the growth and
    saturation level of the magnetic field amplified by the
    magnetorotational instability in the post-collapsed core of
    massive stars.  Our results show that it is to the best advantage
    to resort to ultra-high order methods (e.g., $9^{\rm th}-$order
    Monotonicity Preserving method) to tackle this problem properly,
    in particular in three dimensional simulations.
\end{abstract}

%%%%%%%%%%%%%%%%%%%%%%%%%%%%%%%%%%%%%%%%%%%%%%%%%%%%%%%%%%%%%
%% Text
%%%%%%%%%%%%%%%%%%%%%%%%%%%%%%%%%%%%%%%%%%%%%%%%%%%%%%%%%%%%%

\section{INTRODUCTION}

Every Eulerian MHD code introduces numerical errors during the
integration of an MHD flow because of unavoidable errors resulting
from the spatial and time discretisation of the problem.  These errors
can manifest themselves in two ways.  They can either smear out the
solution (\emph{numerical dissipation}) or introduce phase errors
(\emph{numerical dispersion}).  As the mode of action of numerical
dissipation resembles that of a \emph{physical viscosity} and, for
magnetized flows, also of a \emph{resistivity}, they are commonly
referred to as \emph{numerical viscosity} and \emph{numerical
  resistivity}, respectively (see, e.g., \cite{Laney_1998}, Chap.14,
and \cite{bodenheimer2006}, Chap. 8.3).

A necessary condition for a physically reliable simulation is that the
amount of numerical viscosity and resistivity be sufficiently small.
If this requirement is violated, numerical errors can change the
solution not only quantitatively, but even qualitatively. For example,
\citet{Obergaulinger_et_al_2009} found that the tearing mode (TM)
instability (Furth, Killeen \& Rosenbluth \citeyear{fkr}; FKR63
hereafter) developed in their 2D \emph{ideal} MHD simulations of the
magnetorotational instability \citep[MRI;][]
{Balbus_Hawley__1991__ApJ__MRI} in core-collapse supernovae, although
the TM instability should grow only in resistive MHD.  Thus, it must
have developed due to numerical resistivity (as pointed out by the
authors).

This problem becomes even more exacerbated in relativistic
(magneto)-hydrodynamics, since the jumps of physical variables across
strong shocks are no longer limited in magnitude, and both linearly
degenerate and non-linear eigenfields degenerate when the flow
velocities approach the speed of light
\citep{Mimica_etal__2009__AaA__Afterglow}.

The collapsed core of a massive star is yet another physical
application where viscous and resistive effects can definitive shape
the outcome after core collapse, i.e.\ whether a failed or successful
supernova explosion results. \cite{Abdikamalov:2015} estimate that the
Reynolds number in the gain layer (where neutrino heating is stronger
than neutrino cooling) can be huge ($\sim 10^{17}$), resulting in a
fully turbulent flow in that region.  This turbulence may generate
anisotropic stresses on the flow that definitely help in the supernova
shock revival \citep{Murphy:2013, Couch:2015}. In this context,
convection \citep{Herant:1995, Burrows:1995, Janka:1996,
  Foglizzo:2006}, the growth of the magnetic field induced by the MRI
\citep{Akiyama:2003, Obergaulinger:2006, Cerda:2007, Sawai:2013,  Mosta_2015,
  Sawai:2016, Rembiasz:2016a, Rembiasz:2016b} and its interplay with
buoyancy \citep{Obergaulinger_et_al_2009,
  Obergaulinger:2014,Guilet:2015} are (magneto-)hydrodynamic
instabilities whose numerical treatment crucially depends on the
amount of numerical viscosity and resistivity of the algorithms
employed.

As the magnitude of numerical viscosity and resistivity is \emph{a
  priori} unknown in a given simulation, one has to perform
convergence tests to determine upper limits for these quantities.
However, convergence tests are not always performed in the case of 3D
simulations because of a high computational cost.  Therefore, it would
be valuable to have some way of assessing the importance of numerical
viscosity and resistivity for a given system.  One would also like to
know whether the dominant source of the numerical dissipation are
spatial discretisation errors or time integration errors.

In this paper, we propose a simple ansatz and a corresponding
calibration method to estimate the numerical resistivity and viscosity
of any Eulerian MHD code by investigating the dependence of the
numerical resistivity and viscosity on both the numerical (\ie grid
resolution, Riemann solver, reconstruction scheme, time integrator)
and physical setup of a simulation.  To this end, we performed
simulations of the propagation of (magneto)sonic waves and \alf waves,
and of the TM instability.  By comparing the results of our
simulations with analytical solutions of these resistive-viscous MHD
flow problems and an empirical ansatz for the TM growth rate, we are
able to quantify the magnitude of the numerical resistivity and
viscosity.

We do not consider the effects of numerical dispersion \citep[see,
however,][]{Peterson_Hammett}, because this would be beyond the
scope of this work. Hence, our study should be considered as a first
step to better understand the mode of action of numerical viscosity
and resistivity in MHD simulations, and provide a quantitative measure
of the magnitude of the corresponding errors.

In \secref{sec:ansatz}, we present the key idea of our ansatz to
quantify the numerical viscosity and resistivity of an MHD code, and we
describe the code \aenus,   used for our simulations in
\secref{sec:code}. Although we calibrate the numerical viscosity and
resistivity for a specific code only, the method is general and
independent of \aenus.  As a service to the community, we provide the
data from our tests online\footnote{\url{http://www.uv.es/camap/tmweb/Web\_tm.html}}
to facilitate comparisons with other codes.  In
\secref{sec:simulations}, we present a methodology to compute
numerical viscosity and resistivity based on the results of numerical
simulations of several MHD flows encompassing the propagation of fast
magnetosonic waves, \alf waves, sound waves, and of the TM
instability. In \secref{sec:case} we present an example of the
application of our methodology.  Finally, in \secref{sec:summary}, we
summarise and discuss our results.

%%%%%%%%%%%%%%%%%%%%%%%%%%%%%%%%%%%%%%%%%%%%%%%%%%%%%%%%%%%
%%%%%%%%%%%%%%%%%%%%%%%%%%%%%%%%%%%%%%%%%%%%%%%%%%%%%%%%%%%

\section{NUMERICAL RESISTIVITY AND VISCOSITY}
\label{sec:ansatz}

%-----------------------------------------------------------------------------

\subsection{Numerical Integration of the MHD Equations}

The equations of resistive-viscous (non-ideal) magnetohydrodynamics (MHD) 
can be written as
\begin{align}
\label{eq:cont}
\partial_t \rho + \nabla \cdot ( \rho {\vek v} ) &= 0, \\ 
\label{eq:momentum}
\partial_t( \rho {\vek v }) + 
  \divv \left(\rho {\vek v} \otimes {\vek v} + {\vek T} \right) 
 &= 0, \\
\label{eq:energy}
\partial_{t} e_{\star} +   \divv \left[ 
e_{\star} {\vek v} + {\vek v  \cdot T}
+ \eta \left( {\vek b} \cdot \nabla {\vek b}  
        - \mathsmaller{\frac{1}{2}} \nabla {\vek b}^2\right)  
\right] &= 0, \\ 
\label{eq:induction}
 \partial_t {\vek b}  - \nabla \times [ {\vek v} \times {\vek b}
  +   \eta  (\nabla \times  {\vek b})  ] &= 0,   \\
\divv {\vek b} &= 0,  
\label{eq:divb}
\end{align}
where $\vek v$, $\rho$, $\eta$, and ${\vek b}$ are the fluid velocity,
the density, a uniform resistivity, and the magnetic field,
respectively, expressed in Heaviside-Lorentz units.  The total energy
density, $e_{\star}$, is composed of fluid and magnetic contributions,
\ie
$e_{\star} = \varepsilon + \frac{1}{2} \rho {\vek v}^2 + \frac{1}{2}
{\vek b} ^ 2$
where $\varepsilon$ and $p = p(\rho, \varepsilon, \dots)$ are the
internal energy density and the gas pressure, respectively.  The
stress tensor ${\vek T}$ is given by
\begin{equation}
  {\vek T} = \left[ P + \mathsmaller{\half} {\vek b}^2 
                      + \rho \left( \mathsmaller{\frac{2}{3}} \nu - \xi\right)
                    \divv {\vek v} \right] {\vek I} 
             - {\vek b} \otimes {\vek b} 
             - \rho \nu \left[ \nabla  \otimes {\vek v} + 
                              (\nabla  \otimes {\vek v})^T \right], 
\end{equation}
where ${\vek I}$ is the unit tensor, and $\nu$ and $\xi$ are the
kinematic shear and bulk viscosity, respectively.

The system of partial differential equations (PDEs) given by
Eqs.~(\ref{eq:cont})-(\ref{eq:energy}) is expressed in conservation
form 
\begin{equation}
   \partial_t {\bf U} + \divv \ff ( {\bf U} ) =  0,
\end{equation}
where $ {\bf U} $ is the vector of conservative variables and $\ff$ is
the matrix of the fluxes associated with those variables.  For
simplicity, we do not consider in this work source terms in the
equations.

There exist powerful techniques to integrate numerically hyperbolic
systems of conservation laws including a correct treatment of flow
discontinuities \citep[e.g.][]{LeVeque_Book_1992__Conservation_Laws,
  Toro__1997__Buch__Riemann-solvers,
  Laney_1998,LeVeque_Book_2002__Conservation_Laws}.  Among the most
popular techniques are Eulerian methods, which rely on a numerical
discretisation of the solution (typically in finite volumes) on a
fixed, \ie Eulerian grid. The numerical solution of the discretised
system of PDEs differs from its exact solution by an amount which we
call the {\it numerical error} of the solution. This numerical error
can be interpreted as a sum of
numerical dissipation and 
numerical dispersion  which are  not present in
the original system of hyperbolic equations.  The purpose of this work
is to characterise this numerical dissipation and to assess whether it
can be interpreted as a numerical viscosity or as a numerical
resistivity for magnetised flows.

%------------------------------------------------------------------------------
\subsection{A Simple Example}

To illustrate the concept of numerical viscosity and resistivity, we
present an example of a one-dimensional conservation law for a scalar
quantity without external sources
\begin{equation}
  \partial_t u +  \partial_x f(u) = 0,
\label{eq:scalar}
\end{equation}
where $u (x,t)$ is the conserved variable and $f (u)$ is the flux.
Given its value at $t=0$, $u (0, x) = u_0(x)$, this equation can be
integrated to obtain its solution at any later time.

The numerical solution of Eq.~(\ref{eq:scalar}) can be obtained
discretising the time and spatial derivatives. Hence, the
\emph{numerical} version of Eq.~(\ref{eq:scalar}) reads
\begin{equation}
  (\partial_t u^*)_{\mathrm{num}} +  
      (\partial_x f(u^*))_{\mathrm{num}}  = 0,
\label{eq:scalar51}
\end{equation}
where the subscript "num" means that a given derivative is determined
numerically, and $u^*(x,t)$ is  a function approximating the
solution. 

Let us consider a spatial and time discretisation of the order $r$ and
$q$, respectively. Hence, the numerical approximations of the spatial
and time derivatives differ from the analytical ones by terms of the
order $\mathcal{O}(\Delta x^r)$ and $\mathcal{O}(\Delta t^q)$ or
higher, respectively.  In this case, Eq.~(\ref{eq:scalar51}) reads
\begin{eqnarray}
  \partial_t u^* + \partial_x f(u^*) 
  + f' a_r (\deltx x)^r  \, \frac{\partial^{r+1} u^*} {\partial x^{r+1}}
+ b_q  (\deltx t)^q \frac{\partial^{q+1} u^*}{\partial t^{q+1}}
= \nonumber \\ = \mathcal{O}(\deltx x^{r+1})+ \mathcal{O}(\deltx t^{q+1}) .\quad 
\label{eq:scalargen}
\end{eqnarray}
where $a_r$ and $b_q$ are coefficients that depend on the spatial and
time discretisation, and $f' = {\rm d} f (u)/ {\rm d} u$, whose
analytic expression is not necessarily known.  This equation for $u^*$
differs from Eq.~(\ref{eq:scalar}) in terms which are proportional to
powers of $\Delta x$ and $\Delta t$.  Hence, in the limit
$\Delta x\to 0$ and $\Delta t \to 0$, $u^*$ and $u$ coincide. However,
at finite resolution, the additional terms arising from the
discretisation may change the character of the equation, which, in
certain regimes, may change the hyperbolic system into a parabolic
one.  To show the consequences more explicitly, we consider two
examples.

In the first example, we examine a method with $r=1$ (e.g. using
piecewise constant reconstruction, as in Godunov's method) and a time
integrator with $q>2$. In this case
\begin{equation}
  \partial_t u^* + \partial_x f(u^*)  
 + f'  a_1 \Delta x \, \partial_{xx} u^* 
  + f' a_2  (\Delta x)^2 \, \partial_{xxx} u^* 
= \mathcal{O}(3),
\label{eq:pcrecon}
\end{equation}
where all third or higher order terms are grouped into
$\mathcal{O}(3)$.  The terms proportional to $\partial_{xx} u^*$ and
$\partial_{xxx} u^*$ are usually referred to as \emph{numerical
  dissipation} and \emph{numerical dispersion}, respectively. For
wave-like solutions of the form $\exp[i(\omega t - k x)]$ the
dispersion relation reads
\begin{equation}
  i [\mathcal{R}(\omega)  -k  ( 1 - a_2  (\Delta x)^2 k^2) f'] - 
    [\mathcal{I}(\omega) + f'  a_1 \Delta x \,k^2]  =  \mathcal{O}(3),
\label{eq:dispers}
\end{equation}
where $\mathcal{R}(\omega)$ and $\mathcal{I}(\omega)$ are the real and
imaginary part of $\omega$, respectively.  The dissipative and
dispersive character can be explicitly seen by computing the
dissipation rate $\sigma_{\rm dis}$ and the phase velocity
$v_{\mathrm{ph}}$ of the wave. These quantities are obtained by
identifying the two terms in Eq.~(\ref{eq:dispers}) with the
respective spatial derivate of $u^*$ in Eq.~(\ref{eq:pcrecon}):
\begin{align}
\sigma_{\rm dis} = \mathcal{I}(\omega) 
                \approx& - f'  a_1 (k \Delta x) \,k, \\
v_{\mathrm{ph}} = \frac{\mathcal{R}(\omega)}{k} 
               \approx& ( 1 -  a_2  (k \Delta x)^2) f'.
\end{align}

In the second example, we consider an explicit time integration method
with $q=1$ (e.g. the forward Euler method) and $r>1$. In this case,
keeping only first order terms for simplicity,
\begin{equation}
  \partial_t u^* + \partial_x f(u^*)   
    + f'^2 b_1 \Delta t \, \partial_{xx} u^* 
    + 2f' f'' b_1\Delta t \,(\partial_x u^*)^2 = \mathcal{O}(2),
\end{equation}
where we used that
$\partial_{tt} u^* - f'^2\partial_{xx} u^* - 2f' f'' \,(\partial_x
u^*)^2 = \mathcal{O}(1)$,
to eliminate second order time derivatives.  As in the previous
example, we consider wave-like solutions keeping only terms linear in
the amplitude of the perturbations, which results in the following
dispersion relation:
\begin{equation}
  i [\mathcal{R}(\omega) - k f'] - 
    [\mathcal{I}(\omega) + f' b_1 \Delta t \,k^2f' ]  =  \mathcal{O}(2).
\end{equation}
The resulting error also acts as numerical dissipation (proportional
to $k^2$).

In this work, we focus on the measurement of numerical dissipation in
single-scale problems where the distinction between dissipation and
hyper-dissipation is of minor importance. However, one should bear in
mind that this distinction is important if the estimates presented in
this work are applied to multi-scale problems.

%-----------------------------------------------------------------------------

\subsection{An Ansatz for Numerical Viscosity and Resistivity}
\label{subsec:ansatz}

Following the reasoning of the previous section, one can try to
estimate the importance of the additional terms arising from the
numerical discretisation of the MHD equations. The additional terms in
(\ref{eq:momentum})-(\ref{eq:induction}) are commonly called
\emph{numerical viscosity} and \emph{numerical resistivity}, since
these terms modify the dynamics of the system in a similar way as does
a physical viscosity and resistivity.  This is especially valid for
flux-conservative methods, in which numerical discretisation does not
introduce non-conservative terms in the equations (\ie sources) and
similarities with physical viscosity and resistivity are accentuated.

A detailed analysis of the numerical errors is, in general, a
challenging task.  Here, we will perform an error analysis in a
simplified manner.  We will not discriminate between numerical
dissipation and dispersion, but simply assume that all spatial
discretisation errors and time integration errors only contribute to
numerical dissipation, \ie to numerical viscosity and resistivity. 

Based on the discussion above and the simple tests of the previous
section, we propose an ansatz for the numerical viscosity and
resistivity of an MHD code that depends on the discretisation scheme
and the grid resolution used in a simulation.

In the CGS units, both resistivity and kinematic viscosity have
dimension of $[ \text{cm}^2 \, \s^{-1} ]$, hence their numerical
counterparts must have the same dimension.  The most natural ansatz
for, say, the numerical shear viscosity then has the form
$\nu_{\ast} \propto \vv \timex \LL $, where $\vv$ and $\LL$ are the
characteristic velocity and length of a simulated system,
respectively.  The determination of $\vv$ and $\LL$ is not an easy
task in general, since it is problem dependent, as we will show in the
subsequent sections.

Let us consider a one-dimensional (1D) MHD simulation.
Because numerical errors arise from the spatial ($\deltx x$) and
temporal discretisation ($\deltt t$), these terms should be
proportional to $( \deltx x)^r$ and $( \deltt t)^q$, where $r$ and $q$
depend on the order of the numerical schemes.  Since $\deltx x$ has
dimension $[\text{cm}]$, $(\deltx x)^r$ should be multiplied by
$\LL^{-r}$. The resulting term $(\deltx x / \LL)^r$ has a simple
interpretation: the more zones used to resolve the characteristic
length, the lower the numerical viscosity.  The same argumentation
holds for time integration errors, which should enter the ansatz in
the form $ (\deltt t \vv / \LL)^q$.  Therefore, the ansatz for the
numerical shear viscosity $\nu_{\ast}$ should read
\begin{equation}
\nu_{\ast} = \nn_{\nu}^{\deltx x} \timex \vv \timex \LL \timex 
             \left( \frac{\deltx x}{\LL}\right)^{r} + 
             \nn_{\nu}^{\deltx t} \timex \vv \timex \LL \timex 
             \left( \frac{\vv \deltt t }{\LL}\right)^{q},
\label{eq:nu*}
\end{equation}
where $\nn_{\nu}^{\deltx x}$, $\nn_{\nu}^{\deltt t}$, $r$, and $q$ are
constants for a given numerical scheme.

Using the CFL factor definition for an equidistant grid,
Eq.~(\ref{eq:nu*}) can be rewritten as
\begin{equation}
\nu_{\ast} = \nn_{\nu}^{\deltx x} \timex \vv \timex \LL \timex 
            \left( \frac{\deltx x}{\LL}\right)^{r} +  
            \nn_{\nu}^{\deltt t} \timex \vv \timex \LL \timex 
            \left(\frac{\CFL \deltx x }{\LL}\right)^{q} \timex 
            \left( \frac{ \vv }{\velo_{\textrm{max}} }\right)^{q},
\label{eq:nu*cfl_0}
\end{equation}
where $\velo_{\textrm{max}}$ is the maximum velocity of the system
limiting the timestep. If $\vv = \velo_{\textrm{max}}$,
Eq.~(\ref{eq:nu*cfl_0}) simplifies to
\begin{equation}
\nu_{\ast} = \nn_{\nu}^{\deltx x} \timex \vv \timex \LL \timex 
             \left( \frac{\deltx x}{\LL}\right)^{r} + 
             \nn_{\nu}^{\deltt t} \timex \vv \timex \LL \timex 
             \left(\frac{\CFL \deltx x }{\LL}\right)^{q}. 
\label{eq:nu*cfl}
\end{equation}
Note that time and spatial discretisation contribute to different
derivatives provided $r \ne q$.

The same ansatz should hold for the numerical bulk viscosity
$\xi_{\ast}$ and the resistivity $\eta_{\ast}$, with the coefficients
$\nn_{\eta}^{\deltx x}$, $\nn_{\eta}^{\deltt t}$, and
$\nn_{\xi}^{\deltx x}$ and $\nn_{\xi}^{\deltx t}$, respectively:
\begin{equation}
\xi_{\ast} = \nn_{\xi}^{\deltx x}  \timex \vv \timex \LL \timex
             \bigg( \frac{\deltx x}{\LL}\bigg)^{r} + 
             \nn_{\xi}^{\deltt t} \timex \vv \timex \LL \timex 
             \left( \frac{\vv \deltt t }{\LL}\right)^{q},
\label{eq:xi*}
\end{equation}
\begin{equation}
\eta_{\ast} = \nn_{\eta}^{\deltx x}  \timex \vv \timex \LL \timex
              \bigg( \frac{\deltx x}{\LL}\bigg)^{r} +  
              \nn_{\eta}^{\deltt t} \timex \vv \timex \LL \timex 
              \left( \frac{\vv \deltt t }{\LL}\right)^{q},
\label{eq:eta*}
\end{equation}
where we assume that $r$ and $q$ have the same values as in Eqs.\,(\ref{eq:nu*})--(\ref{eq:nu*cfl}).  Once the unknown
coefficients $\nn$, $r$, and $q$ are determined, the above ansatz can be used to estimate the numerical resistivity and
viscosity in any simulation performed with the same code.  
Throughout this paper, we will differentiate between the
measured order of a numerical scheme, $r$ and $q$, and its theoretically expected value, \ie $r_{\rm th}$ and $q_{\rm
  th}$.  For example, for a 5th-order accurate reconstruction scheme $r_{\rm th} = 5 $, and for a 3rd-order accurate
time integrator $q_{\rm th} = 3$.  However, when fitting simulation data, $r$ and $q$ are always assumed to be fit
parameters and not \emph{a priori} known constants (cf. \tabref{tab:waves}).

In the multidimensional (multi-D) case, the ansatz given by Eqs.~(\ref{eq:nu*}), (\ref{eq:xi*}) and (\ref{eq:eta*}) 
can be generalised. Inspecting Eq.~(\ref{eq:scalargen}) one realizes that in the multi-D case 
similar terms appear for the spatial derivatives in each of the directions and for all possible cross derivatives.
However, the contribution from the time derivative remains the same.
A detailed analysis of the form of these numerical dissipative and dispersive terms is beyond the scope of this work.
Instead, we propose a simple ansatz containing the main features of numerical dissipation in multiple directions. 
The first fact to realize is that
different characteristic length scales apply to the different directions.
For example, in 2D using coordinates $(x,y)$, the relevant quantities
are $\Delta x/\LL_x$ and $\Delta y / \LL_y$, where $\LL_x$ and $\LL_y$
are  characteristic lengths in the respective  direction. Similarly 
there is a characteristic velocity, $\VV_x$ and $\VV_y$, in each direction. As a consequence, dissipation acts differently
for each direction of the grid and it  becomes anisotropic. The diffusion coefficients, $\nu$, $\xi$ and $\eta$, appearing in 
Eqs.~(\ref{eq:momentum})-(\ref{eq:induction}) rely on the assumption of an isotropic fluid \citep[see e.g.][]{Landau_fluids},
therefore numerical dissipation cannot be modeled using these scalar coefficients in the multi-D case. 
However, it is relatively easy to find a prescription for a non-isotropic dissipation generalising the scalar character of the dissipation 
coefficients to 2-tensors. In this way the generalisation of the
scalar kinematic viscosity $\nu_*$ to a tensor $\bm{\nu}_*$ (whose components
are $\nu_*^{ij}$) would imply substitutions
in the MHD equations of the kind
\begin{equation}
\partial_i (\, \nu_* \, \partial_j) \to \partial_i (\, \nu_*^{jk} \, \partial_k),
\end{equation}
and similarly for ${\bm \xi}_*$ and ${\bm \eta}_*$, with components $\xi_*^{ij}$ and $\eta_*^{ij}$. 
Explicit expressions for the case of viscosity can be obtained using a
rank four dynamic viscosity tensor of the form
\begin{equation}
\eta_{iklm} = \rho \left [ \nu_{il} \delta_{km} + \nu_{km} \delta_{il} 
+ (\xi_{lm} - \frac{2}{3}  \nu_{lm}) \delta_{ik} \right ]
\end{equation}
and following the procedure laid out in chapter 5 of \cite{Landau_elasticity}.

We propose an ansatz for these tensorial coefficients,
in which we neglect terms coming from cross derivatives for simplicity, keeping only the contribution 
to the numerical discretisation error in each direction separately. For the 2D case, which can be trivially generalised to 3D,
our ansatz reads
\begin{equation}
\bm{\nu}_* = \left (
\begin{array}{cc}
\nu_{\rm sp}^{xx} & 0 \\
0 & \nu_{\rm sp}^{yy}
\end{array}
\right )
+
\nu_t
\left (
\begin{array}{cc}
1 & 0 \\ 0 & 1
\end{array}
\right ),
\label{eq:visc:2D}
\end{equation}
where
\begin{align}
\label{eq:nu_xx}
\nu_{\rm sp}^{xx} &=  \nn_{\nu}^{\deltx x}  \VV_x \LL_x \left(\frac{\Delta x}{\LL_x} \right)^r,\\
\nu_{\rm sp}^{yy} &=  \nn_{\nu}^{\deltx x}  \VV_y \LL_y \left(\frac{\Delta y}{\LL_y} \right)^r, \\
\nu_{t}          &=  \nn_{\nu}^{\deltx t}   \VV \LL \left( \frac{\VV \deltt t }{\LL}\right)^{q}, 
\label{eq:nu_t}
\end{align}
and similarly for $\xi_*^{ij}$ and $\eta_*^{ij}$.
Note that the temporal contribution to the dissipation coefficients is isotropic and depends on 
the characteristic length and time of the solution ($\LL$ and $\VV$)
instead of on the characteristic scales
along each direction. In this ansatz, we assume that the same algorithm 
is used to compute the derivatives in all spatial directions, and hence the coefficient $\nn_{\nu}^{\deltx x}$
is the same in all components $\nu_{\rm sp}^{ij}$.

One also needs to correctly identify the characteristic velocity, $\VV$, and
length, $\LL$, of the system (or the corresponding quantities in the multi-D case), 
 which may require a good understanding of the
problem  (see 2D simulations of sound waves and TMs  in Sections\, \ref{subsec:2d_waves} and \ref{subsec:tm}, respectively).

To test the robustness of the above ansatz and to determine the
unknown coefficients, we considered four test problems in
resistive-viscous MHD that have analytically known solutions: the
damping of sound waves, Alfv\'{e}n waves, and fast magnetosonic waves,
and the TM instability.  Because slow magnetosonic waves will not be
discussed in the remaining part of this paper, we will simply write
\emph{magnetosonic} waves to denote \emph{fast magnetosonic} waves.

%%%%%%%%%%%%%%%%%%%%%%%%%%%%%%%%%%%%%%%%%%%%%%%%%%%%%%%%%%%
%%%%%%%%%%%%%%%%%%%%%%%%%%%%%%%%%%%%%%%%%%%%%%%%%%%%%%%%%%%

\section{THE CODE}
\label{sec:code}

We used the three-dimensional Eulerian MHD code \textsc{Aenus}
\citep{Obergaulinger__2008__PhD__RMHD} to solve the MHD equations
(\ref{eq:cont})--(\ref{eq:divb}).  The code is based on a
flux-conservative, finite-volume formulation of the MHD equations and
the constrained-transport scheme to maintain a divergence-free
magnetic field \citep{Evans_Hawley__1998__ApJ__CTM}.  Based on
high-resolution shock-capturing methods
\citep[e.g.\,][]{LeVeque_Book_1992__Conservation_Laws}, the code
employs various optional high-order reconstruction algorithms
including a total-variation-diminishing (TVD) piecewise-linear (PL)
reconstruction of second-order accuracy, a third-, \mbox{fifth-,}
seventh- and ninth-order monotonicity-preserving (MP3, MP5, MP7 and
MP9, respectively) scheme \citep{Suresh_Huynh__1997__JCP__MP-schemes},
a fourth-order, weighted, essentially non-oscillatory (WENO4) scheme
\citep{Levy_etal__2002__SIAM_JSciC__WENO4}, and approximate Riemann
solvers based on the multi-stage (MUSTA) method
\citep{Toro_Titarev__2006__JCP__MUSTA} and the HLLD Riemann solver
\citep[]{Harten_JCP_1983__HR_schemes,HLLD}.

We add terms including viscosity and resistivity to the flux terms in
the Euler equations and to the electric field in the MHD induction
equation.  We treat these terms similarly to the fluxes and electric
fields of ideal MHD, except for using an arithmetic average instead of
an approximate Riemann solver to compute the interface fluxes. The
explicit time integration can be performed with Runge-Kutta schemes of
first, second, third, and fourth order accuracy (RK1, RK2, RK3, and
RK4), respectively.

%%%%%%%%%%%%%%%%%%%%%%%%%%%%%%%%%%%%%%%%%%%%%%%%%%%%%%%%%%%
%%%%%%%%%%%%%%%%%%%%%%%%%%%%%%%%%%%%%%%%%%%%%%%%%%%%%%%%%%%

\section{NUMERICAL TESTS}
\label{sec:simulations}

%--------------------------------------------------------------------

\subsection{Wave Damping Tests in 1D}
\label{subsec:waves}

To determine the numerical dissipation of the \textsc{Aenus} code and
to test the ansatz (\ref{eq:nu*}), (\ref{eq:xi*}), and
(\ref{eq:eta*}), we perform a series of numerical tests involving the
propagation of waves in a homogeneous medium. Three kind of waves are
studied, sound waves, Alfv\'en waves, and fast magnetosonic waves.  We
align the propagation direction of the wave with one of the grid
coordinate directions, making the problem 1-dimensional.  We determine
the damping rates of the wave amplitudes, which depend only on the
dissipative terms in the discretised MHD equations, in our case owing
to only numerical dissipation.

To measure the damping rate, we performed numerical simulations
letting the wave cross the simulation box, which has periodic
boundaries, at least $10$ times. The energy of the wave, computed as
an integral of the kinetic energy density over the box, decreases
exponentially with time. We fit a linear function to the logarithm of
this quantity to obtain a measure of the energy damping rate which is
equivalent to twice the amplitude damping rate (see below). To
estimate the different dissipation coefficients of our ansatz, we
exploit the fact that the damping rate of the different kinds of waves
depends differently on numerical viscosity and resistivity.

If not otherwise stated, the simulation box length and the wavevector
are set to $L = 1$ and $k = 2 \pi$, respectively.  An ideal gas
equation of state (EOS) with an adiabatic index $\Gamma = 5/3$ is
used.

%%%%%%%%%%%%%%%%%%%%%%%%%%%%%%%%%%%%%%%%%%%%%%%%%%%%%%%%%%%%%%%%%%%%%%
\begin{table*}
  \caption{
  Wave damping simulations: the columns give (from left to
  right) the series identifier, the wave type, the reconstruction 
  scheme, the Riemann solver, the time integrator, the CFL factor, 
  and the grid resolution. The estimators for $\nn_{\rm tot}^{\deltx x}$, 
  $r$, $\nn_{\rm  tot}^{\deltx t}$, and $q$  (see Eqs.\,\ref{eq:nu*}, 
  \ref{eq:xi*}, and \ref{eq:eta*}) are obtained from linear
  fits to the simulation results. For sound waves, \alfc waves, 
  and magnetosonic waves 
  $\nn_{\rm  tot}^{\deltx x} = (4/3) \nn_{\nu}^{\deltx x} + \nn_{\xi}^{\deltx x}$,\,  
  $\nn_{\rm  tot}^{\deltx x} = \nn_{\nu}^{\deltx x} +  \nn_{\eta}^{\deltx x}$,
  and 
  $\nn_{\rm  tot}^{\deltx x}= (4/3) \nn_{\nu}^{\deltx x} + \nn_{\xi}^{\deltx x}  + (3/8)
  \nn_{\eta}^{\deltx x}$, respectively. The estimators $\nn_{\rm
    tot}^{\deltx t}$ are defined analogously. }
\begin{center}
\begin{tabular}{cccccccccccc}
\tableline
series & wave & Reco &  Riemann &  time & CFL &  resolution  &   $\nn_{\rm  tot}^{\deltx x}$ &  $r$ &  $\nn_{\rm  tot}^{\deltx t}$  &  $q$
\\  \tableline
\#S1 & sound & PL & HLL & RK4& $0.01$ & $64 \dots  1028$ & $ 14.3 \pm  0.7$ &  $ 3.049 \pm 0.009 $ & - & - 
 \\
\#S2 & sound & MP5 & LF & RK4 & $0.01$ & $8 \dots  256$ &  $ 42.9  \pm  2.3 $ & $ 4.957  \pm 0.013  $  & - & - 
 \\
\#S3 & sound & MP5 & HLL & RK4 & $0.01$ & $8 \dots  256$ &  $ 43.4  \pm  2.5$ &$  4.961 \pm 0.014$ & - & - 
 \\
\#S4 & sound & MP5 & HLLD & RK4 & $0.01$ & $8 \dots  256$ &  $ 42.7  \pm  2.2   $ & $ 4.956 \pm 0.013 $& - & - 
 \\
\#S5 & sound & MP7 & HLL & RK4 & $0.01$ & $8 \dots  64$ &  $ 302  \pm  20$ & $ 6.897 \pm 0.021 $ & - & - 
\\
\#S6 & sound & MP9 & HLL & RK4 & $0.01$ & $8 \dots  32$ &  $ \  \; 830  \pm   340$ & $ 8.42 \pm 0.15$  & - & - 
\\
\#S7 & sound & MP9 & HLL & RK3 & $0.5$ & $8 \dots$  256  & - & -  & $ 1.492  \pm  0.013 $  & $2.985 \pm 0.002 $ 
\\
\#S8 & sound & MP9 & HLL & RK3 & $0.1  \dots  0.9$ &  $64$   & - & -   & $ 2.45  \pm  0.17 $  & $  \; \; 2.95 \pm 0.01$   
\\
\#S9  &   sound  &  MP9 &  HLL &  RK4 &  $0.5$ &   $ 8 \dots 32 $ &  $ - $  &   $ - $ &  $   71  \pm  32 $ &   $ 5.5  \pm  0.2    $   
\\
\tableline
\#A1 & Alfv\'en & MP5 & LF & RK4 & $0.01$ &  $8 \dots  256$ &    $ 42  \pm  3 $ & $ 4.95  \pm 0.02 $    & - & - 
\\
\#A2 & Alfv\'en & MP5 & HLL & RK4 & $0.01$ &  $8 \dots  256$ &   $ 42.6  \pm  2.1 $ & $\ \  4.96 \pm 0.01 $     & - & - 
\\
\#A3 & Alfv\'en & MP5 & HLLD & RK4 & $0.01$ &  $8 \dots  256$ &   $ 42  \pm 3  $ & $ 4.95 \pm 0.02 $     & - & - 
\\
\#A4 & Alfv\'en & MP7 & HLL & RK4 & $0.01$ &  $8 \dots  128$ &  $ \ \  44  \pm  53$ & $  6.19 \pm 0.03 $   & - & - 
\\
\#A5 & Alfv\'en & MP9 & HLL & RK4& $0.01$ &  $8 \dots  64$ &   $ 1190  \pm  190$ & $ 8.57 \pm 0.06$    & - & - 
\\
\#A6 & Alfv\'en & MP9 & HLL & RK3 & $0.8$ &  $16 \dots  128$   & - & - &  $  0.86 \pm  0.08   $  & $  2.949   \pm  0.022   $  
\\
\#A7 & Alfv\'en & MP9 & HLL & RK4 & $0.8$ &  $8 \dots  64$   & - & - &  $   7.6 \pm  2.5  $  & $     5.18       \pm  0.10 $ 
\\
\#A8 & Alfv\'en & MP5 & HLL & RK3 & $0.5$ &  $5 \dots  1024$   & - & - &  -& -
\\
\tableline
\#MS1 & magnetosonic & MP5 &  HLL &  RK4 & $0.01$ &  $8 \dots  128$ &  $ 40  \pm 3 $ &$ 4.95  \pm  0.02 $   & - & - 
\\
\#MS2 & magnetosonic & MP7 &  HLL &  RK4 & $0.01$ &  $8 \dots  64$ &  $ 288  \pm 20  $ & $ 6.903 \pm 0.023 $ & - & - 
\\
\#MS3 & magnetosonic & MP9 &  HLL &  RK4 & $0.01$ &  $8 \dots  32$ & $ 1970  \pm 160 $ & $ 8.82  \pm 0.03$  & - & - 
\\
\#MS4 & magnetosonic & MP9 &  HLL &  RK3 & $0.1 \dots  0.9$ &  $64$ & - & - & $ 1.77   \pm 0.06    $  & $  2.977    \pm  0.007   $  
\\
\#MS5 & magnetosonic & MP9 &  HLL &  RK4 & $0.2 \dots  0.9$ &  $64$ & - & - & $4.3  \pm  0.8  $  & $    4.834        \pm 0.013  $ 
\\
\tableline
  \end{tabular}
\label{tab:waves}
\end{center}
 \end{table*}

\begin{table*}
  \caption{
    Wave damping simulations performed with the MP5 
    reconstruction scheme, the HLL Riemann solver, and the 
    RK3 time integrator to identify the characteristic velocity  
    and length  of the system. 
    The CFL factor was set to $0.01$ to guarantee negligible time 
    integration errors. The columns give (from left to right) the
    series identifier, the wave type, the initial pressure, density 
    and magnetic field, and the wavelength $\lambda$.
    $\nn_{\rm  tot}^{\deltx x}$  is defined as in \tabref{tab:waves}
    for the corresponding wave type. The exponent $\alpha$ is obtained 
    for each simulation series by means of the fitting functions given 
    by Eqs.\ \eqref{eq:soundfit}, \eqref{eq:lambda_fit_sound}, 
    \eqref{eq:magsonfit}, \eqref{eq:lambda_fit_alf}, and
    \eqref{eq:ln_cmag}, respectively.  }
\begin{center}
\begin{tabular}{cccccccc}
\tableline
series & wave & $p_0$ & $\rho_0$  &  $b_0$& $\lambda$ &  $ \nn_{\rm  tot}^{\deltx x} $&  $\alpha$
\\ 
 \tableline
\#cS1 & sound & $1 \dots 10^{4}$& $1$ & $0$ & $1$ & $46.2 \pm 2.3$ & $ 0.993 \pm 0.003 $  
 \\
\#cS2 & sound & $1$ & $10^{-4} \dots 10$ & $0$ & $1$ & $46.2 \pm 2.3$ & $ 0.993 \pm 0.003 $ 
 \\
\#cS3 & sound & $1$ & $1 $ & $0$ & $0.1 \dots 20$ & $46.2 \pm 2.4 $ & $ 0.9899  \pm 0.0024 $ 
 \\
 \tableline
\#cA1 & Alfv\'en & $10^{-1}$ & $1$ & $10^{-3} \dots 20$ & $1$ & $34.8 \pm 2.4$ & $0.945\pm 0.015$  
 \\
\#cA2 & Alfv\'en & $2 \times 10^{-3} \dots 10^{7} $ & $1$ & $ $1$ $ & $1$ & $34.8 \pm 2.4$ & $0.945 \pm 0.015$  
 \\
\#cA3 & Alfv\'en & $2 \times 10^{-3}$ & $10^{-4} \dots 10^{4}$ & $1 $ & $1$ & $34.8 \pm 2.4$ & $ 0.945 \pm 0.015 $ 
 \\
\#cA4 & Alfv\'en & $2 \times 10^{-3}$ & $1$ & $1$ & $0.1 \dots 10$ & $44 \pm 2$ & $-1.0003 \pm 0.0003 $ 
 \\
 \tableline
\#cMS1 & magnetosonic & $1$ & $1$ & $10^{-4} \dots 10^{3}$ & $1$ & $40 \pm 3$ & $0.997 \pm 0.006$  
 \\
\#cMS2 & magnetosonic & $10^{-4} \dots 10^{4}$ & $1$ & $1$ & $1$ & $40 \pm 3$ & $0.997 \pm 0.006 $   
 \\
\#cMS3 & magnetosonic & $1$ & $10^{-3} \dots 10^4$ & $1$   & $1$ & $40 \pm 3$ & $0.997 \pm 0.006 $ 
 \\
 \tableline
  \end{tabular}
\label{tab:length}
\end{center}
 \end{table*}

%%%%%%%%%%%%%%%%%%%%%%%%%%%%%%%%%%%%%%%%%%%%%%%%%%%%%%%%%%%%%%%%%%%%%%

\subsubsection{Sound Waves}
\label{subsec:sound_waves_1d}

We measured the numerical shear and bulk viscosity of the
\textsc{Aenus} code using sound waves.  We set the background density
and pressure to $\rho_0 = p_0 = 1$, and imposed a perturbation of the
form
\begin{align}
\velo_{1x}(x,t=0) &= \epsilon  \sin(k  x), \\
\rho_1 (x,t=0)    &= \frac{\velo_{1x}(x,0)}{\cs} \rho_0, \\
p_1 (x,t=0)       &= \frac{\velo_{1x}(x,0)} {\cs} \Gamma p_0,
\end{align}
where $\cs = \sqrt{ \Gamma p_0 / \rho_0 }$ is the sound speed.  The
amplitude of the velocity perturbation is set to $\epsilon = 10^{-5}$,
which is small enough to prevent wave steepening \citep[cf.][]{Shore}
within the time of our simulations.

In the presence of (numerical) viscosity, the wave is damped with
time.  For a plane wave
$\velo_{x1}(x,t) = \hat{\velo}_{x1} \exp[ i (kx - \omega t)]$ one
finds from the dispersion relation
\begin{equation}
\omega = \frac{ -i (4 \nu /3 + \xi) k^2}{2} \pm 
         k \cs \sqrt{1 - \frac{k^2  \rho_0 (4 \nu /3 + \xi)^2 } {4 \Gamma p_0}}.
\end{equation}
In the weak damping approximation, \ie if
\begin{equation}
 \frac{k^2 \rho_0\, (4 \nu /3 + \xi)^2 }{4 \Gamma p_0} \ll 1,
\end{equation}
the phase velocity remains constant and the solution can be
written as
\begin{equation}
 v_x(x,t) = \hat{v}_{1x} e^{ -\dd_s  t }  e^{i k (x \mp \cs t)},
\label{sound1}
\end{equation}
where the sound damping coefficient $\dd_s$ is defined as 
\begin{equation}
\dd_s = \frac{k^2}{2} \left( \frac{4}{3}\nu  + \xi \right).
\label{eq:dds}
\end{equation}

The sound wave propagates with a constant speed and its amplitude
decreases with time. Performing simulations with different values of
the physical shear and bulk viscosity we found an excellent agreement
between the analytical (Eq.\,\ref{sound1}) and the numerical solution
\citep[see][for details]{Rembiasz}.

%------------------------------------------------------------------------------
\begin{figure*}
 \centering
 \includegraphics[width=0.45\textwidth]{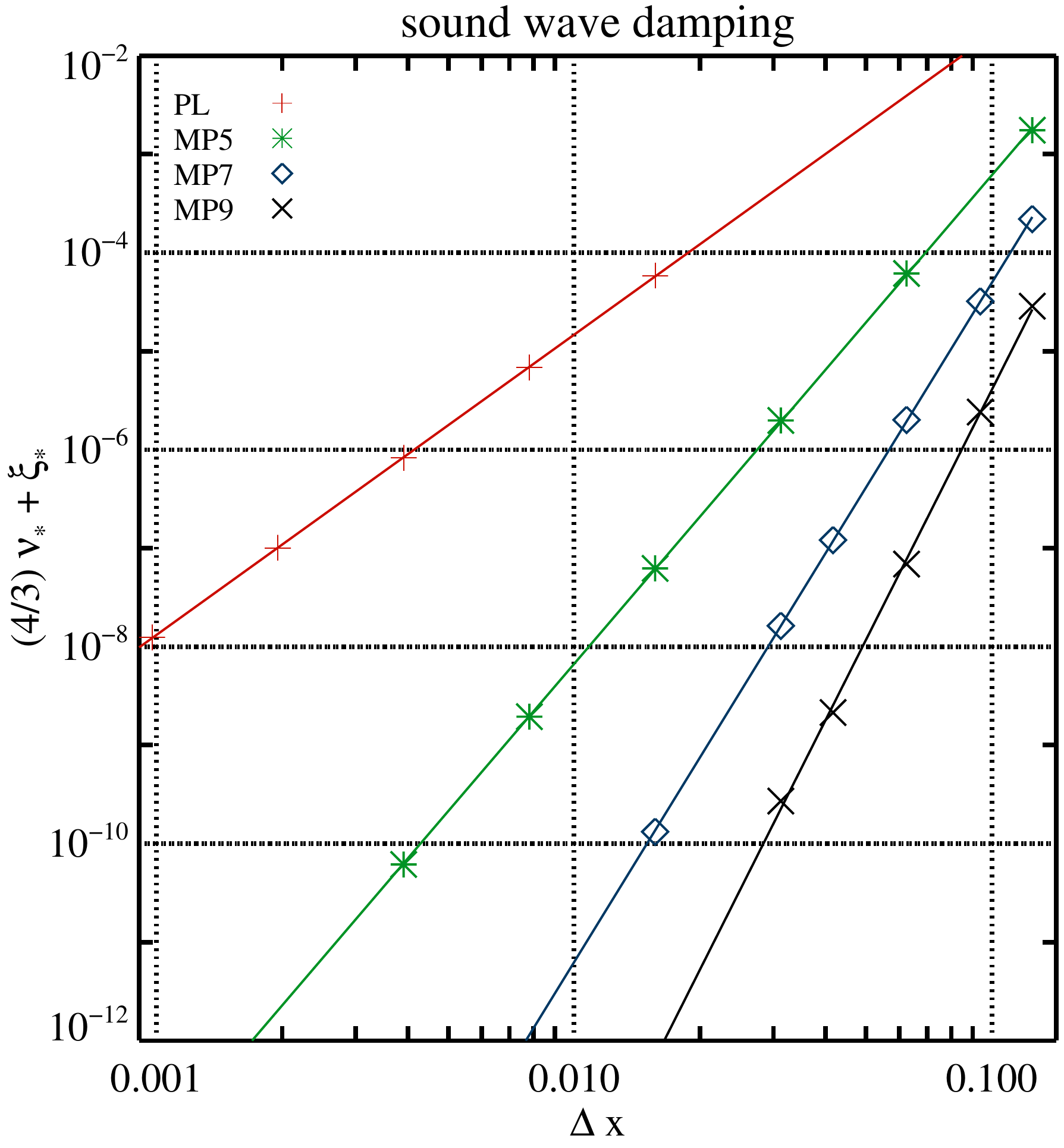} \phantom{MM}
 \includegraphics[width=0.45\textwidth]{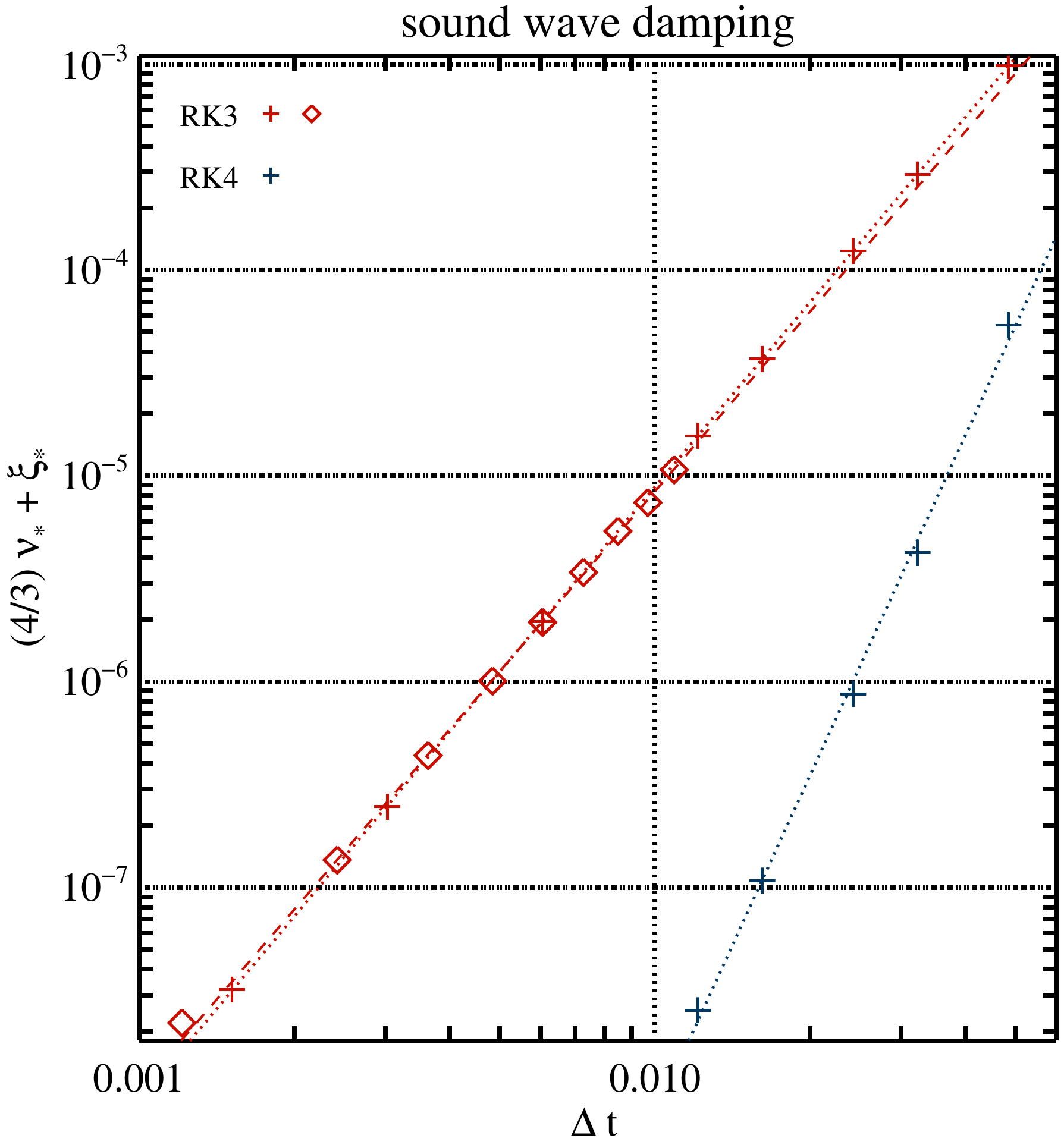}\\
 \includegraphics[width=0.45\textwidth]{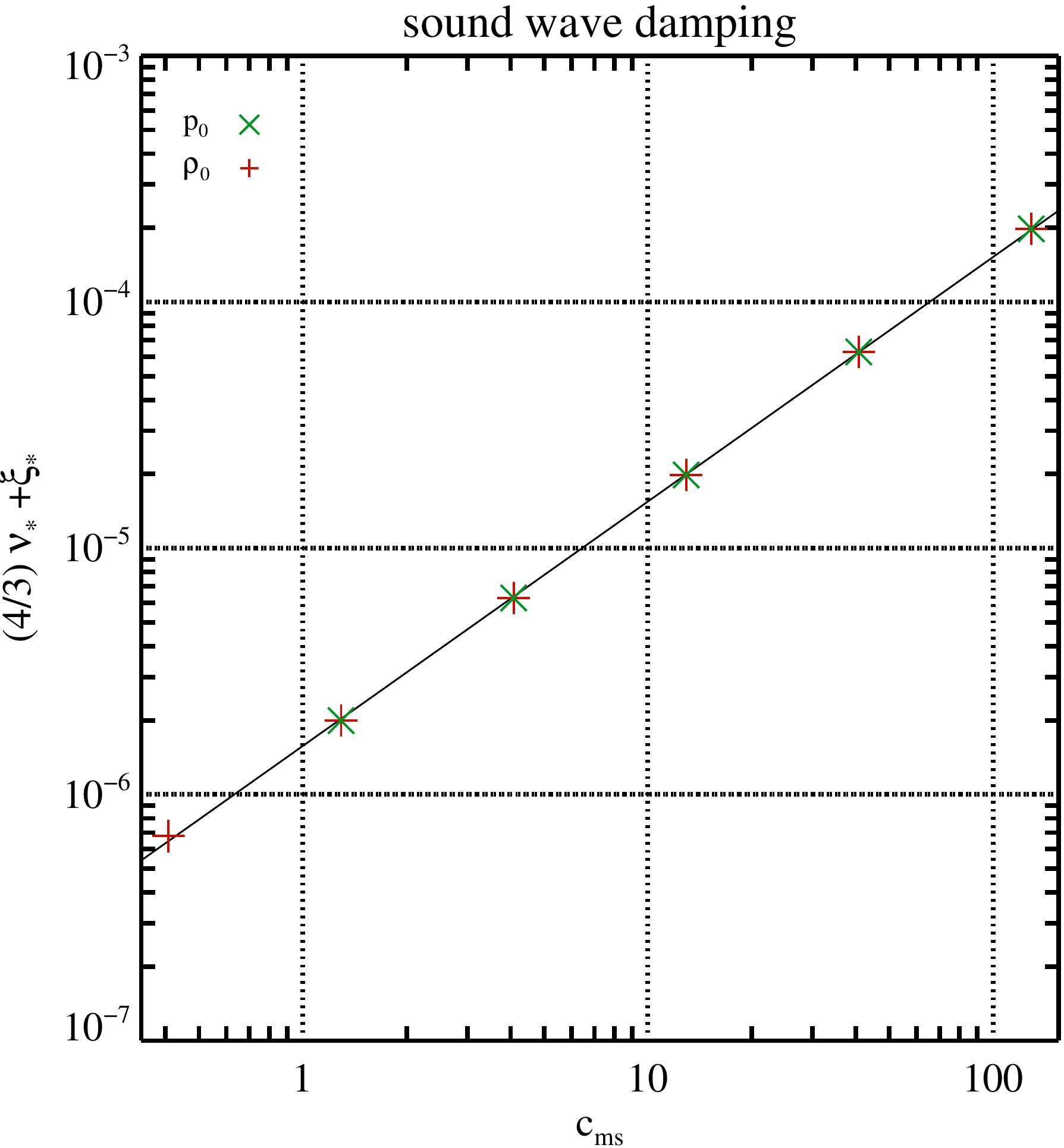} \phantom{MM}
 \includegraphics[width=0.45\textwidth]{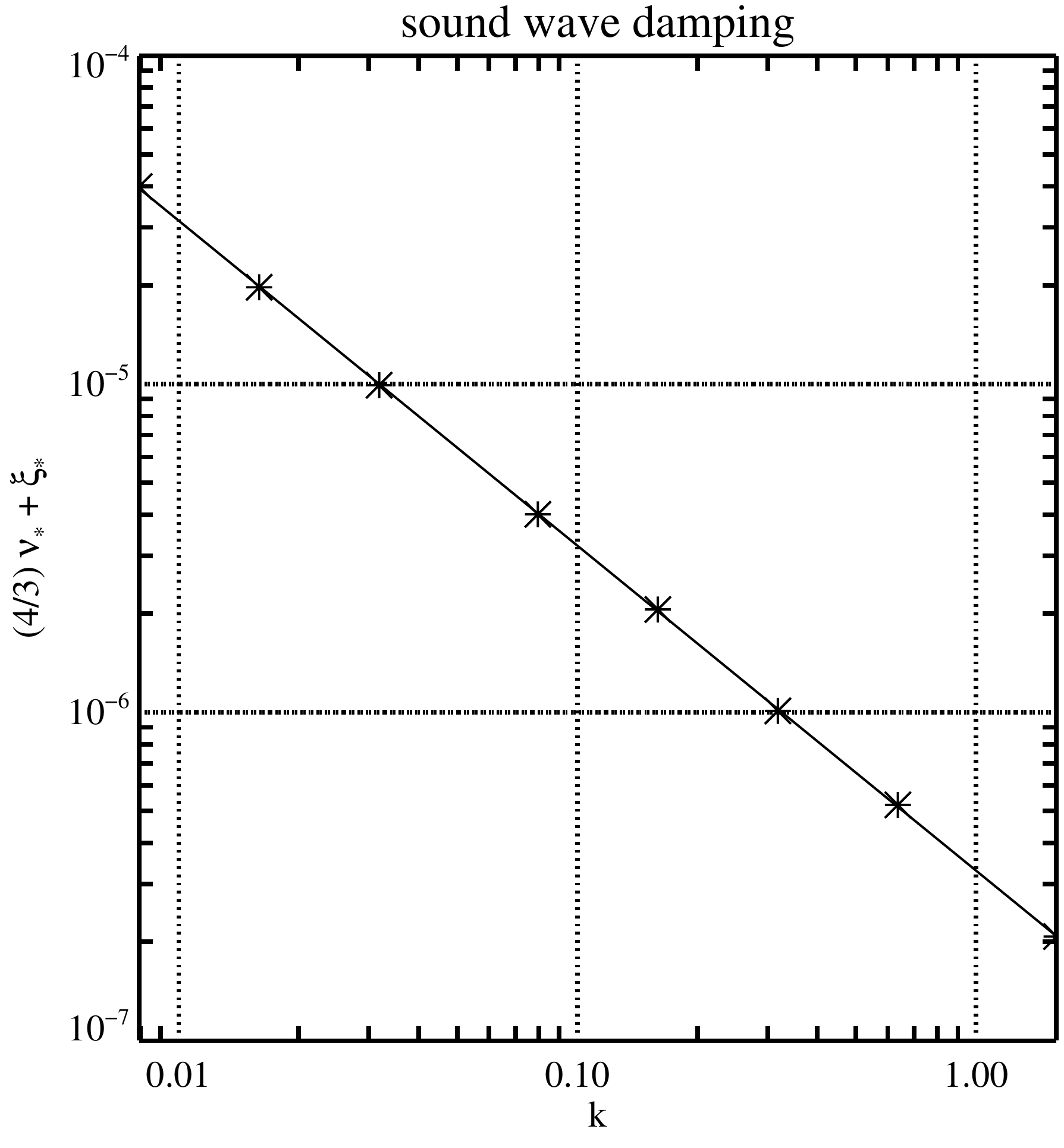}
 \caption{
   Numerical dissipation of sound waves. 
   \emph{Upper left panel:} dependence on the grid resolution 
   $\Delta x$ for different reconstruction schemes 
   (PL/\tr{\#S1}, MP5/\tr{\#S3}, MP7/\tr{\#S5}, and MP9/\tr{\#S6}), 
   the RK4 time integrator, and $\CFL = 0.01$.
   \emph{Upper right panel:} 
   dependence on the size of the time step 
   size $\Delta t$  for the RK3 (red; series \tr{\#S7} and \tr{\#S8}) and RK4 (blue; series \tr{\#S9}) time integrators
in simulations done with    the MP9  scheme (so that spatial reconstruction
errors are negligible). Different time steps were obtained by putting $\CFL = 0.5$ and varying the grid resolution (plus symbols, dotted lines),
or,  additionally for the RK3 integrator, by keeping resolution fixed (i.e.\ $64$ zones) and varying the CFL factor (diamonds, dashed line).
   \emph{Lower left panel:} dependence on the sound speed for two 
   simulations series in which we kept the background density $\rho_0$
   constant but varied the background pressure $p_0$ (\tr{\#cS1},
   green crosses), and in which we kept $p_0$ constant but varied 
   $\rho_0$ (\tr{\#cS2}, red plus signs).
   \emph{Lower right panel:} dependence on the wavenumber 
   $k = 2\pi/\lambda$, \ie on the box size (\tr{\#cS3}). 
   The results shown in the two lower panels were obtained using $32$
   zones, the MP5 scheme, the RK3 time integrator, and $\CFL = 0.01$.
   Straight lines are fits to the simulation results. }
  \label{fig:sound}
\end{figure*}
%------------------------------------------------------------------------------

With simulation series \tr{\#S1, \#S3, \#S5,} and \tr{\#S6} (\tabref{tab:waves};
upper left panel of \figref{fig:sound}), we investigated the influence
of the reconstruction scheme on the numerical dissipation.  To keep
the contribution of the time integration errors as small as possible,
we set $\CFL = 0.01$.  For every simulation, we measure the damping
rate $\dd_{\mathrm{s} \ast}$ from the decay of the kinetic energy,
from which we compute the numerical dissipation of the code according
to Eq.\,(\ref{eq:dds}) as
\begin{equation}
\frac{4}{3} \nu_{\ast} + \xi_{\ast} = \frac{2 }{k^2} \dd_{\mathrm{s} \ast},
\end{equation}
where the right hand side is given by the simulation results.  Thus,
in the case of sound waves, one cannot determine $\nu_{\ast}$ and
$\xi_{\ast}$ separately from the value of $\dd_{\mathrm{s} \ast}$, but
only a linear combination of both quantities.  For every simulation
series (\ie reconstruction scheme), we fit the function
\begin{equation}
  \ln\bigg( \frac{4} {3} \nu_{\ast} + \xi_{\ast} \bigg) 
   =  r \ln (\deltx x) + d,
\label{eq:bla0}
\end{equation}
where $r$ is the measured order of convergence of the scheme. From the
fit parameter
\begin{equation}
d = \ln \left[\left(  \frac{4}{3} \nn_{\nu}^{\deltx x} +
    \nn_{\xi}^{\deltx x}\right) \frac{\vv}{\LL^{r-1}}\right],
\end{equation}
we can compute
$\frac{4}{3} \nn_{\nu}^{\deltx x} + \nn_{\xi}^{\deltx x}$ if both the
characteristic speed and the length of the system are known. As we
will show below, for this test $\LL=\lambda=1$ ($\lambda=2\pi/k$
  being the wavelength) and $\vv=\cs = 1.291$.  The results presented
in \tabref{tab:waves} and the upper right panel of \figref{fig:sound}
show that all schemes have an exponent $r$ close to the theoretical
order $r_{\rm th}$ of the method.

The results of the simulation series \tr{\#S2, \#S3,} and \tr{\#S4}
(\tabref{tab:waves}) show that the LF, HLL and HLLD Riemann solvers
damp sound waves very similarly.

With the simulation series  \tr{\#S7}--\tr{\#S9}, we determine the
 contribution of the RK3  (\tr{\#S7} and \tr{\#S8}) and RK4 (\tr{\#S9}) time integrators to the numerical dissipation.
 To keep the contribution of the spatial discretisation errors as low
 as possible, we use the MP9 reconstruction.  We vary the timestep
 either by varying the grid resolution (keeping $\CFL =0.5$; series
 \tr{\#S7}  and \tr{\#S9}) or by varying the CFL factor (series \tr{\#S8}).  According to
 Eq.\,(\ref{eq:nu*cfl}), both approaches should be equivalent.  In both
 cases, the RK3 scheme performs very close to third order accuracy, whereas the order of the RK4 integrator is higher
 than expected.  We attribute the overperformance of RK4 in this test
 to the fact that it is not a TVD scheme since we have not computed the
 time-reversed operator $\tilde{L}$ as suggested in the \cite{SO88} and
 in \cite{Suresh_Huynh__1997__JCP__MP-schemes} for time integration
 schemes with order larger than three. However, the
 overperformance of RK4 in this test is very likely a fortunate coincidence (see
 Sec.\,\ref{sec:ms_waves}, where it is not the case).
  The estimators $\frac{4}{3} \nn_{\nu}^{\deltt t} + \nn_{\xi}^{\deltt t}$ are obtained by employing a fitting procedure analogous to the
 one for $\frac{4}{3} \nn_{\nu}^{\deltx x} + \nn_{\xi}^{\deltx x}$ (see \tabref{tab:waves} and the upper
 right panel of \figref{fig:sound}).

The natural characteristic speed of this flow problem should be the
sound speed ($\vv = \cs$). To test this hypothesis, we performed
simulations varying the background pressure (series \tr{\#cS1}) or the
density (\tr{\#cS2}) within the range given in \tabref{tab:length}.  The
results were fitted with the function
\begin{equation}
  \ln\bigg( \frac{4}{3}  \nu_{\ast} + \xi_{\ast} \bigg)  
  =  \alpha  \ln( \cs ) + d.
\label{eq:soundfit}
\end{equation}
From the fit, we obtain the value of $\alpha$, expecting $\alpha = 1$,
and from the offset $d$, we determine
$\nn_{\rm tot}^{\deltx x} = \nn_{\rm tot}^{\deltx x} = (4/3)
\nn_{\nu}^{\deltx x} + \nn_{\xi}^{\deltx x}$
(\tabref{tab:length}). The table and the lower left panel of
\figref{fig:sound} clearly show that the sound speed is indeed the
characteristic speed of the system.

To determine the characteristic length of the system (the natural
candidate being the wavelength $\lambda$), we performed the simulation
series \tr{\#cS3} varying the wavelength $\lambda$ (and the size of the
simulation domain accordingly, \ie $L = \lambda$). The results were
fitted with the function
\begin{equation}
   \ln\bigg( \frac{4}{3} \nu_{\ast} +  \xi_{\ast}\bigg)  = \alpha \ln (\lambda) + d,
\label{eq:lambda_fit_sound}
\end{equation}
expecting again $\alpha = 1$.  Table~\ref{tab:length} and the lower
right panel of \figref{fig:sound} confirm our hypothesis. The figure
shows that the numerical viscosity term is proportional to $k^{-1}$,
\ie $D_{s*}/k^2 \propto k^{-1}$ (see Eq.\,\ref{eq:dds}).

%%%%%%%%%%%%%%%%%%%%%%%%%%%%%%%%%%%%%%%%%%%%%%%%%%%%%%%%%%%%%%%%%%%%%%

\subsubsection{\alf Waves}
\label{sec:alfven_waves}

With the help of \alf wave simulations, we determine a linear
combination of the numerical shear viscosity and resistivity of the
code. We set the background magnetic field and density to
$b_{0x} = \rho_0 = 1 $, the pressure to $p_0 = 2 \times 10^{-3} $, and
the transversal velocity to $\velo_{0y}=\velo_0=0$.  We imposed a
perturbation of the form
\begin{equation}
 b_{1y}(x,0) = \epsilon \sin (k x), 
\end{equation}
\begin{equation}
\velo_{1y}(x,0) = - \frac{b_{y1}}{ \sqrt{ \rho_0} }.
\end{equation}
In ideal MHD, an \alf wave propagates with a constant amplitude at the
\alf speed $\ca = b_{0x} / \sqrt{\rho_0}$.  In the presence of
viscosity and resistivity, the wave amplitude decreases with time.  In
the weak damping approximation, \ie for
$k^4 ( \nu + \eta )^2 /( 4 \ca^2 ) \ll 1$, the velocity evolution
reads \citep[for the derivation, see][]{Campos}
\begin{equation}
 \velo_y(x,t) = \velo_0 e^{ -\dda t } e^{i k ( x \mp \ca t)},
\label{eq:alf_vy_final}
\end{equation}
where the \alf damping rate is defined as
\begin{equation}
 \dda = \frac{k^2}{2} ( \eta + \nu ) .
\label{eq:dda}
\end{equation}
We verified Eq.\ (\ref{eq:dda}) with the help of numerical
simulations, and also checked that the bulk viscosity does not
influence the damping coefficient \citep[see][for details]{Rembiasz}.

%------------------------------------------------------------------------------
\begin{figure*}
  \centering
  \includegraphics[width=0.45\textwidth]{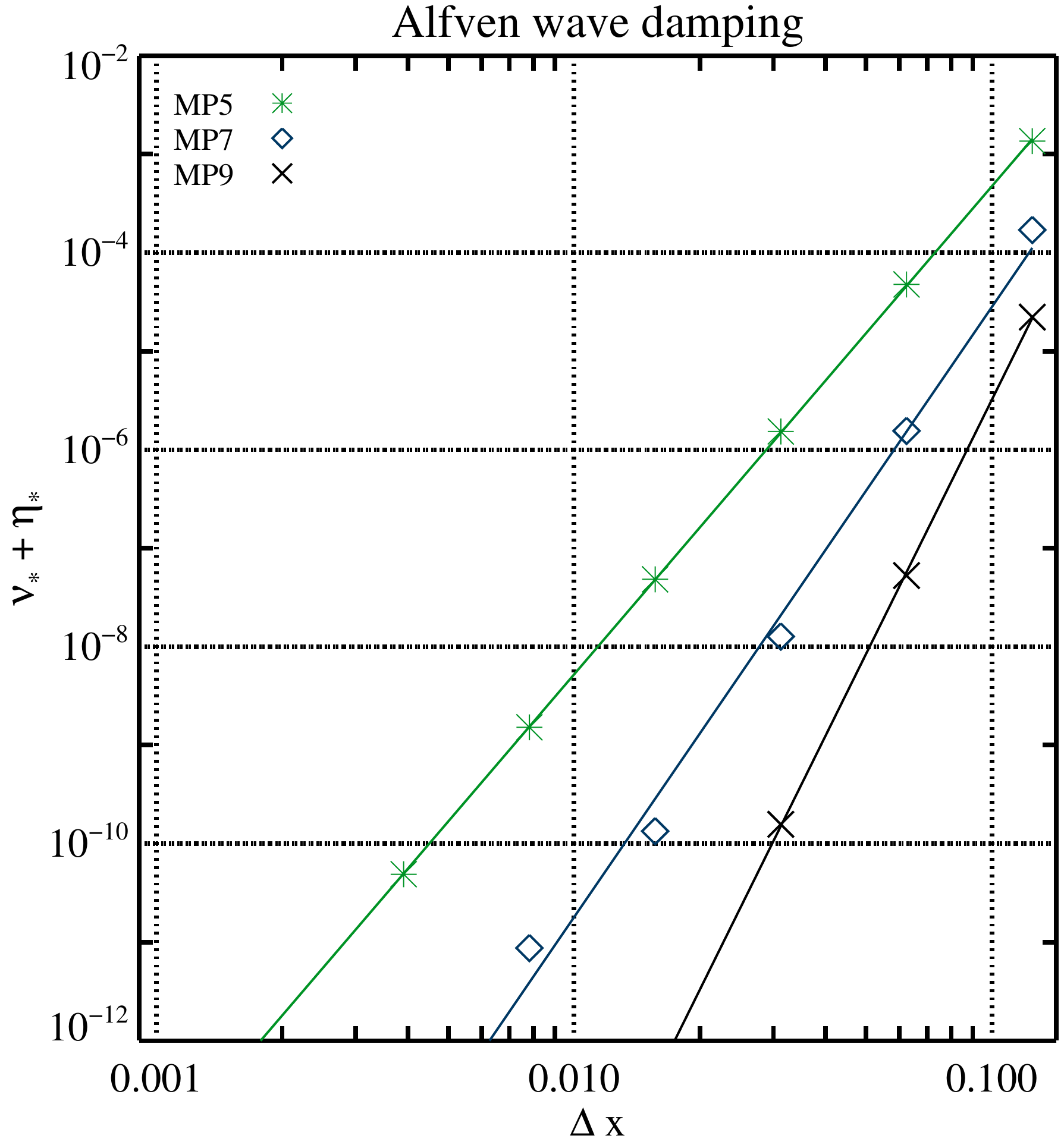} \phantom{MM}
  \includegraphics[width=0.45\textwidth]{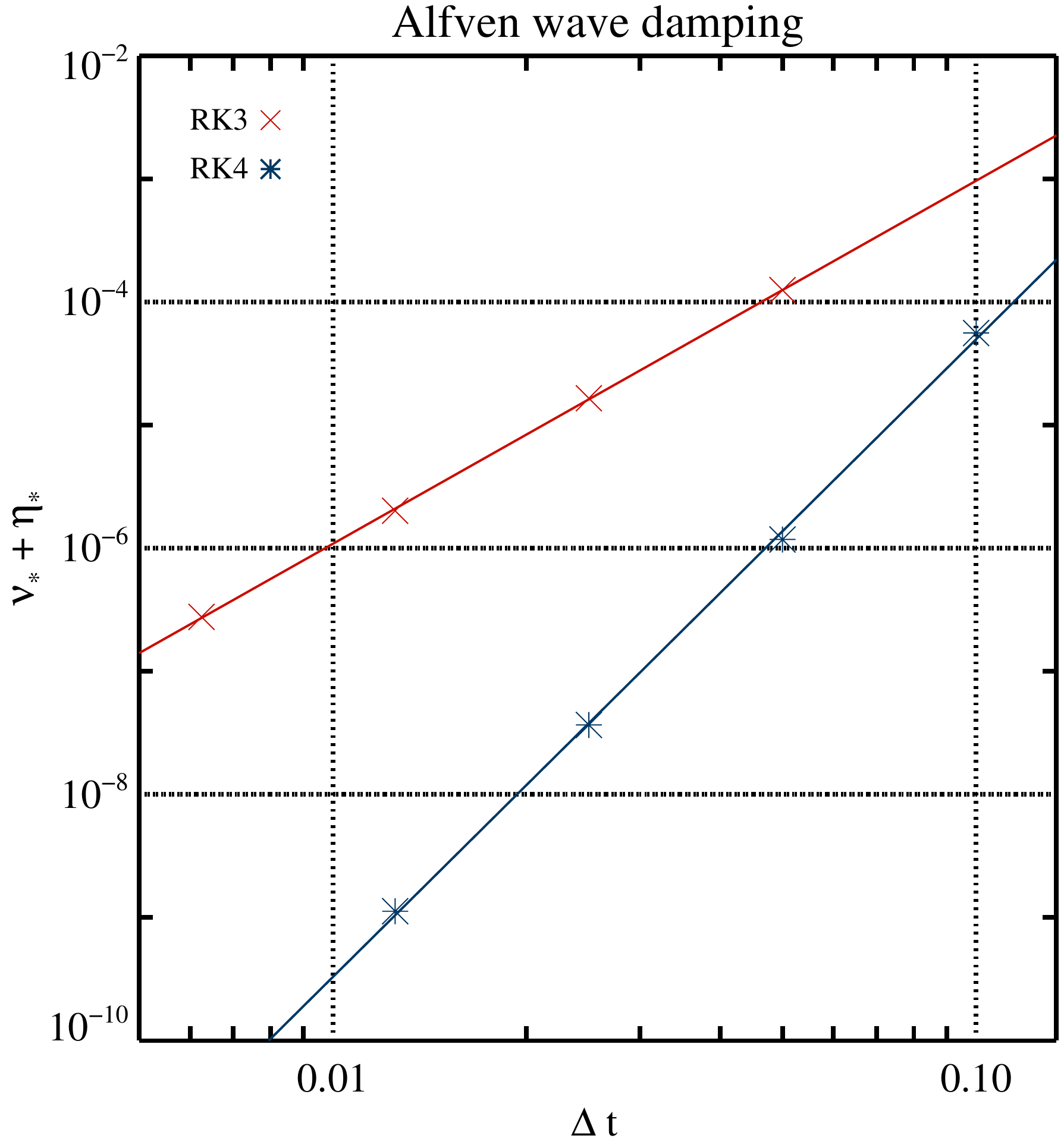}
  \includegraphics[width=0.45\textwidth]{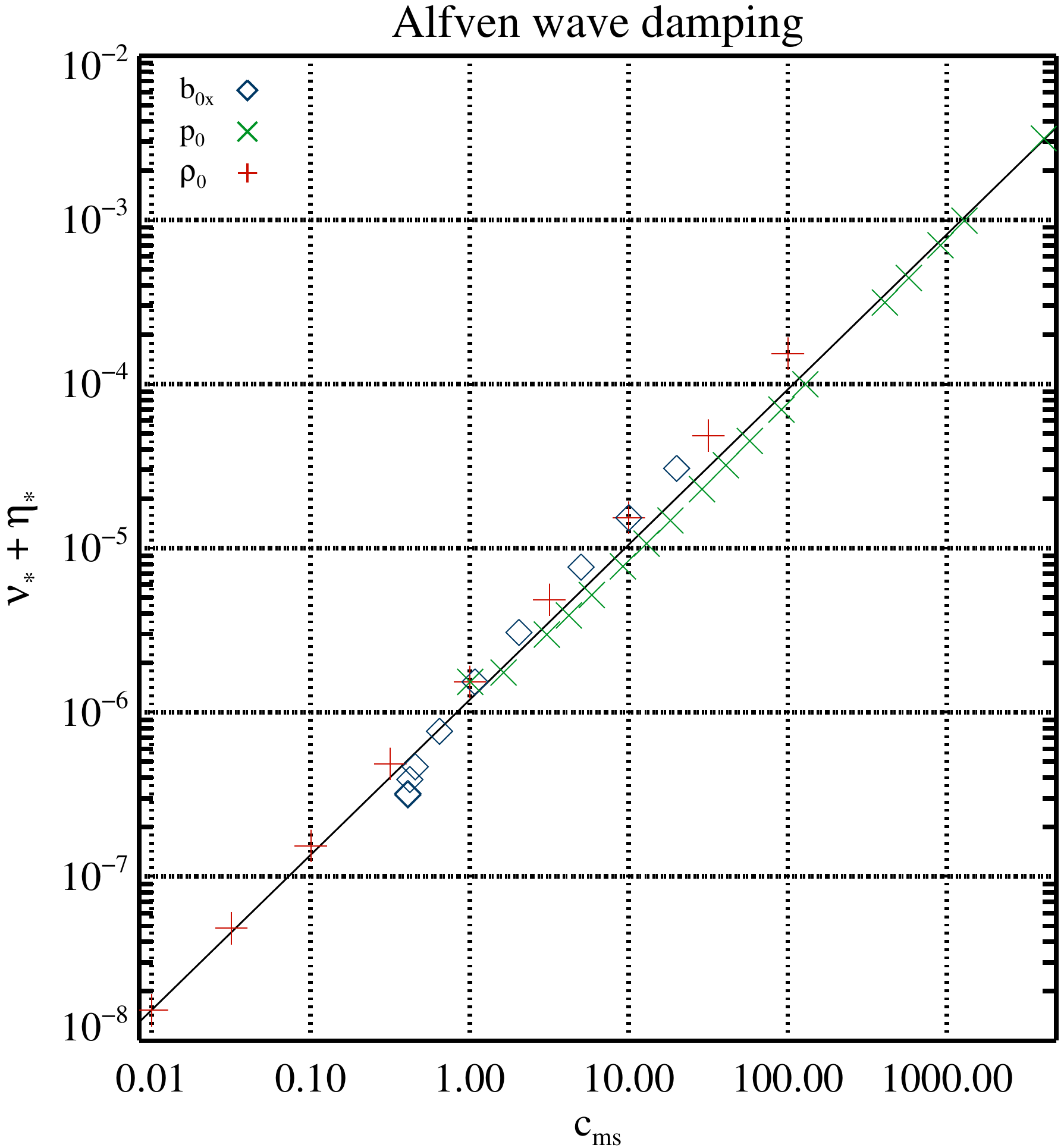} \phantom{MM}
  \includegraphics[width=0.45\textwidth]{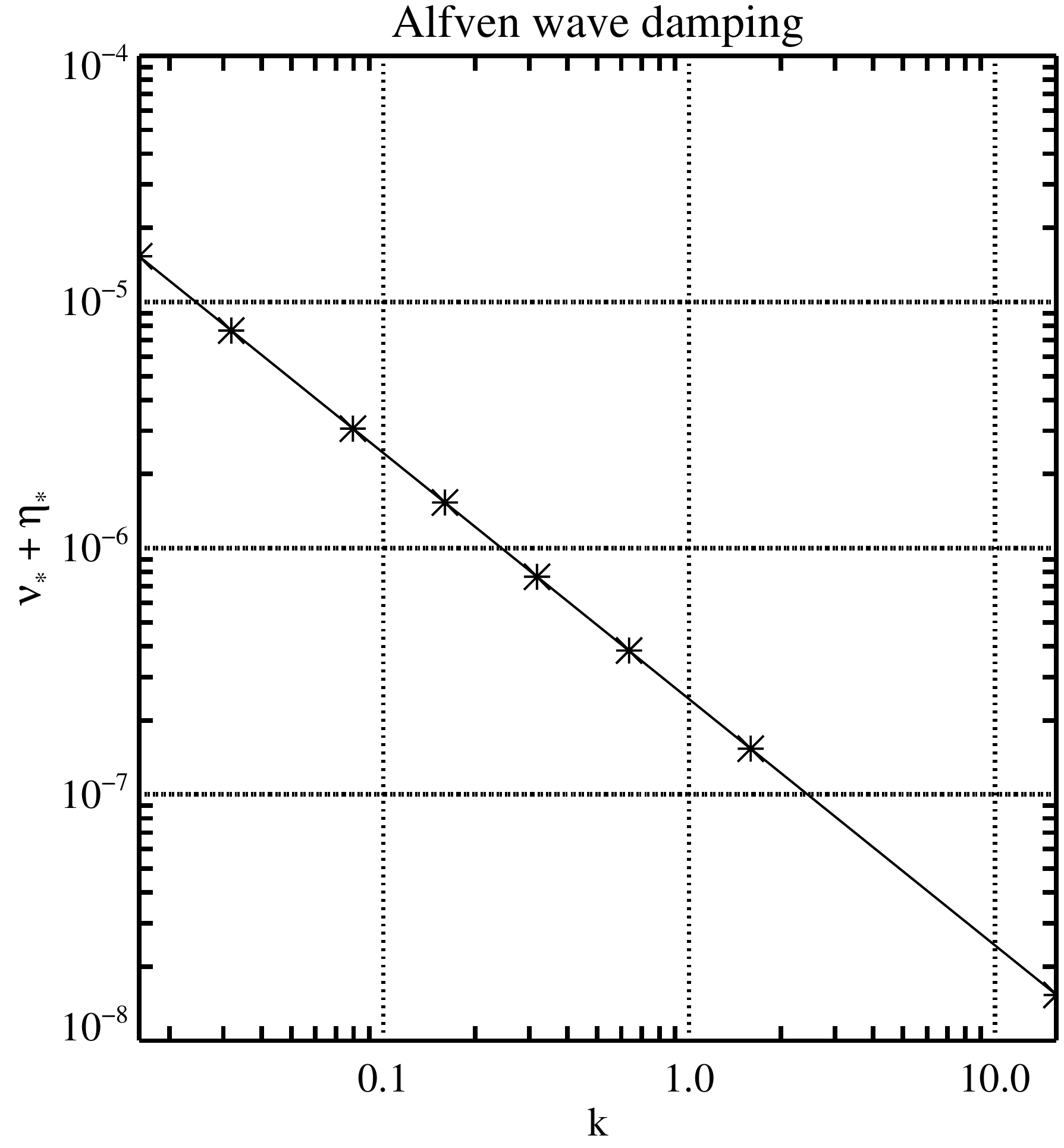}
  \caption{
    Numerical dissipation of \alf waves. 
    \emph{Upper left panel:} dependence on the grid resolution 
    $\Delta x$ for three different reconstruction schemes 
    (MP5/\tr{\#A2}, MP7/\tr{\#A3}, and MP9/\tr{\#A4})  
    using RK4 and $\CFL = 0.01$.
    \emph{Upper right panel:} dependence on the time step size 
    $\Delta t$ (changing the grid resolution but keeping $\CFL = 0.8$) 
    for two different time integrators (\tr{\#A6}, and \tr{\#A7})
    using the HLL Riemann solver and MP9.
    \emph{Lower left panel:} dependence on the fast magnetosonic speed
    for three simulation series varying the background magnetic field
    strength $b_{0x}$ (\tr{\#cA1}, blue diamonds), the background pressure
    $p_0$ (\tr{\#cA2}, green crosses), and the background density $\rho_0$
    (\tr{\#cA3}, red plus signs) keeping all other parameters
    constant. 
    \emph{Lower right panel:} dependence on the wavenumber 
    $k = 2\pi/\lambda$, \ie on the box size \tr{\#cA4}. 
    Straight lines are fits to the simulation results. }
\label{fig:alf}
\end{figure*}
%--------------------------------------------------------------------------------

In the simulation series \tr{\#A2, \#A4,} and \tr{\#A5} (see \tabref{tab:waves})
we compared the influence of the MP5, MP7, and MP9 reconstruction
schemes on the numerical shear viscosity $ \nu_{\ast} $ and
resistivity $\eta_{\ast}$.  For every simulation, we measured the
decrease of the kinetic energy, from which we determined a linear
combination of the numerical shear viscosity and resistivity
\begin{equation}
  \nu_{\ast} + \eta_{\ast} = \frac{2 }{k^2} \dd_{\mathrm{A}\ast}.
\label{eq:alf_lin}
\end{equation}
The simulation results are fitted with the function 
\begin{equation}
  \ln (  \nu_{\ast} + \eta_{\ast} ) = r \ln ( \deltx x) + d,
  \label{eq:alffit}
\end{equation}  
where $r$ is the numerically measured order of accuracy of
the reconstruction scheme. From the fit parameter $d$ and
Eqs.\,(\ref{eq:nu*}), (\ref{eq:eta*}), and (\ref{eq:alf_lin}) we
determined $\nn_{\nu}^{\deltx x} + \nn_{\eta}^{\deltx x}$.
\footnote{The characteristic velocity and length of the system was set
  to $\vv = 1$ and $\LL=\lambda$, respectively.  See later in this
  subsection for an extended discussion.}
Table~\ref{tab:waves} and the upper left panel of \figref{fig:alf} show
that all methods have an order of convergence close to the theoretical
expectation.

According to the results of the simulation series \tr{\#A1, \#A2,} and \tr{\#A3}
(\tabref{tab:waves}), the numerical dissipation of the LF, HLL, and
HLLD Riemann solvers are also very similar for \alf waves.

With the simulation series \tr{\#A6} and \tr{\#A7} (upper right panel of
\figref{fig:alf}), we assessed the contribution to the numerical
dissipation of the RK3 and RK4 time integrators, respectively.  We set
$\CFL = 0.8$ and changed the timestep by varying the grid resolution.
The results presented in \tabref{tab:waves} and the upper right panel
of \figref{fig:alf} show that the RK3 time integrator performs at its
theoretical order, whereas the order of the RK4 integrator is  once again (like in the sound wave tests)  higher
than expected.  

The characteristic velocity $\VV$ for the \alf wave test problem can
be inferred from simulation series \tr{\#cA1, \#cA2,} and \tr{\#cA3}, in which
we varied the magnetic field, pressure, and density, respectively. We
find that the logarithm of the numerical dissipation determined from
these simulation data can be fitted (see lower left panel of
\figref{fig:alf}) by the function
\begin{equation}
  \ln ( \nu_{\ast} + \eta_{\ast} ) = \alpha \ln (\cmag) + d,
  \label{eq:magsonfit}
\end{equation}
where $\alpha$ and $d$ are fitting parameters, and $\cmag$ is the fast
magnetosonic speed, which is is defined as
\begin{equation}
  c_{\mathrm{ms}} = \sqrt{ \frac{1}{2} \left(  \ca^2 + \cs^2 + \sqrt{ (\ca^2
      + \cs^2 )^2 - 4  \ca^2 \cs^2 \cos^2 \theta } \right)},
\label{eq:ms_full}
\end{equation}
where $\theta$ is the angle between the perturbation wave vector and
the background magnetic field.  For a wavevector parallel to the
background field ($\theta=0$)
\begin{equation}
  \cmag = \mathrm{max}\{\ca, \cs \}.
\end{equation}
The values of the fitting parameter $\alpha$, which are given in
\tabref{tab:length}, confirm that the characteristic velocity $\VV$ is
the fast magnetosonic speed (not the \alf speed as one could have
presumed), both in the flow regime where $\VV$ is dominated by the
\alf speed and the sound speed.

The simulation series \tr{\#cA4} (\tabref{tab:length}) shows that the
characteristic length of the \alf wave simulations is, as for the
sound wave test, the wavelength, and that the numerical dissipation
can be fitted by (see \tabref{tab:length} and the lower right panel of
Fig.\,\ref{fig:alf})
\begin{equation}
  \ln( \nu_{\ast} + \eta_{\ast} ) = \alpha \ln{\lambda} + d\,.
\label{eq:lambda_fit_alf}
\end{equation}

\begin{figure}%
  \centering
  \includegraphics[width=0.45\textwidth]{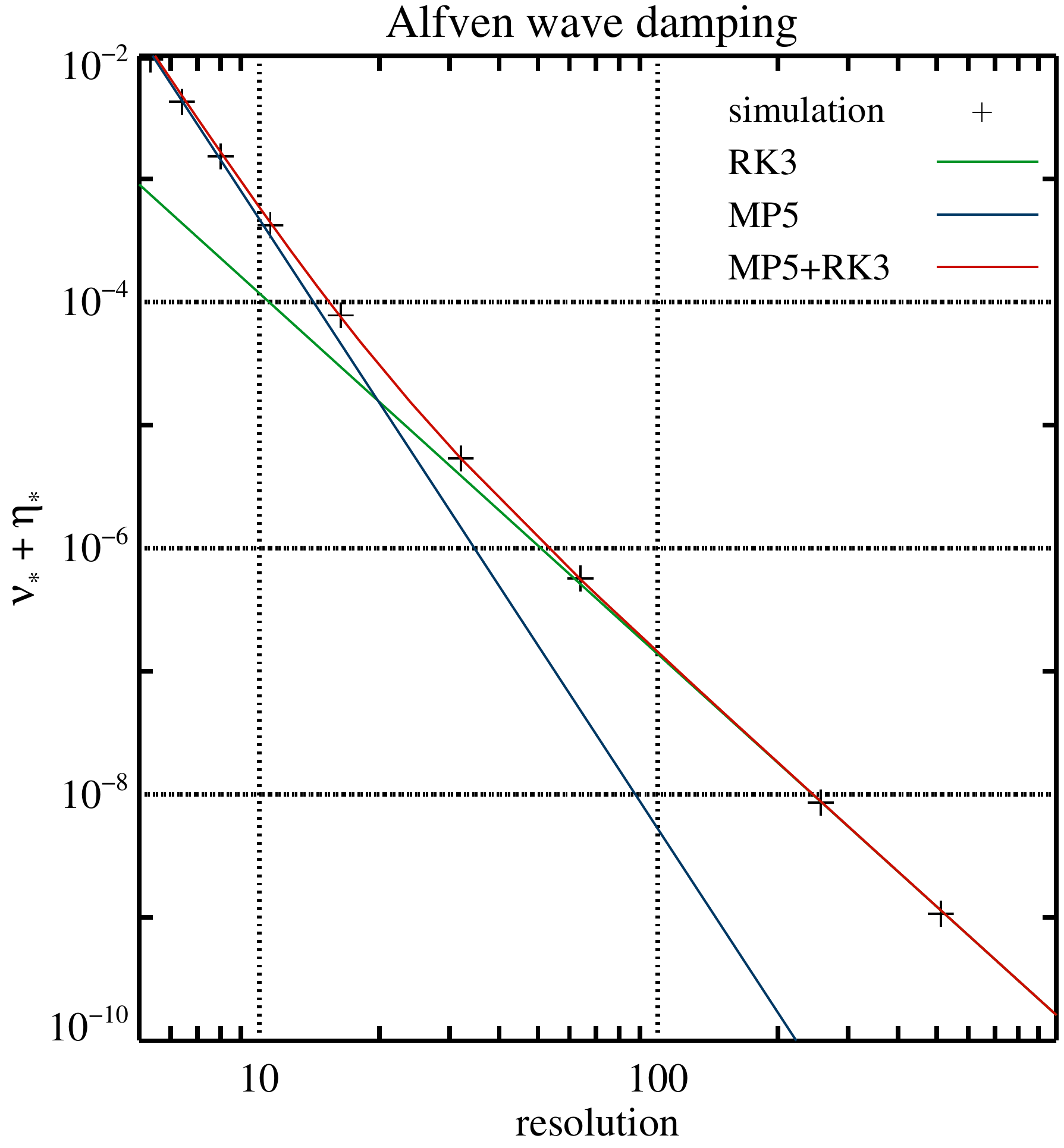}
  \caption{Numerical dissipation as a function of grid
    resolution for the MP5 reconstruction scheme and the RK3 time
    integrator with $\CFL =0.5$. The simulation results are marked
    with black plus signs.  The straight lines depict the predicted
    numerical dissipation of the RK3 time integrator (green), the MP5
    scheme (blue), and the sum of both types of numerical dissipation
    (red). For a grid resolution $\le 16$ zones, \alf waves are mainly
    damped by spatial discretisation errors, and for $\ge 128$ zones
    by time integration errors. In the intermediate regime
    ($16 \ldots 128$ zones), both types of errors add linearly (like
    proper scalars).}
  \label{fig:alf_add}
\end{figure}

Finally, to investigate whether the errors resulting from spatial
discretisation and time integration are additive, we performed
simulations \tr{\#A8}, in which both types of errors should contribute
non-negligibly to the numerical dissipation.  Figure \ref{fig:alf_add}
shows the simulation results together with the expected numerical
dissipation of the RK3 integrator (green), the MP5 scheme (blue), and
the sum of both contributions (red).  As the figure shows, the errors
add linearly.

%----------------------------------------------------------------------------
%%%
%----------------------------------------------------------------------------

%%%%%%%%%%%%%%%%%%%%%%%%%%%%%%%%%%%%%%%%%%%%%%%%%%%%%%%%%%%%%%%%%%%%%%

\subsubsection{Magnetosonic Waves} 
\label{sec:ms_waves}

%----------------------------------------------------------------------------
\begin{figure*}
  \centering
  \includegraphics[width=0.33\textwidth]{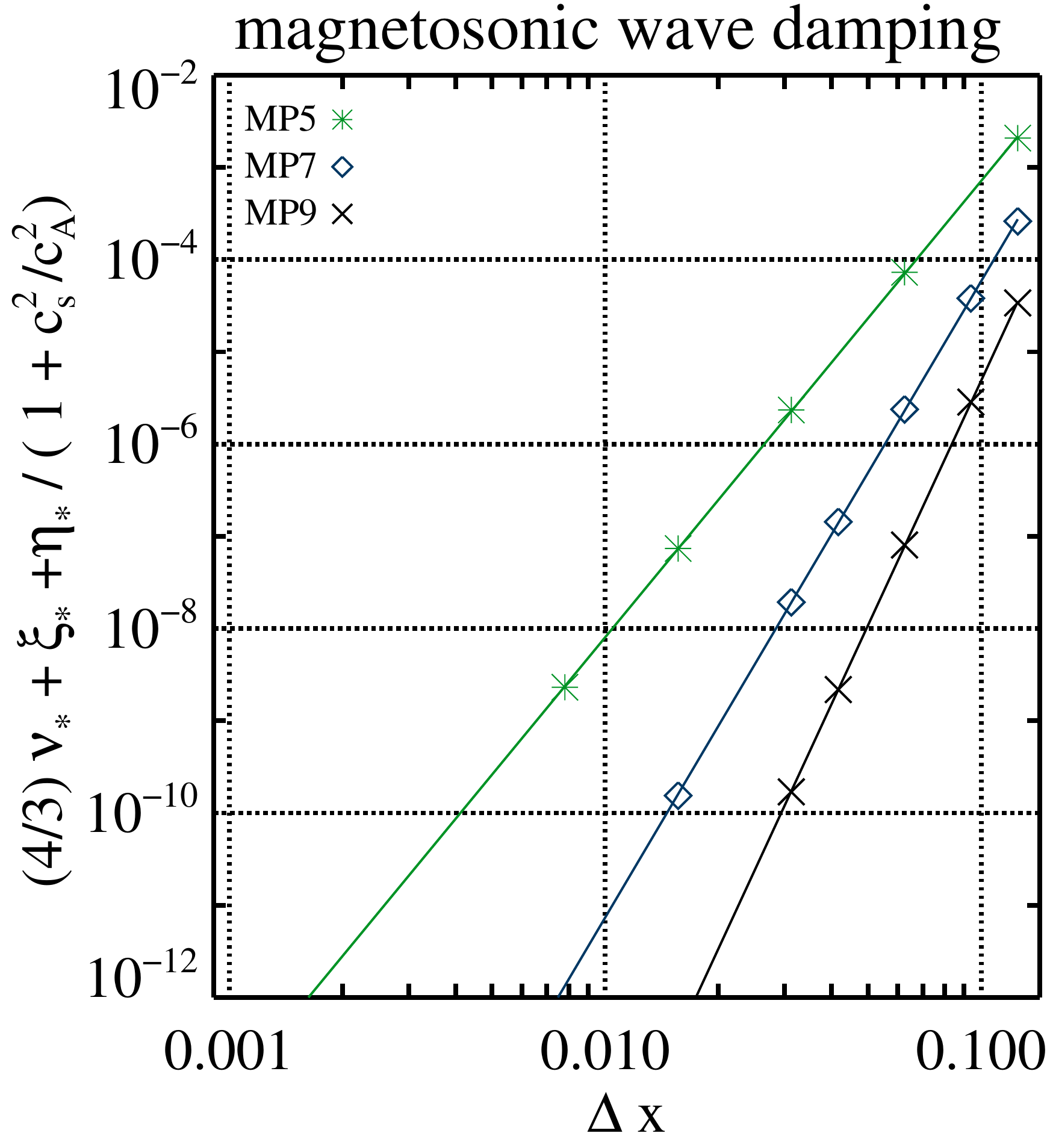} 
  \includegraphics[width=0.33\textwidth]{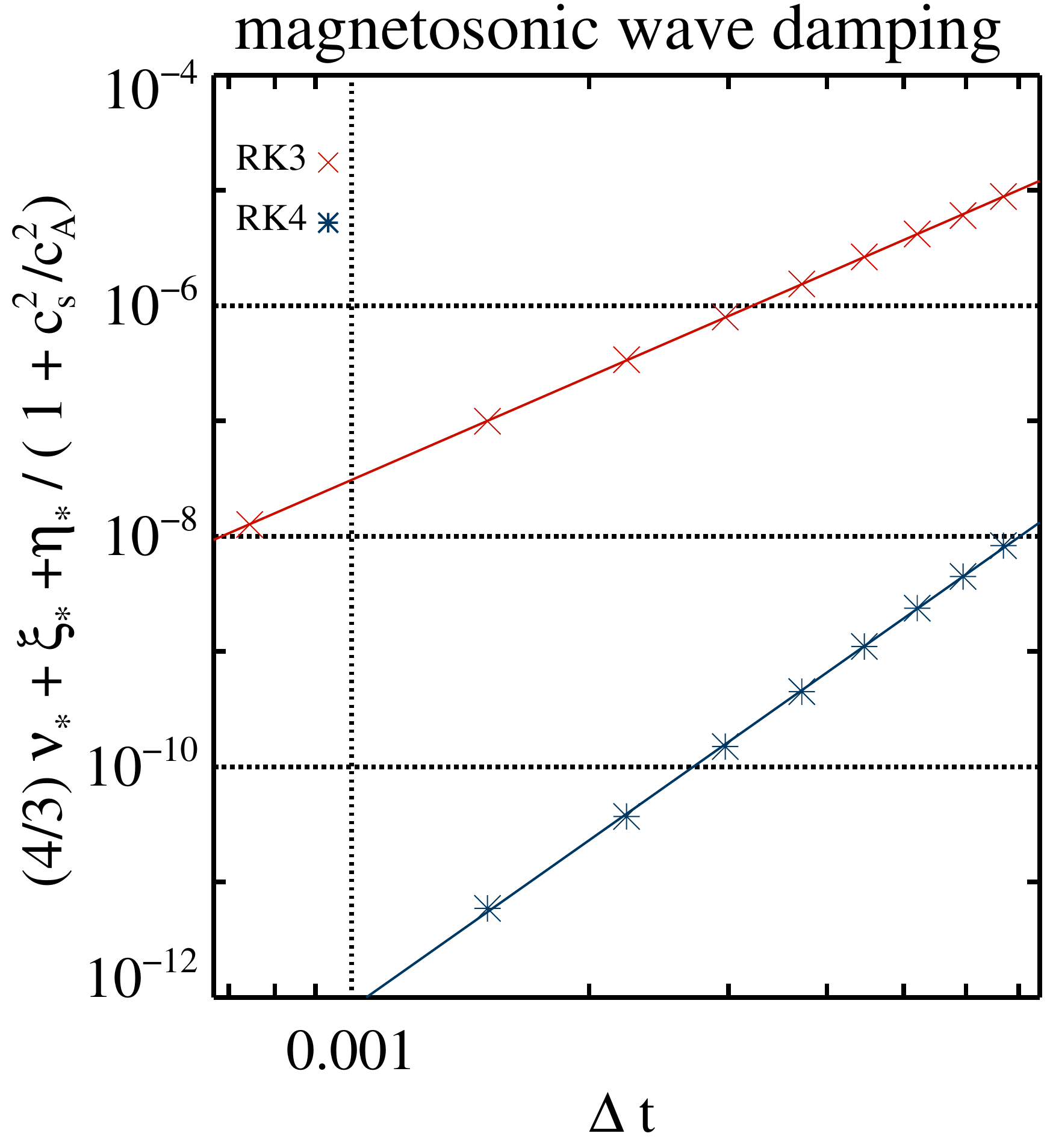} 
  \includegraphics[width=0.33\textwidth]{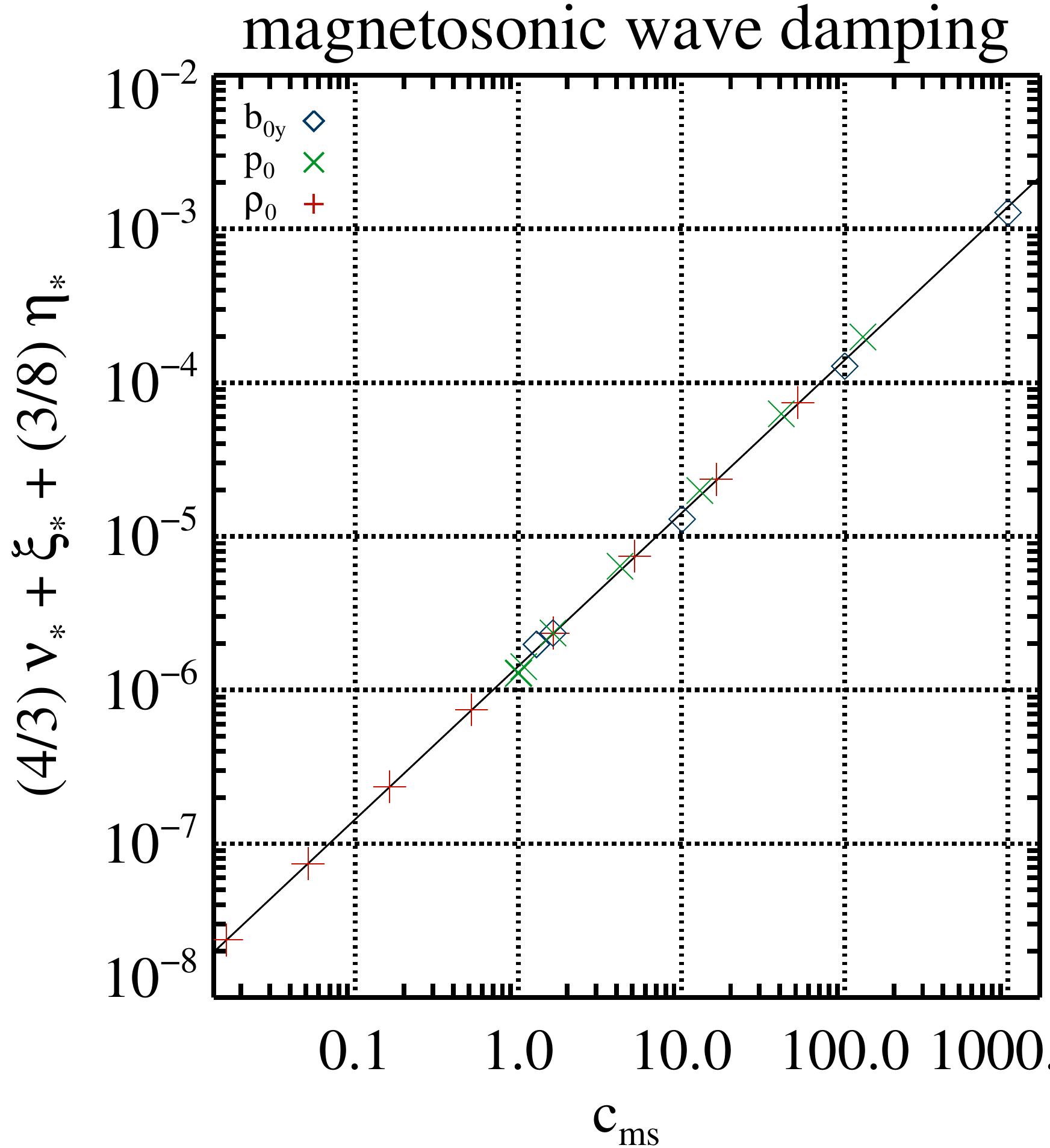}
  \caption{
    Numerical dissipation of magnetosonic waves.  
    \emph{Left panel:} dependence on the grid resolution $\Delta x$ 
    for three different reconstruction schemes 
    (MP5/\tr{\#MS1}, MP7/\tr{\#MS2}, and MP9/\tr{\#MS3})  
    using a HLL Riemann solver, RK4, and $\CFL=0.01$.  
    \emph{Middle panel:} dependence on the time step size $\Delta t$ 
    for two different time integrators (RK3/\tr{\#MS4} and RK4/\tr{\#MS5)}
    using the HLL Riemann solver, MP9, and a grid resolution of 
    64 zones.
    \emph{Right panel:} dependence on the fast magnetosonic speed for 
    simulations varying the background magnetic field strength $b_{0y}$ 
    (\tr{\#cMS1}, blue diamonds), the background pressure $p_0$
    (\tr{\#cMS2}, green crosses), and the background density $\rho_0$
    (\tr{\#cMS3}, red plus signs) keeping all other parameters
    constant. 
    Straight lines of are fits to the simulation results.}
  \label{fig:mag}
\end{figure*}
%----------------------------------------------------------------------------

From the simulations of magnetosonic waves, we determine the numerical
resistivity and viscosity of the \aenus\, code.  If not otherwise
stated, the background pressure, density, and magnetic field strength
are set to $p_0 = \rho_0 = b_{0y} = 1$ and
${\bf b_0} = b_{0y} {\bf \hat{y}}$, respectively.  We perturb the
background by a magnetosonic wave of the form
\begin{align}
\label{eq:vx_1d}
\velo_{x1}(x,0) &= \epsilon  \sin(k_x  x), \\
\velo_{y1} (x,0)&= \velo_{x1} \frac{k_x^2 b_{0x} b_{0y} }{ b_{0x}^2 k_x^2 -
                   \rho_0   \omega^2},  \\
b_{y1} (x,0)    &= \velo_{x1} \frac{k_x b_{0y} \omega \rho_0}{ \rho_0
                   \omega^2 - b_{0x}^2 k_x^2 }, \\
\rho_1 (x,0)    &= \velo_{x1} \frac{ k_x \rho_0}{ \omega}, \\
e_1 (x,0)       &= \velo_{x1} \frac{ k_x p_0 \Gamma}{\omega(\Gamma -1)},
\end{align}
where $e_1$ is the total specific energy of the wave. The velocity
amplitude $\epsilon = 10^{-5}$, and the wave's angular frequency
$\omega$ is given by
\begin{equation}
  \omega^2 = k^2 c_{\rm ms}^2.
\label{eq:cms_original}
\end{equation}
For $\theta =\pi/2$ (a value chosen in almost all simulations) the
magnetosonic speed reads (see Eq.\,\ref{eq:ms_full})
\begin{equation}
  \cmag = \frac{\omega}{k} = \sqrt{ \frac{ {\bf b_0}^2 +
                \Gamma p_0}{\rho_0}}.
\end{equation}
In the presence of viscosity or resistivity, the wave will be damped
with time, \ie the $x$ component of the wave velocity will decrease as
\begin{equation}
  \velo_x(x,t) = \epsilon\,e^{ - \ddms t} e^{i k ( x \mp \cmag t) },
\end{equation}
where the damping coefficient for a fast magnetosonic wave propagating
in the direction  perpendicular to the background magnetic field
is \citep[for the derivation, see][]{Campos}
\begin{equation}
\ddms =  \frac{k^2}{2} \left( \frac{4}{3} \nu + \xi + 
                              \frac{\eta} {1 + \cs^2 / \ca^2} \right) .
\label{eq:ddms}
\end{equation}
We verified this equation numerically \citep[see][for
details]{Rembiasz}.

We also performed simulations \tr{\#MS1, \#MS2,} and \tr{\#MS3}
(\tabref{tab:waves}) to investigate the influence of the MP5, MP7, and
MP9 reconstruction schemes.  From the measured kinetic energy damping,
we determined a linear combination of the numerical resistivity, shear
viscosity, and bulk viscosity, \ie
\begin{equation}
  \frac{4}{3} \nu_{\ast} + \xi_{\ast} +  \frac{3}{8} \eta_{\ast} =
  \frac{2}{k^2}  \dd_{{\rm ms}\ast}.
\label{eq:mag_ln}
\end{equation}
We fitted the simulation results with the function
\begin{equation}
  \ln \left( \frac{4}{3} \nu_{\ast} + \xi_{\ast} + \frac{3}{8} \eta_{\ast} 
      \right) = r\ln ( \deltx x) + d,
\end{equation} 
where the fit parameter $r$ is the numerically measured order of the
reconstruction scheme.  From the fit parameter $d$, and with Eqs.\
(\ref{eq:nu*}), (\ref{eq:xi*}), (\ref{eq:eta*}), and
(\ref{eq:mag_ln}), we determined the combination of coefficients
$ \frac{4}{3} \nn_{\nu}^{\deltx x} + \nn_{\xi}^{\deltx x} +
\frac{8}{3} \nn_{\eta}^{\deltx x}$
(see \tabref{tab:waves} and left panel of \figref{fig:mag}).

Using simulation series \tr{\#MS4} and \tr{\#MS5}, we studied the contribution
of the RK3 and RK4 time integrators to the numerical dissipation (see
\tabref{tab:waves} and middle panel of \figref{fig:mag}).  We find
that the RK4 integrator performs at a higher order than theoretically
expected.  Again, we point out that probably due to the non-TVD
preserving property of our implementation of RK4, it overperformes in
this test (see Sect.\ \ref{sec:alfven_waves}).

To determine the characteristic speed, we performed simulation series
\tr{\#cMS1, \#cMS2} and \tr{\#cMS3} varying background magnetic field strength,
pressure, and density, respectively (\tabref{tab:length}).  Hardly
surprisingly, the characteristic speed is the fast magnetosonic speed
(see bottom panel of \figref{fig:mag}), which is confirmed
quantitatively with the help of the fit function
\begin{equation}
  \ln \left(  \frac{4}{3} \nu_{\ast} + \xi_{\ast} +  \frac{3}{8}
  \eta_{\ast} \right) = \alpha \ln (\cmag) + d.
\label{eq:ln_cmag}
\end{equation}
As expected, the fit (see \tabref{tab:length}) is consistent with
$\alpha=1$ within the measurement errors. The value of $d$ can be used
to estimate
$ \frac{4}{3} \nn_{\nu}^{\deltx x} + \nn_{\xi}^{\deltx x} +
\frac{3}{8}\nn_{\eta}^{\deltx x} $.
In the asymptotic regime $b_0 \ll p_0$, the numerical damping is
independent of the magnetic field strength, while it is proportional
to the field strength for $b_0 \gg p_0$.

%%%%%%%%%%%%%%%%%%%%%%%%%%%%%%%%%%%%%%%%%%%%%%%%%%%%%%%%%%%%%%%%%%%%%%

\subsubsection{Estimation of Numerical Resistivity and Viscosity}

So far, we measured the numerical damping for three wave types
separately.  For each type of a wave, the damping coefficient depends
on a linear combination of the resistivity, shear viscosity and bulk
viscosity (see Eqs.\,\ref{eq:dds}, \ref{eq:dda}, and \ref{eq:ddms}).
This gives a system of three linearly independent equations with three
unknowns, which has a unique solution.
 
%----------------------------------------------------------------------------
\begin{figure*}
  \centering
 \includegraphics[width=0.33\textwidth]{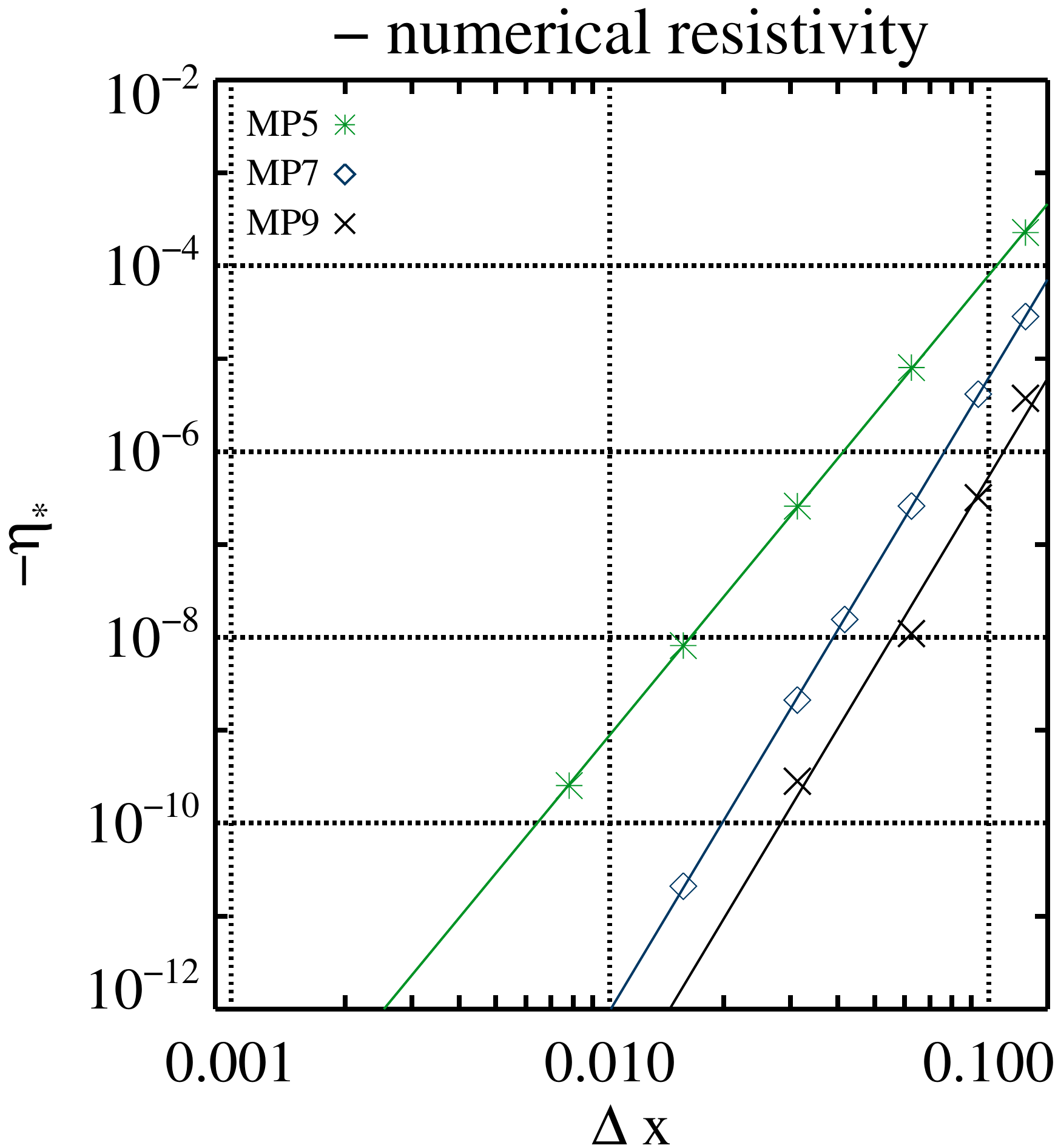} 
 \includegraphics[width=0.33\textwidth]{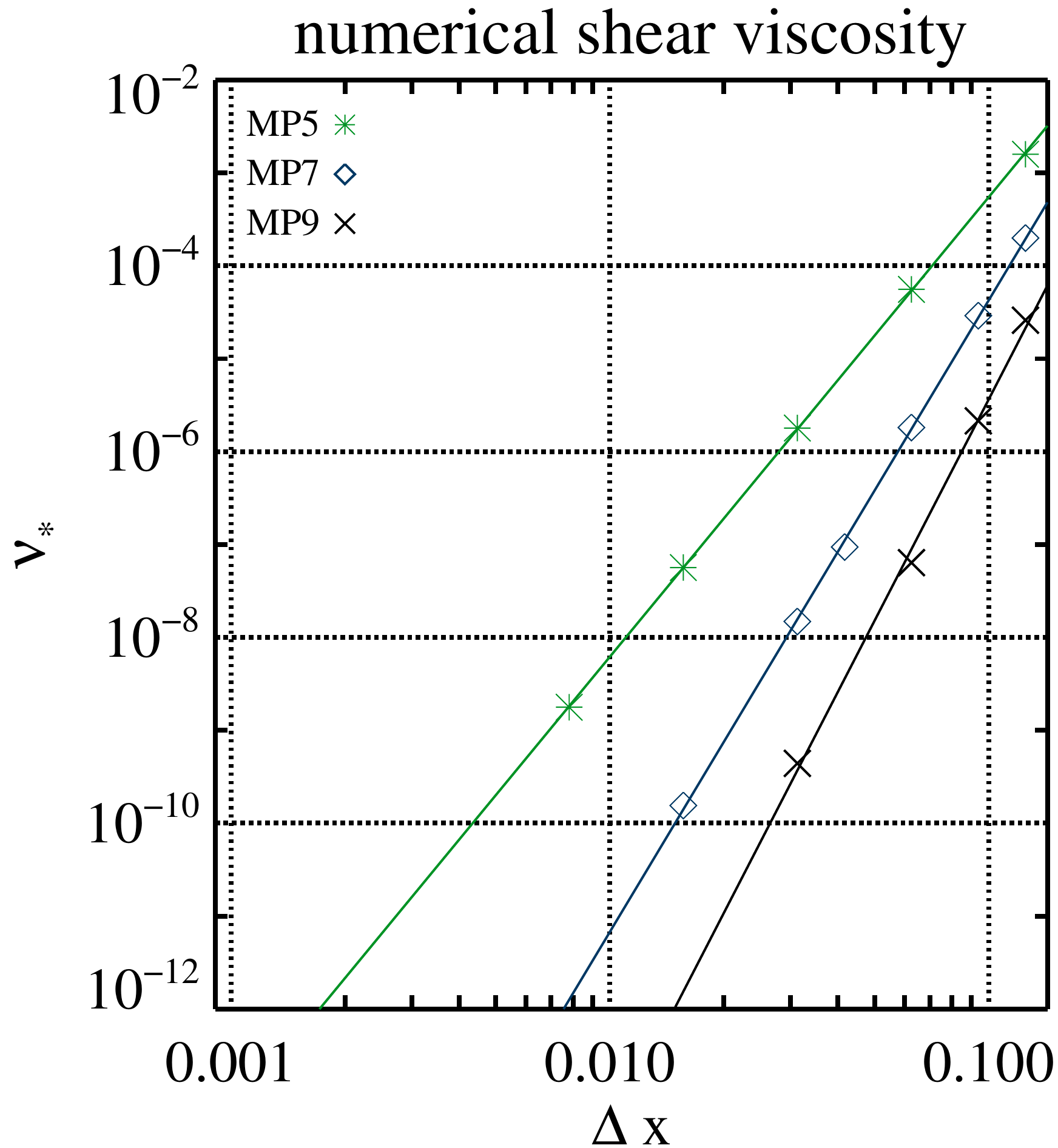} 
 \includegraphics[width=0.33\textwidth]{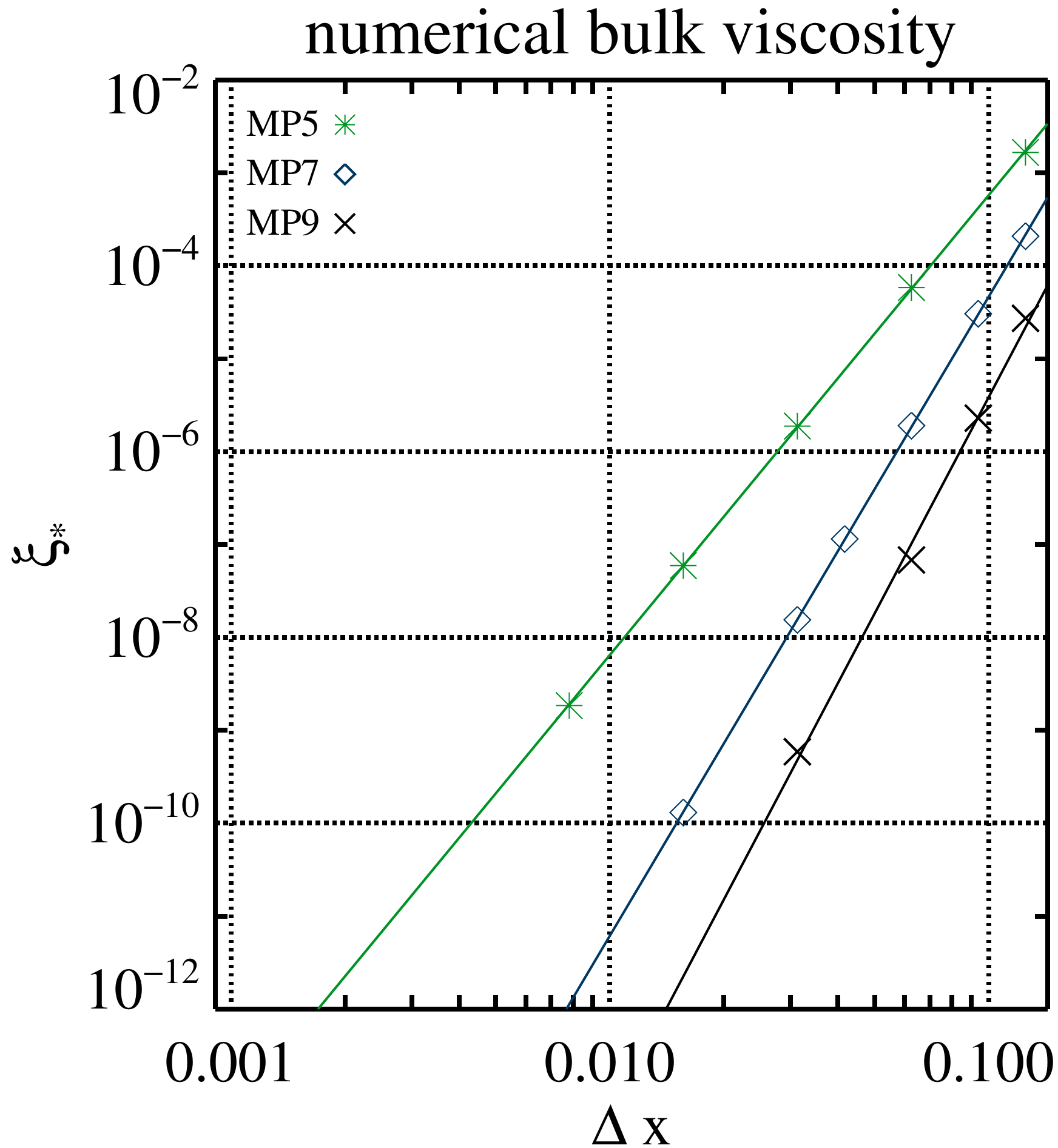} 
 \caption{Fits of the numerical resistivity (left), shear viscosity
   (center), and bulk viscosity (right) computed from the combined
   damping rates of sound waves, \alf waves, and magnetosonic waves
   for three different reconstruction schemes. Note that the numerical 
   resistivity is negative.}
  \label{fig:resvis}
\end{figure*}
%----------------------------------------------------------------------------
 
If we consider series of simulations in which time discretisation
errors are negligible (as those in upper left panels of
Figs.~\ref{fig:sound}, \ref{fig:alf}, and \ref{fig:mag}), at a fixed
grid resolution $k\Delta x = const.$ and for a given numerical method
the numerical viscosity and resistivity should be the same according
to our ansatz (Eqs.\, \ref{eq:nu*}, \ref{eq:xi*}, and \ref{eq:eta*}).
Therefore, the damping rates of the three propagating waves can (in
principle) be used to compute the numerical viscosity and resistivity.

In Fig.~\ref{fig:resvis} we present the resistivity, shear viscosity,
and bulk viscosity for three different reconstruction schemes (MP5,
MP5 and MP9) as a function of grid resolution. In all simulations we
used the HLL Riemann solver, a RK4 time integrator, and $\CFL=0.01$.
Fitting a power law to the data allows us to compute the coefficients
entering in the ansatz given by Eqs.\,(\ref{eq:nu*}), (\ref{eq:xi*}),
and (\ref{eq:eta*}).  From the fit parameters, which can be found in
Tab.~\ref{tab:resvis}, we compute the exponent $r$ appearing in the
ansatz independently for resistivity and viscosity.  For all three
reconstruction schemes, (i) the values of $r$ are close to the
expected order of convergence, $r_{\rm th}$, except for a small
deviation in the case of MP9, and (ii) the value of the numerical
viscosity is significantly larger than the absolute value of the
numerical resistivity.

Our most striking result is that the value of the numerical resistivity $\eta_*$ is negative, its absolute value being
about one order of magnitude smaller (and even two orders for MP9) than the value of the numerical viscosity (see
Tab.~\ref{tab:resvis}).  The value of the resistivity coefficient $ \nn_{\eta}^{\deltx x} $ obtained with MP9 (providing
the most accurate result) is compatible with a non-negative (very small) numerical resistivity, while the values of the
numerical shear viscosity and bulk viscosity are positive and very similar. The resulting damping rate, which is a
combination of resistivity and viscosity, prevented an amplification of the wave amplitude in all three systems
studied. Taken together these facts suggest that the numerical viscosity must be considerably larger than the numerical
resistivity of the code.  Hence, we conjecture that there are large systematic uncertainties that prevent us from
properly measuring the numerical resistivity of the code in all three wave propagation tests.

Given that the dissipation is dominated by viscosity rather than resistivity, we had to turn to a completely different
system in order to study whether our ansatz for numerical resistivity is a valid one, and if true, whether the results
are consistent with a positive value of the resistivity (see Sec.\,\ref{subsec:tm}).

%----------------------------------------------------------------------------
\begin{table*}
\centering
    \begin{tabular}{ccccccc}
        \tableline
reconstruction & $ \nn_{\eta}^{\deltx x} $ & $r_\eta$ & $\nn_{\nu}^{\deltx x} $ 
        & $r_\nu$ & $ \nn_{\xi}^{\deltx x} $ & $r_\xi$   \\ \tableline
MP5 & $ -7.0  \pm  0.5$ &$  4.94 \pm 0.02$ & $49 \pm 3$& $4.95 \pm 0.02$ 
    & $ 51 \pm   4$ & $4.95 \pm 0.02$\\
MP7 & $ -41  \pm  3$ & $ 6.81 \pm 0.03 $& $270 \pm 50$& $6.80\pm 0.06$ 
    & $354 \pm  18$ & $6.881 \pm 0.017$\\ 
MP9 & $ \ \ -3  \pm  6$ & $ 6.8 \pm 0.6$  &$300\pm200$& $7.9\pm 0.3$  
    & $200 \pm 200$ & $7.7 \pm 0.3$\\
        \tableline
    \end{tabular}
    \caption{
      Values of the parameters in our ansatz for the spatial 
      dependence of the numerical viscosity (Eqs.\,\ref{eq:nu*}  
      and \ref{eq:xi*}) and numerical resistivity (Eq.\,\ref{eq:eta*}) 
      based on a fit of the results shown in Fig.~\ref{fig:resvis}. 
      Note that we obtain three different estimates of the exponent
      $r$ from fits of resistivity ($r_\eta$), shear viscosity
      ($r_\nu$), and bulk viscosity ($r_\xi$). }
\label{tab:resvis}
\end{table*}
%----------------------------------------------------------------------------

%%%%%%%%%%%%%%%%%%%%%%%%%%%%%%%%%%%%%%%%%%%%%%%%%%%%%%%%%%%%%%%%%%%%%%

\subsubsection{Waves with a Background Velocity}
\label{sec:background_velocity}

\begin{figure}
  \centering    
  \includegraphics[width=0.45\textwidth]{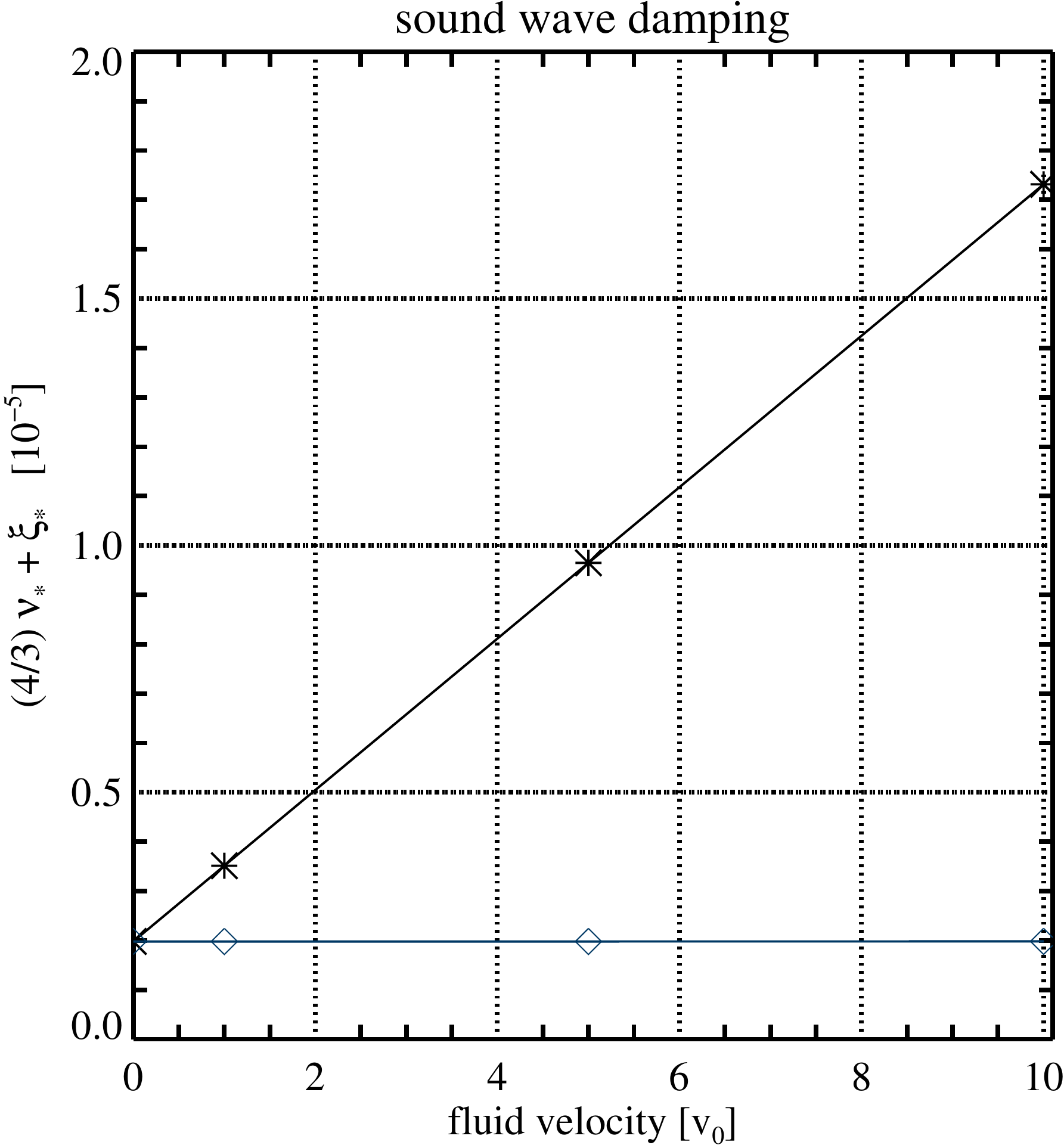}
  \caption{
    Numerical dissipation as a function of the flow velocity parallel
    (black asterisks) and perpendicular (blue diamonds) to the
    direction of the propagation of the sound wave. The simulations
    were performed with 32 zones, MP5, RK3 (with $\CFL = 0.1$), and
    the HLL Riemann solver. Other solvers (LF and HLLD) exhibit a very
    similar behaviour.}
  \label{fig:sound_vx_vy}
\end{figure}

So far, in the wave damping simulations, we have set the background
velocity to zero. To test whether a background velocity affects the
numerical damping (\ie by modifying the characteristic speed of the
flow), we repeated the damping test for the sound waves, \alfc waves,
and magnetosonic waves with a non-zero background velocity.  All
simulations were performed with the MP5 reconstruction scheme and the
RK3 time integrator (with $\CFL = 0.1$).  For all three Riemann
solvers and all types of waves we observed the same behaviour, \ie the
component of the background flow velocity that is perpendicular to the
propagation of the wave ($\velo_{0y}$) does not affect the numerical
dissipation, whereas the parallel component ($\velo_{0x}$) does.
Figure \ref{fig:sound_vx_vy} shows the numerical dissipation for some
exemplary simulations of the sound wave damping test done with the HLL
Riemann solver.  Based on these simulations, we conclude that the
characteristic speed of the flow is given by the sum of (the parallel
component of) the flow velocity and the fast magnetosonic speed or, in
other words, by the fast magnetosonic eigenvalue of the ideal MHD
equations.

%----------------------------------------------------------------------------
%----------------------------------------------------------------------------

\subsection{ Sound waves in  2D} 
\label{subsec:2d_waves}

So far, we have studied  wave propagation problems in 1D.  However,
it is well known that unless a genuinely multi-D reconstruction
algorithm is used \citep[as proposed by,
\eg][]{Colella_et_al_2011,McCorquodale_Colella_2011,Zhang_et_al_2011,Buchmueller_Helzel_2014},
the order of a reconstruction scheme can be reduced in simulations
involving more than one spatial dimension.  Our code \textsc{Aenus}
employs several independent one-dimensional reconstruction steps (one
per dimension).  Thus, the convergence rate may be degraded to
second order in multi-D simulations, \ie well below that of
1D applications.

We studied this aspect with the help of 2D simulations (Tab.\
\ref{tab:2d_waves}) of sound wave propagation
in  a box of size $L_x \times L_y$ with periodic boundary conditions in both directions.
We set the background density and pressure to $\rho_0 = p_0 = 1$, and imposed a perturbation of the
form
\begin{align}
  \label{eq:vx_2d}
\velo_{1x}(x,y,t=0) &=   \epsilon  \sin (\theta ) \sin ( k_x x + k_y y ), \\
\velo_{1y}(x,y,t=0) &=   \epsilon  \cos (\theta ) \sin ( k_x x + k_y y ), \\
\rho_1 (x,t=0)    &= \frac{ \epsilon }{\cs} \rho_0  \sin ( k_x x + k_y y ), \\
p_1 (x,t=0)       &= \frac{\epsilon } {\cs} \Gamma p_0 \sin ( k_x x + k_y y ) 
\end{align}
where $\Gamma = 5/3$,  $\epsilon = 10^{-5}$,
 $\theta$ is an angle between the $x-$axis and the wavevector, ${\vek k} = (k_x, k_y), $
 where
\begin{align}
  \label{eq:kx_2d}
k_x &=  \frac{ 2 \pi  } { L_x }, \\
k_y &=  \frac{ 2 \pi  } { L_y }. %,
  \label{eq:ky_2d}
\end{align}
The wavelength is given by
\begin{equation}
  \lambda = \frac{2 \pi}{k} = \frac{ 2 \pi}{ \sqrt{ k_x^2 + k_y^2 } }  = \frac{ L_x L_y} { \sqrt{ L_x^2 +  L_y^2} }.
  \label{eq:lambda_2d}
\end{equation}

In all 2D simulations, we set $L_x = 1$, like in all 1D sound wave
simulations (but series \tr{\#cS3}) and use a uniform grid, i.e.\
$\Delta x = \Delta y$.  Note that the 1D expressions are
  recovered in the limit $L_y\rightarrow \infty$.
We determine numerical damping from the kinetic energy of sound waves 
whose time evolution is
\begin{equation}
E_{kin}(t)  = E_{kin}(0)  e^{-2 \dd_s t} 
\end{equation}
in an analogous manner as described in Sec.\ \ref{subsec:sound_waves_1d}. 
In the case of  scalar constant bulk and shear viscosities, the damping rate  $\dd_s$ is given by (analogically to the 1D case, Eq.\ \ref{eq:dds}) 
\begin{equation}
\dd_s = \frac{k^2}{2} \left( \frac{4}{3}\nu  + \xi \right)= \frac{k_x^2 + k_y^2}{2} \left( \frac{4}{3}\nu  + \xi \right).
\label{eq:dds_2d}
\end{equation}
However, as already mentioned in Sec.\ \ref{subsec:ansatz} (see Eq.~\ref{eq:visc:2D}),
numerical viscosities have a tensorial character in a multi-D simulation, hence the damping rate is given by
\begin{align}
\dd_s &= \frac{1}{2} {\bf k}^T   \left( \frac{4}{3} \bm{\nu}_{\ast}  + \bm{\xi}_{\ast} \right) {\bf k}  = \nonumber \\
 &=  \frac{1}{2} \Bigg[ k_x^2  \left( \frac{4}{3}\nu_{\rm sp}^{xx}  + \xi_{\rm sp}^{xx} \right) 
+ k_y^2  \left( \frac{4}{3}\nu_{\rm sp}^{yy}  + \xi_{\rm sp}^{yy} \right) %+ \nonumber \\
%&
+ k^2 \left( \frac{4}{3}\nu_t  + \xi_t \right)  \Bigg].
\label{eq:dds*_2d}
\end{align}

To be able to determine (linear combinations of)  
%estimators 
$\nn_{\nu}^{\deltx x},\nn_{\xi}^{\deltx x},  \nn_{\nu}^{\deltx x}$ and $\nn_{\xi}^{\deltx x} $
(defined in Eqs.\ \ref{eq:nu*} and \ref{eq:xi*}), i.e.\
  $\nn_{\rm  tot}^{\deltx x} = (4/3) \nn_{\nu}^{\deltx x} +  \nn_{\xi}^{\deltx x}$ and
  $\nn_{\rm  tot}^{\deltx t} = (4/3) \nn_{\nu}^{\deltx t} +
  \nn_{\xi}^{\deltx t}$ (Tab.\ \ref{tab:2d_waves}),
we first need to identify the characteristic velocities $\vv_x,  \vv_y$ and $\vv$,
and lengths  $\LL_x, \LL_y$ and $\LL$ of the system. 
From our studies in 1D, we infer that the former must be the sound speed, which is homogenous and isotropic 
in the whole system and, thereby, $\vv_x = \vv_y = \vv = \cs$.
Moreover, we postulate that $\LL_x = 2 \pi / k_x$ and  $\LL_y = 2 \pi
/ k_y$, since for the reconstruction scheme in
each  dimensional sweep,  this 2D sound wave problem reduces to a 1D wave propagation 
(e.g.\ Eq.\,\ref{eq:vx_2d} reduces to Eq.\,\ref{eq:vx_1d} for $y = \rm{const.}$).
The characteristic (physical)  time scale of the system is $\TT = \cs / \lambda$.
And since the time integration errors must be proportional to  $\Delta t / \TT$, we conclude that
$\LL = \lambda$.
Therefore,  for our system, ansatzes for  $\nu_{\rm sp}^{xx}$ and $\nu_{t}$ (see Eqs.\  \ref{eq:nu_xx} and  \ref{eq:nu_t})  read
\begin{align}
\nu_{\rm sp}^{xx}&=  \nn_{\nu}^{\deltx x} \cs \frac{2 \pi}{k_x}   \left(\frac{k_x \, \Delta x }{2 \pi} \right)^r , \\
\nu_{t}         &=  \nn_{\nu}^{\deltx t}   \cs \lambda  \left( \frac{ \cs \, \deltt t }{ \lambda }\right)^{q}.
\end{align}
The ansatzes for the other components of numerical viscosities have an
analogous form. Finally, the damping rate for 2D wave simulations is
\begin{equation}
    \begin{split}
\dd_s &=   2 \pi^2   \cs \left\{   \left[ (4/3) \nn_{\nu}^{\deltx x} +  \nn_{\xi}^{\deltx x} \right] \left[ \left( \frac{k_x}{2 \pi} \right)^{r+1} (\Delta x)^r +  \left( \frac{k_y}{2 \pi} \right)^{r+1} ( \Delta y)^r  \right] \vphantom{ \left[ (4/3) \nn_{\nu}^{\deltx t} + \nn_{\xi}^{\deltx t}  \right]  \left[    \left( \frac{ k }  { 2 \pi }   \right)^{q+1}  \CFL^q  ( \Delta x )^q \right] } \right.   \\
 & + \left. 
 \left[ (4/3) \nn_{\nu}^{\deltx t} + \nn_{\xi}^{\deltx t}  \right]  \left[    \left( \frac{  k } { 2 \pi }   \right)^{q+1}  \CFL^q  ( \Delta x )^q \right] 
 \right\},
\label{eq:ds_2d_full}
    \end{split}
\end{equation}
which for an equidistant grid,  i.e.\ $\Delta x = \Delta y$, and $L_x = 1$, further simplifies to
\begin{equation}
    \begin{split}
\dd_s &=   2 \pi^2   \cs \left\{   \left[ (4/3) \nn_{\nu}^{\deltx x} +  \nn_{\xi}^{\deltx x} \right] 
\left[ (\Delta x)^r \left(1 +  L_y^{-r-1} \right) \right] \vphantom{ \left[ (4/3) \nn_{\nu}^{\deltx t} + 
\nn_{\xi}^{\deltx t}  \right]  \left[    \left(  1 +  L_y^{-2}    \right)^{(1+q)/2} \CFL^q  ( \Delta x )^q \right] } \right.  \\
 & + \left. 
 \left[ (4/3) \nn_{\nu}^{\deltx t} + \nn_{\xi}^{\deltx t}  \right]  \left[    \left(  1 +  L_y^{-2}    \right)^{(1+q)/2}  \CFL^q  ( \Delta x )^q \right] 
 \right\}.
\label{eq:ds_2d_particular}
    \end{split}
\end{equation}

\begin{figure}%%[h]
\centering
  \includegraphics[width=0.45\textwidth]{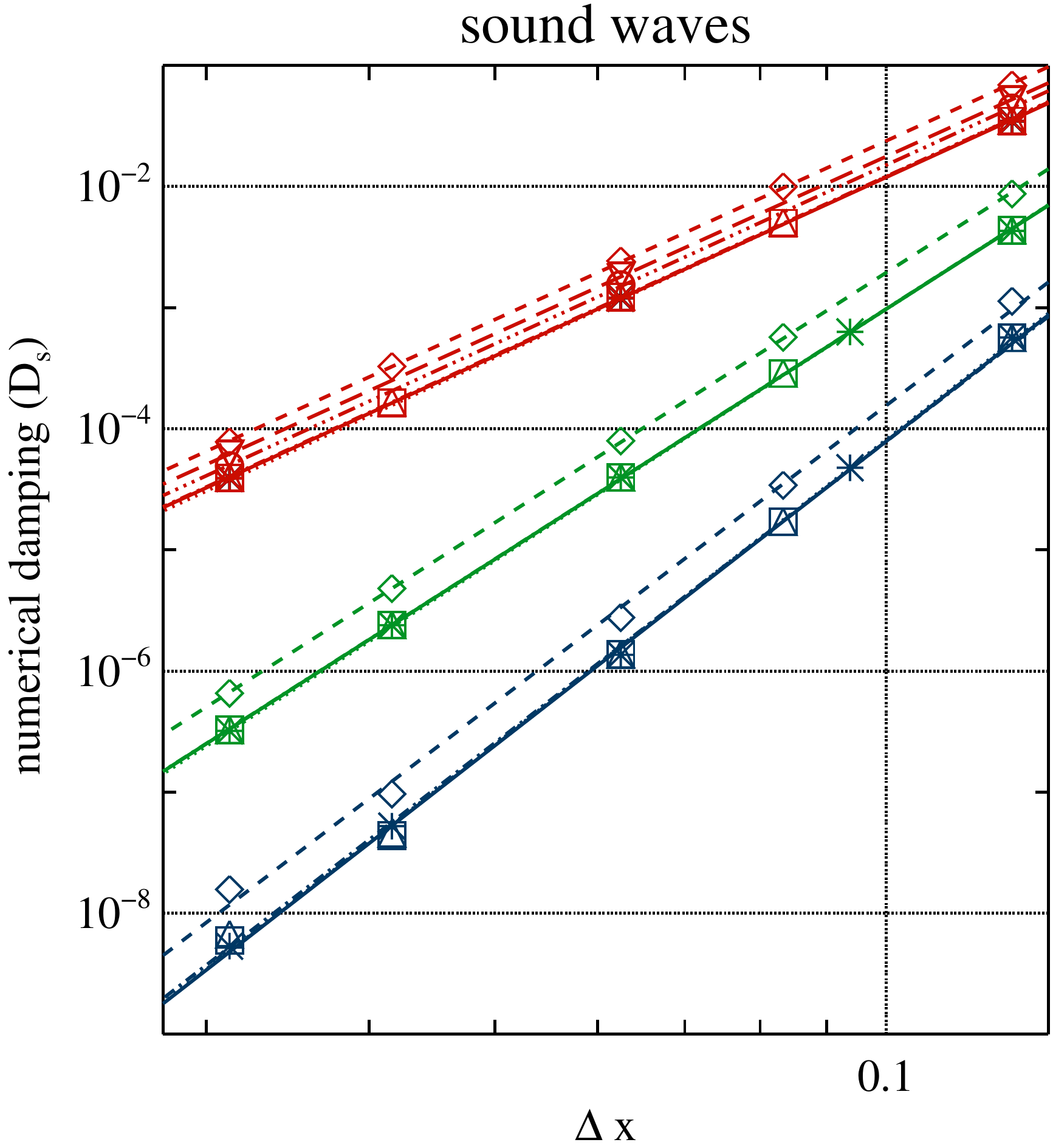}
  \includegraphics[width=0.45\textwidth]{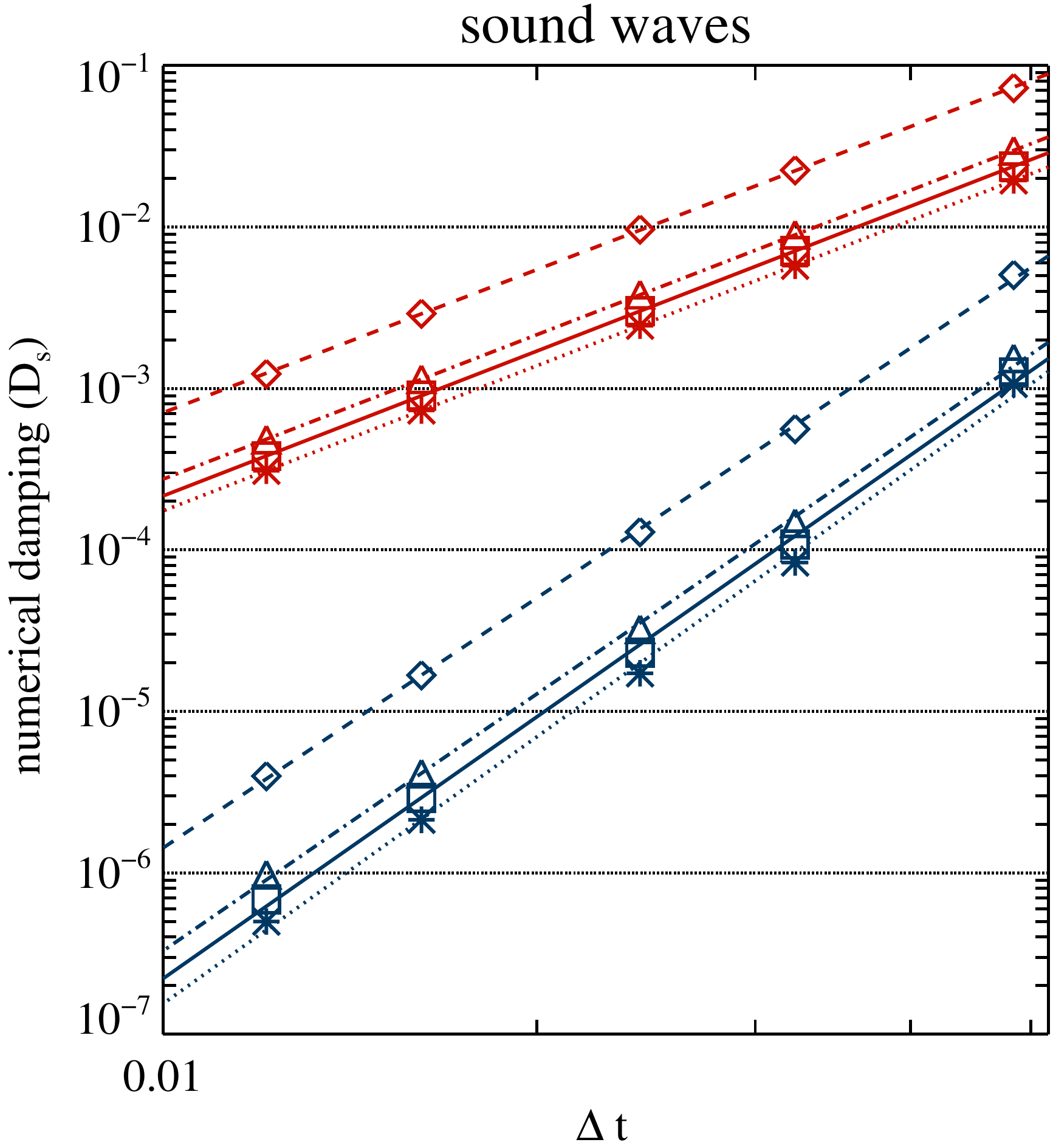}
  \caption{Numerical damping in sound  wave simulations performed with the HLL Riemann solver
in  1D (asterisks, dotted lines), and  in 2D with  $L_y = 1$  (diamonds, dashed lines) , $L_y = 1.125$  (downwards triangles, dash-dot-dot-dot line),  $L_y = 1.25$  (circles, long-dashed line),  $L_y =  2$ (upwards triangles, dashed-dotted lines) and  $L_y = 3$ (squares, solid lines). 
\emph{Top:} simulations done with the RK3 time integrator with $\CFL = 0.01$ (so that the time integration errors are negligible) and
the MP5 (red), MP7 (green) and  MP9 (blue)  reconstruction schemes.
\emph{Bottom:}  simulations done with the MP9 reconstruction scheme (so that time discretisation errors are negligible) and  the RK3 (red) and  RK4( blue) time integrators with $\CFL = 0.5$.
See also Tab.\ \ref{tab:2d_waves}.
}
\label{fig:2d_waves}
\end{figure}

According to the above equation, for simulations in which numerical
dissipation is dominated either by spatial reconstruction
errors (series \tr{\#LS1}--\tr{\#LS11} from Tab.\ \ref{tab:2d_waves})
or by time integration errors (series
\tr{\#LS12}--\tr{\#LS17}), the dissipation rate should respectively be
$\left(1 + L_y^{-r-1} \right) $ and
$\left( 1 + L_y^{-2} \right)^{(1+q)/2} $ times larger than in the
corresponding 1D simulations  (with the same $L_x = 1)$.
Note that this difference between 1D and 2D simulations is due to the small change in the
  value of $\lambda$.  Deviations from this
  expected value would be indicative of differences in the dissipation
  coefficients between 1D and 2D simulations.  
For the MP5 reconstruction scheme, assuming $r = 5$,  and boxes 
with  $L_y = 1, 1.125$ and $1.25$,  the dissipation should be  respectively  $2, 1.49$ and $1.26$  times larger than in the 1D case, whereas
in  boxes with $L_y = 2$ and $3$, the numerical dissipation rate
should be basically equal to the 1D case (i.e.\,merely greater by a
factor  of $1.016$ and $1.001$, respectively).
The upper panel of  \figref{fig:2d_waves} depicts  (in red)
the damping rates  in simulation series  \tr{\#LS1}, \tr{\#LS4},
\tr{\#LS5}, \tr{\#LS6},  and \tr{\#LS9} (\tabref{fig:2d_waves}) performed with the
MP5 reconstruction scheme in 2D boxes of those sizes.
The ratios of these damping rates to the damping rates in 1D 
(simulation series \tr{\#S3}  from Tables  \ref{tab:waves} and \ref{tab:2d_waves},     marked with asterisks in the figure)
are in a very good agreement with the above estimates.
Similarly, we expect twice higher dissipation rates in simulations done with MP7 (series \tr{\#LS2}) and MP9 (series \tr{\#LS3}) 
reconstruction schemes in boxes with $L_y = 1$ than in their 1D counterparts (simulation series \tr{\#S5} and \tr{\#S6}, respectively),
and basically equal (to the 1D case) dissipation rates in simulations
with  $L_y = 2$ and  $L_y = 3$ (simulation series \tr{\#LS7}, \tr{\#LS8},\ \tr{\#LS10},  and \tr{\#LS11}). Indeed, dissipation rates presented in the upper panel of \figref{fig:2d_waves} exhibit this behaviour.

In the above analysis, we implicitly assumed that  $r,\, \nn_{\nu}^{\deltx x}$, etc.\ are equal in 1D and 2D simulations,
so  the previous analysis only provides a consistency  check. However, these coefficients can actually be measured in 2D simulations
and can be compared with the coefficients obtained for the 1D case.
In Table \ref{tab:2d_waves}, we  present estimators for these quantities determined  in the 2D simulations  from dissipation rates with 
the help of Eq.\ (\ref{eq:ds_2d_particular}) in an analogous way as described in Sec.\  \ref{subsec:sound_waves_1d}.
The estimators are indeed equal within the measurement errors for each reconstruction scheme, i.e.\ MP5, MP7 and MP9,
in 1D and in 2D simulations.  This signifies that our  ansatzes (\ref{eq:visc:2D})--(\ref{eq:nu_t})
are correct at least for 2D wave simulations for the spatial reconstruction errors.

The bottom panel of Fig.\ \ref{fig:2d_waves} depicts dissipation rates in 
simulations series \tr{\#LS12}--\tr{\#LS17}  (\tabref{tab:2d_waves}) performed  with the MP9 reconstruction scheme (so that spatial discretisation errors are negligible), $\CFL = 0.5$, and RK3 (red) or RK4 (blue) time integrators in 2D boxes with $L_y = 1$, $2$ and $3$ as well as in 1D (series \tr{\#S7}  and \tr{\#S9}).
The estimators for  $q$ and $\nn_{\rm tot}^{\deltx t}$ determined from these data are presented in  \tabref{tab:2d_waves}.
The RK3  time integration scheme once again (like in 1D)   has its  theoretical order, i.e.\ $q \approx 3$, whereas the RK4 time integrator
once again overperforms by one unit the expected order, i.e. $q \approx 5$. The
estimators for  $\nn_{\rm tot}^{\deltx t}$ and $q$ for the RK3 scheme are
very similar in 1D and 2D simulations, whereas
for the RK4 scheme, there is a discrepancy, which
cannot be explained by the measurement errors
(i.e.\ $\nn_{\rm  tot}^{\deltx t}$  ranges from $17 \pm 3$ to $71 \pm 32$, and $q$ from  $5.14 \pm 0.05$ to $5.5 \pm 0.2$).
Note, however, that this discrepancy is insignificant in the
considered range of $\Delta t$  as there is  a clear correlation
between $q$ and $\nn_{\rm tot}^{\deltx t}$, i.e.\ the larger $q$, the
larger $\nn_{\rm tot}^{\deltx t}$, leading  to very similar
predictions for the dissipation rate as we show in the next paragraph.
Therefore, we conclude that   our ansatzes (\ref{eq:visc:2D})--(\ref{eq:nu_t}) are valid  for time integration errors in 2D and that both RK3 and RK4
time integration schemes perform (basically) identically in 1D and 2D simulations (with various box sizes).

Based on Eq.\ (\ref{eq:ds_2d_particular}), we can make the following estimates for simulations where time integrator errors are dominant.
For simulations done with RK3, assuming $q = 3$ (and equal $\nn_{\rm tot}^{\deltx t}$), in a box with
$L_y = 1,2$ and $3$ (series \tr{\#LS12}, \tr{\#LS14} and \tr{\#LS16}, respectively) numerical dissipation should be respectively  $4, 1.56$ and $1.23$ times greater than 
in 1D simulations (series \tr{\#S7}).  
For analogous simulations done with RK4 (series \tr{\#LS13}, \tr{\#LS15} and \tr{\#LS17}, respectively), assuming $q = 5$ (and equal $\nn_{\rm tot}^{\deltx t}$), the dissipation rates
should respectively be $8, 1.95, 1.37$  greater than in the 1D case (series \tr{\#S9}). As can be seen in the bottom panel of Fig.\  \ref{fig:2d_waves}, these predictions
agree very well with our simulation results.

%%%%%%%%%%%%%%%%%%%%%%%%%%%%%%%%%%%%%%%%%%%%%%%%%%%%%%%%%%%%%%%%%%%%%%
\begin{table*}
  \caption{
Sound  wave damping in 2D simulations (series \tr{\#LS}). Additionally, for the reader's convenience,
we repeat some 1D simulations (series \tr{\#S}) from Tab.\, \ref{tab:waves}.
The columns give (from left to
  right) the series identifier,  the $L_y$ box length ($L_x = 1$), the reconstruction 
  scheme, the Riemann solver, the time integrator,  and the CFL factor. 
In all 2D simulations,  an uniform grid is used (i.e. $\Delta y = \Delta x$) and
the number of  grid zones $N_x$ per $L_x$ is in the range from $8$ to $32$.
The estimator for $\nn_{\rm tot}^{\deltx x}$, 
  $r$, $\nn_{\rm  tot}^{\deltx t}$, and $q$  (see Eqs.\,\ref{eq:nu*}
and   \ref{eq:xi*}) is obtained from linear
  fits to the simulation results. For sound waves, 
  $\nn_{\rm  tot}^{\deltx x} = (4/3) \nn_{\nu}^{\deltx x} +  \nn_{\xi}^{\deltx x}$ and
  $\nn_{\rm  tot}^{\deltx t} = (4/3) \nn_{\nu}^{\deltx t} +  \nn_{\xi}^{\deltx t}$.}
\begin{center}
\begin{tabular}{ccccccccccc}
\tableline
series & $L_y$ & Reco &  Riemann &  time & CFL &   $\nn_{\rm  tot}^{\deltx x}$ &  $r$ &  $\nn_{\rm  tot}^{\deltx t}$  &  $q$
\\  \tableline
\#LS1 &  $1$  & MP5 & HLL & RK3& $0.01$ &  $  37.1 \pm  2.6 $ & $  4.90  \pm 0.02   $   &  $ - $ &  $ - $
 \\
\#LS2  &  $1$  & MP7 & HLL & RK3 & $0.01$  & $ 273  \pm  21  $ & $  6.86 \pm  0.03 $ &  $ - $ &  $ - $
 \\
\#LS3  &  $1$  & MP9 & HLL & RK3 & $0.01$  & $ 440  \pm 300  $ &  $  8.2 \pm  0.2 $ &  $ - $ &  $ - $
 \\
\#LS4  &  $1.125$  & MP5 & HLL & RK3 & $0.01$ & $  33.7 \pm  4.4  $ &  $  4.86 \pm  0.02  $ &  $ - $ &  $ - $
 \\
\#LS5  &  $1.25$  & MP5 & HLL & RK3 & $0.01$ &  $ 37  \pm  7 $ &  $  4.90  \pm  0.04  $ &  $ - $ &  $ - $
 \\
\#LS6  &  $2$  & MP5 & HLL & RK3 & $0.01$ & $   37.2 \pm  2.7 $ & $ 4.90   \pm 0.02   $ &  $ - $ &  $ - $
 \\
\#LS7  &  $2$  & MP7 & HLL & RK3 & $0.01$ & $ 276   \pm 24   $ & $  6.86  \pm  0.03 $ &  $ - $ &  $ - $
\\
\#LS8  &  $2$  & MP9 & HLL & RK3 & $0.01$ & $  570 \pm 370   $ &  $  8.3  \pm  0.2 $ &  $ - $ &  $ - $
\\
\#LS9  &  $3$  & MP5 & HLL & RK3 & $0.01$ & $   37.1 \pm  2.5  $ & $  4.90  \pm 0.02  $ &  $ - $ &  $ - $
\\
\#LS10  &   $3$  & MP7 & HLL & RK3 & $0.01$ & $  277  \pm  24  $ &  $  6.86  \pm  0.03 $ &  $ - $ &  $ - $
\\
\#LS11 &   $3$  & MP9 & HLL & RK3 & $0.01$ & $  680 \pm 360  $ &  $  8.35  \pm  0.18 $ &  $ - $ &  $ - $
\\
\tableline
\#S3 &   $ -$  & MP5 & HLL & RK4 & $0.01$ & $  43.4\pm  2.5  $ &  $   4.962 \pm  0.0141 $ &  $ - $ &  $ - $
\\
\#S5 &   $ - $  & MP7 & HLL & RK4 & $0.01$ & $  302 \pm  20  $ &  $  6.897  \pm  0.021 $ &  $ - $ &  $ - $
\\
\#S6 &   $ -$  & MP9 & HLL & RK4 & $0.01$ & $  830 \pm 340  $ &  $  8.42  \pm  0.15 $ &  $ - $ &  $ - $
\\
\tableline
\#LS12  &   $1$  & MP9 & HLL & RK3 & $0.5$ &  $ - $  &     $ - $  &  $ 1.28 \pm  0.04$ &  $   2.94  \pm  0.01           $  
\\ 
\#LS13  &   $1$  & MP9 & HLL & RK4 & $0.5$ & $ - $  &  $ - $ & $  17 \pm  3  $ &  $5.14  \pm  0.05              $  
\\
\#LS14  &   $2$  & MP9 & HLL & RK3 & $0.5$ &  $ - $  &     $ - $  &  $ 1.4  \pm  0.2  $ &  $   2.970  \pm    0.005         $  
\\ 
\#LS15  &   $2$  & MP9 & HLL & RK4 & $0.5$ & $ - $  &  $ - $ &  $ 31  \pm 9  $ &  $   5.3  \pm  0.1           $  
\\
\#LS16  &   $3$  & MP9 & HLL & RK3 & $0.5$ &  $ - $  &     $ - $  & $  1.56 \pm  0.01  $ &  $  2.978   \pm    0.002         $  
\\ 
\#LS17  &   $3$  & MP9 & HLL & RK4 & $0.5$ & $ - $  &  $ - $ &  $  46  \pm 17  $   &  $   5.38  \pm  0.13           $  
\\
\tableline
\#S7  &   $ - $  & MP9 & HLL & RK3 & $0.5$ & $ - $  &  $ - $ & $   1.492  \pm  0.013  $ &  $ 2.985  \pm  0.002    $  
\\
\#S9  &   $ -$  & MP9 & HLL & RK4 & $0.5$ & $ - $  &  $ - $ & $   71  \pm  32 $ &  $ 5.5  \pm  0.2    $  
\\
\tableline
  \end{tabular}
\label{tab:2d_waves}
\end{center}
 \end{table*}

%%%%%%%%%%%%%%%%%%%%%%%%%%%%%%%%%%%%%%%%%%%%%%%%%%%%%%%%%%%%%%%%%%%%%%

\subsection{Tearing Mode Tests}
\label{subsec:tm}

%----------------------------------------------------------------------------
\begin{figure*}%%%[h]
\centering
  \includegraphics[width=0.85\textwidth]{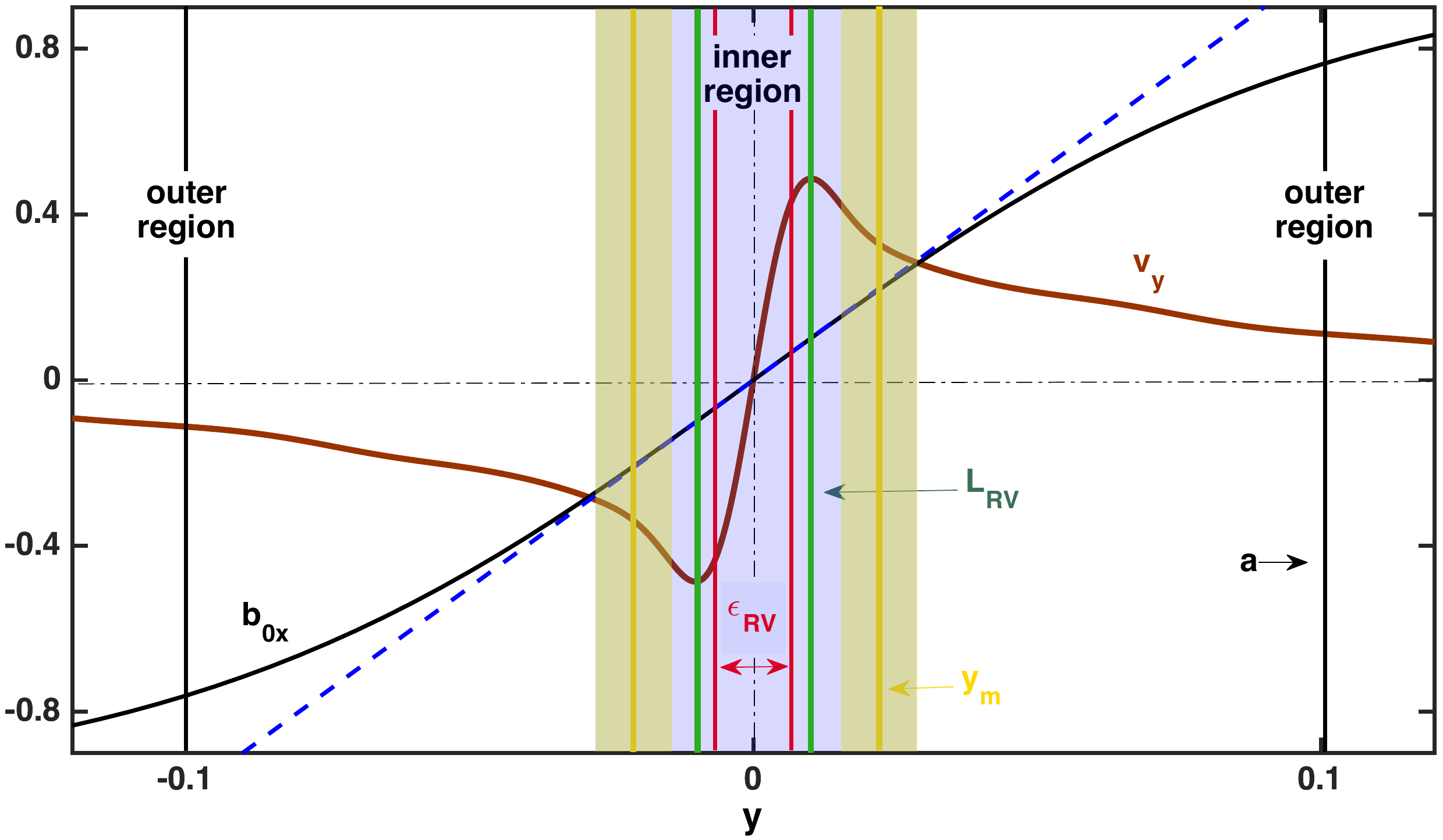}
  \caption{
    Background magnetic field, $b_{0x}$, (black; Eq.\,\ref{eq:b0_tan})
    and the amplitude (rescaled for a better visibility) of the
    velocity perturbation $v_y$ (brown; Eq.\,\ref{eq:bibi1}) of
    a TM. Only the central part of the computational domain
    is shown, \ie $y\in [-0.15,0.15]$.  In the outer region 
    (white) and the inner region (blue) 
    $v_y$ is given by Eqs.\,(\ref{eq:v1y_out}) and
    (\ref{eq:Phi_integral}), respectively.  In an intermediate region
    (shaded yellow) both solutions are valid and must be matched at a
    certain matching point, $y_{\mathrm{m}}$, within this region
    (yellow vertical line). Black, green, and red vertical lines
    respectively mark, the shear width $a$, the width of the
    resistive viscous layer $L_{\rm RV}$
    (Eq.\,\ref{eq:epsilon_rv_org_main}), and the point where
    $y = \epsilon_{\rm RV}$ (defined in Eq.\,\ref{eq:epsilon_rv_org}).
    In the inner region, the Taylor expansion
    $ b_{0x}(y) \approx b_0 y / a $ (Eq.\,\ref{eq:b_taylor}) is used to
    obtain the analytical solution (see Appendix \ref{app:tm_ana}).
    The blue dashed line depicts the function $b_0 y / a$ (with
    $b_0 = 1$) to illustrate the regime of the validity of the Taylor
    expansion. }
\label{fig:tm_theory}
\end{figure*}%%
%----------------------------------------------------------------------------

The TM instability is a resistive MHD instability that can develop in
\emph{current sheets}, where, as a direct consequence of Amp\`{e}re's
law, the magnetic field changes direction. TMs dissipate magnetic
energy into kinetic energy and subsequently into thermal energy,
disconnect and rejoin magnetic field lines, thereby changing the
topology of the magnetic field. The linear theory of TM was
extensively studied, in the context of plasma fusion physics, in a
seminal paper by FKR63.  TMs are of great relevance in astrophysics,
(e.g.~in the magnetopause or magnetotail of the solar wind, in flares
or coronal loops of the Sun, and in the flares of the Crab pulsar
\citep[cf.][]{Priest_Forbes,Pucci_Velli}.  They have also been
suggested to be a terminating agent of the MRI \citep[][but see
\citealt{Rembiasz:2016a}  who observed an MRI termination by the
  Kelvin-Helmholtz instability in their 3D MRI
  simulations]{Balbus_Hawley__1991__ApJ__MRI,Latter_et_al,Pessah}.

In this section, we present a test involving TMs, for which we know
how the reconnection rate depends on the relevant parameters
(resistivity, viscosity, etc.).  By performing numerical simulations
of viscous, but non-resistive MHD flows at different grid resolutions
with various numerical methods, we developed a method to measure the
numerical resistivity of MHD codes.

%%%%%%%%%%%%%%%%%%%%%%%%%%%%%%%%%%%%%%%%%%%%%%%%%%%%%%%%%%%%%%%%%%%%%%

\subsubsection{Theory} 
\label{sec:tm_theory}

In this section, we sketch how to analytically obtain a growth rate
and an instability criterion of the TM instability, leaving out all
technical details which can be found in Appendix \ref{app:tm_ana}.
Many of the results presented here were already obtained by FKR63, or
they are different limits of expressions found in that work.

Consider a two dimensional flow in the $x$-$y$ plane of constant
background density $\rho_0=1$ threaded by a magnetic field
\begin{equation}
  b_{0x} = b_0 \tanh( y / a), 
\label{eq:b0_tan}
\end{equation}
where $b_0$ is the magnetic field strength and $a$ defines the shear
length (see Fig.\,\ref{fig:tm_theory}). This magnetic field
configuration gives rise to a current sheet at $| y / a| \lesssim 1$.
To balance the resulting magnetic pressure gradient, one can introduce
either a gas pressure gradient, so that
$\nabla_y ( p + b_{0x}^2/2) = 0$ (\emph{pressure equilibrium}
configuration), or an additional magnetic field component, so that
$\nabla_y ( b_{0x}^2/2 + b_{0z}^2/2) = 0$ (\emph{force-free}
configuration). Both equilibrium configurations are stable in ideal
MHD, but are TM unstable in resistive MHD.

FKR63 derived the instability criterion and the growth rate using the
linearised resistive-viscous MHD equations in the incompressible
limit, which read
\begin{eqnarray}
  \partial_t {\bf b} &=& \nabla \times ({\bf v_1} \times {\bf b_0}) + \eta
  \nabla^2 {\bf b},
\label{eq:dt_b} \\
  \rho_0 \partial_t {\bf v_1} &=& - \nabla p + (\nabla \times {\bf b_1})
  \times {\bf b_0} + (\nabla \times {\bf b_0}) \times {\bf b_1}
                                \nonumber \\ & &+ \rho_0 \nu
  \nabla^2 {\bf v_1}, 
\label{eq:dt_v}  \\
  \nabla \cdot {\bf v} &=& 0, \label{eq:div_v}  \\
  \nabla \cdot {\bf b} &=& 0, \label{eq:div_b}
\end{eqnarray}
where ${\bf v}={\bf v_0}+{\bf v_1}$, ${\bf b}={\bf b_0}+{\bf b_1}$ and
we denote background and perturbed quantities with subscripts "$0$"
and "$1$", respectively.  In the incompressible limit,
$|{\bf v_1}| \ll c_s$ holds, which was used to obtain the linearised
equations.  To simplify the notation, we omit hereafter the subscript
``$1$'' for the velocity perturbations, because the background flow is
assumed to be at rest.

FKR63 solved the above equations using a WKB ansatz, \ie
\begin{align}
\velo_y (x,y,t) &= \velo(y) e^{i k x + \gamma t}, \label{eq:wkb1} \\
  b_{1y}(x,y,t) &= b_1(y) e^{i k x + \gamma t}, \label{eq:wkb2}
\end{align}
where $k$ is the wavevector in the $x$ direction, and $\gamma$ is the
growth rate of the TM instability.  This ansatz is justified only if
the time dependence of the background magnetic field can be neglected.
This is the case when the diffusion time scale is much larger than the
instability time scale, \ie $ a^2 / \eta \gg \gamma^{-1}$.  The \alf
crossing time must be sufficiently short too, \ie
$a / \ca \ll \gamma^{-1}$, which is equivalent to considering
instantaneous propagation of \alf waves through the system.  Combining
both conditions, we have
\begin{equation}
  \frac{  a^2 } { \eta } \gg \gamma^{-1} \gg  \frac{   a  }{ \ca}. 
  \label{eq:timescales_main}
\end{equation}

Among other cases, FKR63 also considered perturbations whose
wavelengths in the $x$ direction are comparable to (but smaller than)
the shear width, \ie
\begin{equation}
  k \lesssim  a^{-1} . \label{eq:k_delta_main} 
\end{equation}
For such perturbations, the wavevectors may differ from that of the
fastest growing mode appreciably, and it is possible to set up a
numerical test in which, for a given grid resolution, both the
magnetic shear layer and the TM are well resolved.

FKR63 solved the TM problem in the limit (\ref{eq:k_delta_main}) with
a so-called \emph{boundary layer analysis} (BLA; for details see
Appendix \ref{app:tm_ana}).  They define an inner region 
(see Fig.\ref{fig:tm_theory}) 
where resistive effects are important. We call this region the
\mbox{\it resistive} layer of width $L_{\mathrm{R}}$ or, if the flow
is also viscous, the \mbox{\it resistive-viscous} layer of width
$L_{\mathrm{RV}}$.  Far away from this layer, $|y| \gg L_{\mathrm{R}}$
or $|y| \gg L_{\mathrm{RV}}$, there is an outer region
(see Fig.~\ref{fig:tm_theory})
where resistivity can be ignored and the ideal MHD equations are
valid.  The layer width $L_{\mathrm{R}}$ or $L_{\mathrm{RV}}$ can be
expressed in terms of the physical parameters of the system (see
Appendix \ref{app:tm}).  The velocity perturbation $v_y (x,y,t)$ of
the WKB ansatz (Eq.\,\ref{eq:wkb1}) exhibits two extrema at
$y = \pm L_{\mathrm{R}}$ or $y = \pm L_{\mathrm{RV}}$
(see Fig.~\ref{fig:tm_theory}).
Because the location of these extrema can be determined from our
simulation results, the layer width \mbox{$L_{\mathrm{R}}$} or
\mbox{$L_{\mathrm{RV}}$} is an appropriate quantity to compare
simulation and theory.

In the BLA, the inner solution of the resistive MHD equations in a
linearised background,  \ie
\begin{equation}
   b_{0x}(y) \approx b_0  y / a . 
\label{eq:b_taylor}
\end{equation}
(which holds for $|y| \ll a$; blue dashed line in
Fig.~\ref{fig:tm_theory}), is matched with the ideal MHD solution in
the outer region at some matching point $|y_{\mathrm{m}}|$. The
coordinate $y_{\mathrm{m}}$ has to
fulfill the condition $ L_{\mathrm{R}} \ll |y_{\mathrm{m}}| \ll a$ (or
$ L_{\mathrm{RV}} \ll |y_{\mathrm{m}}| \ll a$), which implies that
these transition region can exist only if
\begin{equation}
L_{\mathrm{R}} \ll a \qquad\text{or}\qquad L_{\mathrm{RV}} \ll a.
\label{eq:scales_main}
\end{equation}

In the inviscid limit, analytic TM solutions of the resistive MHD
equations can be obtained.  For the background magnetic field given by
Eq.\,(\ref{eq:b0_tan}), TM will grow if
\begin{equation}
 a\, k< 1      \ \ \ \mathrm{(instability)}.
\end{equation}
In this case, the growth rate of the TM is
\begin{equation}
\gamma =     1.82 \eta^{3/5}  \left( \frac{  b_0 k}{\sqrt{
    \rho_0} } \right)^{2/5} a^{-6/5} \left(\frac{1}{a\, k} -
       a\, k \right)^{4/5},
\label{eq:gr_ideal_main}
\end{equation}
and the width of the resistive layer is according to our definition
\begin{equation}
L_{\mathrm{R}}=  1.40 \eta^{2/5} \left( \frac{   
\sqrt{\rho_0} }{b_0 k} \right)^{2/5}   a^{1/5}
\left( \frac{1}{a\, k} - a\, k\right)^{1/5}
\label{eq:epsilon_r_main}
\end{equation}
(see equations \ref{eq:gr_ideal} and \ref{eq:yLR}, respectively).  

The growth rate given by Eq.\ (\ref{eq:gr_ideal_main}) is of little
value for our purpose, since it is obtained in the inviscid
limit. However, as we have argued in section~\ref{sec:ansatz}, both
numerical viscosity and resistivity are an unavoidable result of the
discretisation of the equations.  Hence, if we want to use TMs to
measure numerical resistivity, we have to use an approach which takes
into account numerical viscosity as well.  FKR63 also considered the
resistive-viscous case for Prandtl numbers
\begin{equation}
\Pm \equiv \frac{ \nu }{ \eta }   \gtrsim 1,
\label{eq_prandtl}
\end{equation}
in the limit $k \ll a^{-1}$.  Based on their approach, we obtained the
TM growth rate for wavenumbers $k \lesssim a^{-1}$ (see
Eq.\,\ref{eq:gr_visc})
\begin{equation}
 \gamma \approx 0.84 \eta^{5/6} \nu^{-1/6} \left( \frac{b_0
   k}{\sqrt{\rho_0}} \right)^{1/3}
 a^{-4/3} \left( \frac{1}{ a\, k }- a\, k \right),
\label{eq:gr_visc_main}
\end{equation}
and the width of the resistive-viscous layer 
\begin{equation}
L_{\mathrm{RV}} \sim (\eta \nu)^{1/6} \left( a \frac{\sqrt{\rho_0}}{
  b_0 k  } \right)^{1/3}.
\label{eq:epsilon_rv_org_main}
\end{equation}
These expressions should be useful to set up a test to measure
numerical resistivity. However, as we will show in the next sections,
it is difficult to find a region in the numerical parameter space
where Eq.~(\ref{eq:gr_visc_main}) holds, \ie where
Eqs.~(\ref{eq:k_delta_main}), (\ref{eq:timescales_main}),
(\ref{eq:scales_main}), and (\ref{eq_prandtl}) are fulfilled.

%%%%%%%%%%%%%%%%%%%%%%%%%%%%%%%%%%%%%%%%%%%%%%%%%%%%%%%%%%%%%%%%%%%%%%

\subsubsection{Numerical Simulations of Physical TM} 
\label{sec:phys_tm}

To demonstrate the possibility of using a TM simulation to measure
numerical resistivity, we first set up the test using physical
resistivity. This allows us to estimate the reconnection rate as a
function of physical resistivity and viscosity.

Our numerical experiment is based on \citet{llvb}, who were mainly
interested in the non-linear phase of TM, \ie the formation of
magnetic islands and the onset of turbulence. Since we want to study
the exponential growth of a single TM in detail, we modified their
setup for our purposes.  We used a computational box of size
$[-L_x,L_x] \times [-L_y,L_y]$, where $L_x = L_y = \pi/3$, with
periodic and open boundary conditions in $x$ and $y$ direction,
respectively.  We set the density and pressure to $\rho_0 = p_0 = 1$,
and used an ideal-gas EOS with $\Gamma = 5/3$.  In the expression for
the background field, \Eqref{eq:b0_tan}, we set $b_0 = 1$ and $a =
0.1$.

We tested both the pressure equilibrium and force-free configurations
and found that only the latter is suitable for our numerical
experiments \citep[see][for details]{Rembiasz}.  To obtain the
force-free configuration we set
\begin{equation}
  b_{0z} = \frac{ b_0 }{ \cosh(  y /a ) }.
\end{equation}

We note that our initial perturbations differ from those of
\cite{llvb}. As those authors only perturbed the velocity, the TM
instability is triggered promptly for high resistivities only
($\eta \ge 10^{-5}$).  Instead, we perturb both the velocity and the
magnetic field based on an analytic solution of the TM:
\begin{align}
  \velo_{y}(x,y,t=0) &=  \velo (y) \sin(k x),  \label{eq:bibi1}  \\
     b_{1y}(x,y,t=0) &=  b_1(y) \cos(k x) \label{eq:bibi2}.
\end{align}
The function $\velo(y)$ is given by Eqs.\ (\ref{eq:v1y_out}) and
(\ref{eq:Phi_integral}) for $|y|\ge y_{\mathrm{m}}$ and
$|y|\le y_{\mathrm{m}}$, respectively, where $y_{\mathrm{m}}$ is the
matching point (typically $y_{\mathrm{m}} = 2 L_{ \mathrm{RV}}$).
\citet{Landi_et_al_2015} used similar perturbations, \ie
eigenfunctions of the TM, in their studies of what they called ``an
ideal TM''.  This ideal case is a solution of the TM problem in a
regime first studied by FKR63, but different from the one we consider
here.

The function $b_1(y)$ in Eq.\,(\ref{eq:bibi2}) is given by
Eq.\,(\ref{eq:b1y_out}) for $|y|\ge y_{\mathrm{m}}$, and it is
constant for $|y|\le y_{\mathrm{m}}$, \ie
$b_1(y \le y_{\mathrm{m}}) = b_1(y_{\mathrm{m}})$ according to the
\emph{constant $\psi$ approximation} (see Appendix \ref{app:tm}).  The
remaining perturbations $v_{1x}(x,y,t=0)$ and $b_{1x} (x,y,t=0)$ are
determined from the divergence free conditions
$\divv {\bf b } = \divv {\bf v} = 0$.  To reduce the computational
cost, we chose the value of $k$ in such a way that exactly one TM fits
into the box, \ie $k=3$.

\begin{figure}
  \centering
  \includegraphics[width=0.49\textwidth]{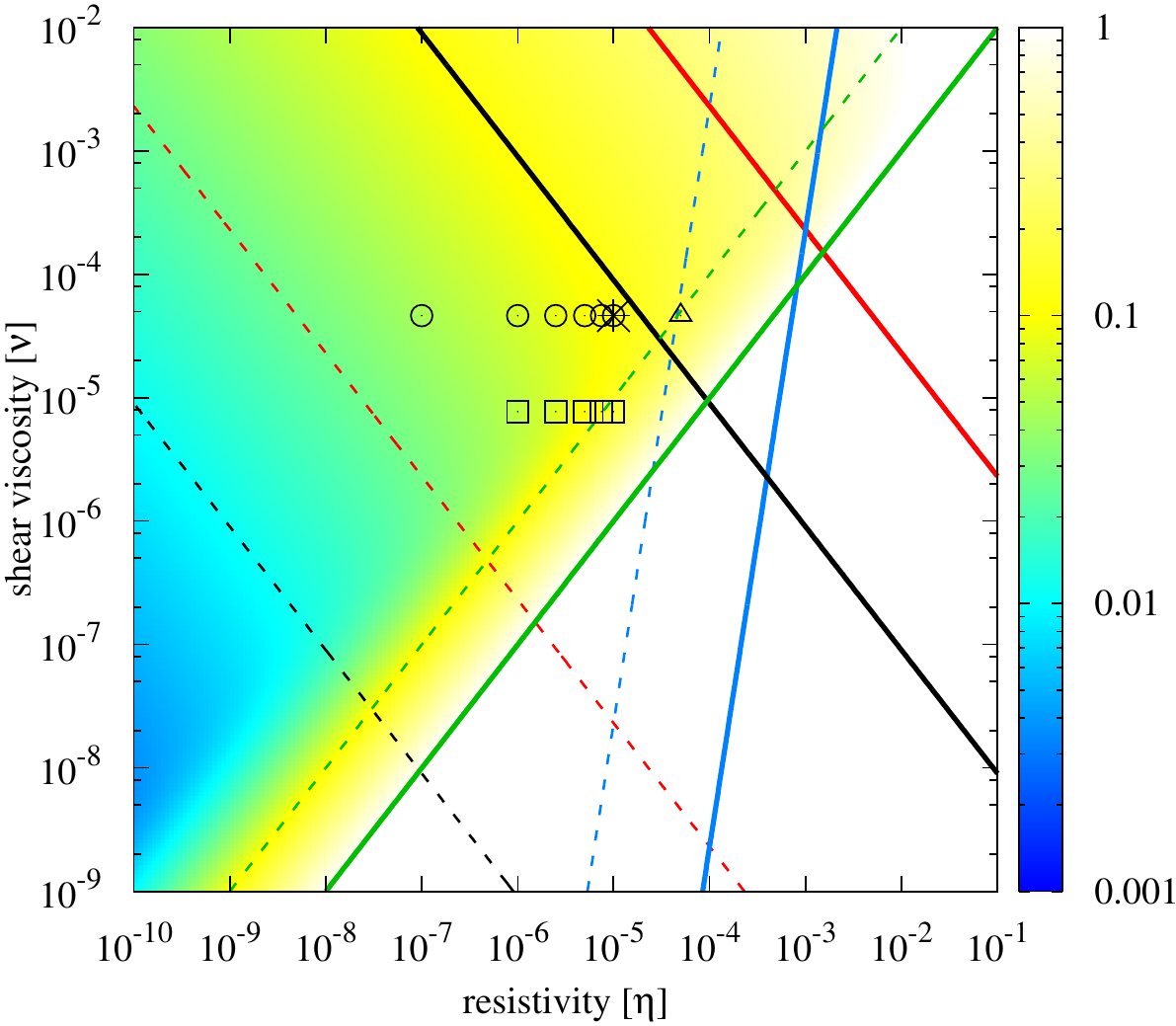}
  \caption{Parameter space of our numerical TM simulations and
    validity limits of the analytical expression for the growth rate
    $\gamma$ given by Eq.\,(\ref{eq:gr_visc_main}).  The color map
    gives the value of the quantity
    $(C_1^{-2}+C_2^{-2}+C_3^{-2} + 0.01P_{\rm m}^{-2})^{1/2}$, and the
    red, blue, and black solid and dashed lines show the locations
    where $\mathcal{C}_1$, $\mathcal{C}_2$, and $\mathcal{C}_3$ (see
    Eqs.\,\ref{eq:C1} - \ref{eq:C3}) have a constant value of $10$ and
    $100$, respectively. Solid and dashed green lines show the
    locations where $P_{\mathrm m}$ has a value $0.1$ and $1$,
    respectively.  Symbols indicate the parameter values obtained for
    the reference simulation \tr{\#Rf} (asterisk), and the
      simulations \tr{\#TMa} (circles), \tr{\#TMc} (squares), and
    \tr{\#TMc} (triangles). In the blue shaded region the conditions
    given by Eqs.\,(\ref{eq:C1}) - (\ref{eq:C3}) are satisfied best.}
\label{fig:parspace}
\end{figure}

To compare the results of our TM simulations with the analytical
predictions of Eqs.~(\ref{eq:gr_visc_main}) and
(\ref{eq:epsilon_rv_org_main}) for the TM growth rate and the width of
the resistive viscous layer, respectively, we must ensure that we are
in the regime of applicability of these analytical predictions, \ie
Eqs.~(\ref{eq:k_delta_main}), (\ref{eq:timescales_main}),
(\ref{eq:scales_main}), and (\ref{eq_prandtl}) should be fulfilled.
The first condition (Eq.~\ref{eq:k_delta_main}) is ensured by our
choice of $k$ and $a$.  The other conditions can be written as
\begin{align}
\mathcal{C}_1 \equiv &\, \gamma a^{2} \eta^{-1} \gg 1, \label{eq:C1}\\
\mathcal{C}_2 \equiv &\, a^{-1} \ca \gamma^{-1} \gg 1, \label{eq:C2}\\
\mathcal{C}_3 \equiv &\, a L_{\mathrm RV}^{-1} \gg 1, \label{eq:C3}\\
P_{\mathrm m} \equiv &\, \nu \eta^{-1} \gtrsim 1. \label{eq:Pmmin}
\end{align}
We plot iso-contours of these four quantities in the $\eta$-$\nu$
plane (see Fig.~\ref{fig:parspace}) to locate the region where the
analytical expressions are valid.

%----------------------------------------------------------------------------
%----------------------------------------------------------------------------

We first discuss the results of a simulation with $\eta=10^{-5}$ and
$\nu=10^{-4}$, which we call {\it reference model}  (\tr{\#Rf})
and which is marked by an asterisk in Fig.\,\ref{fig:parspace}. The
\emph{first condition} (Eq.\,\ref{eq:C1}) is only marginally satisfied
for the reference model ($\mathcal{C}_1 \approx 25$).  To improve the
situation, one should decrease the resistivity and viscosity, \ie one
should increase the grid resolution. This would place the model
towards the lower left corner of Fig.~\ref{fig:parspace}, where all
the conditions are better satisfied. Therefore, we are limited here by
the numerical resolution that we can afford. In the following, we
present simulations with numerical resistivities and viscosities as
low as $10^{-7}$, corresponding to values of $\mathcal{C}_1$ in the
range $1< \mathcal{C}_1 < 100$, which are {\em marginally} consistent
with (Eq.\,\ref{eq:C1}).  As a result, the diffusion timescale is only
about ten times larger than the TM e-folding time, \ie we observe
diffusion of the background solution within the duration of the
simulation.  We circumvent this problem by solving instead of the
proper induction equation (\ref{eq:induction}) a modified (physically
incorrect) version for a constant resistivity, $\eta$:
\begin{equation}
\label{eq:trick}
  \partial_t {\bf b} = \nabla \times ({\bf v} \times {\bf b}) + \eta
  \nabla^2 ( {\bf b } - {\bf b_0}) .
\end{equation}
Thereby, the background magnetic field, ${\bf b_0}$, does not suffer
from diffusion by resistivity.

The \emph{second condition} (Eq.\ref{eq:C2}) yields
$\mathcal{C}_2 \approx 403 \gg 1$ for the reference model, \ie the
\alf crossing time is sufficiently small compared to the growth time
scale of the TM instability.

The \emph{third condition} (Eq.\,\ref{eq:C3}) is
$\mathcal{C}_3 \approx 10$ for the reference model, \ie it is only
roughly fulfilled.  This condition is the most challenging one to be
met in numerical simulations, because the size of the
resistive-viscous layer $L_{\mathrm RV} $ has to be much smaller than
the width $a$ of the magnetic shear layer. This can be achieved again
by decreasing viscosity and resistivity, but since
$L_{\mathrm RV} \propto (\eta\nu)^{-1/6}$
(Eq.\,\ref{eq:epsilon_rv_org_main}) it is necessary to decrease
$\eta\nu$ by six orders of magnitude to decrease $L_{\mathrm RV}$ by a
factor of 10. Thus, if we aim for $\mathcal{C}_3 \approx 100$, we need
$\sim 10^4$ grid points for each box dimension to resolve the
resistive-viscous layer with $\sim 10$ grid points.

The \emph{fourth condition} (Eq.\,\ref{eq:Pmmin}) is satisfactorily
fulfilled, since $P_{\mathrm m} = 10 \gtrsim 1$.

\begin{figure*}%%  
  \includegraphics[width=0.45\textwidth]{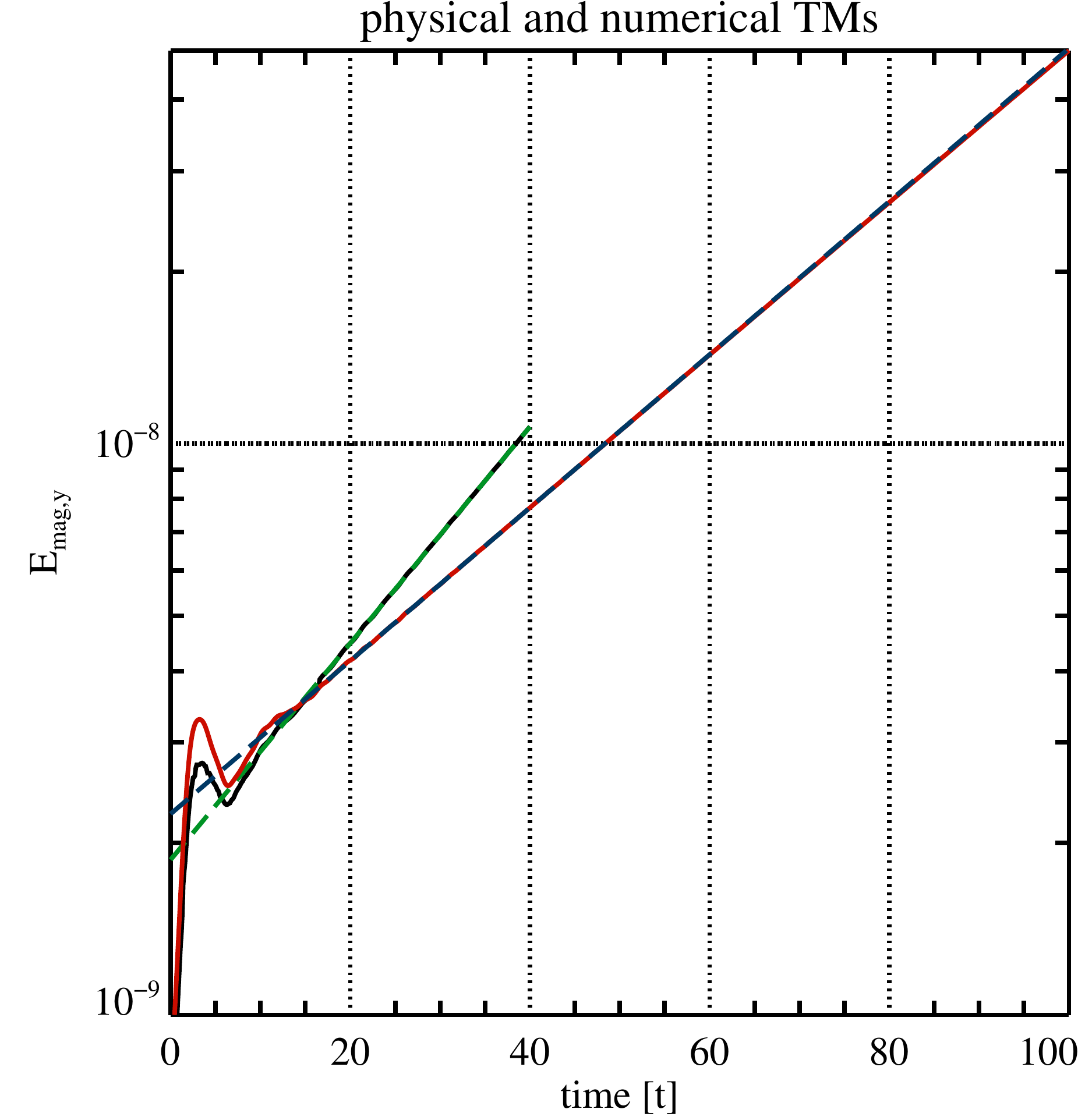}  \phantom{MM}
  \includegraphics[width=0.45\textwidth]{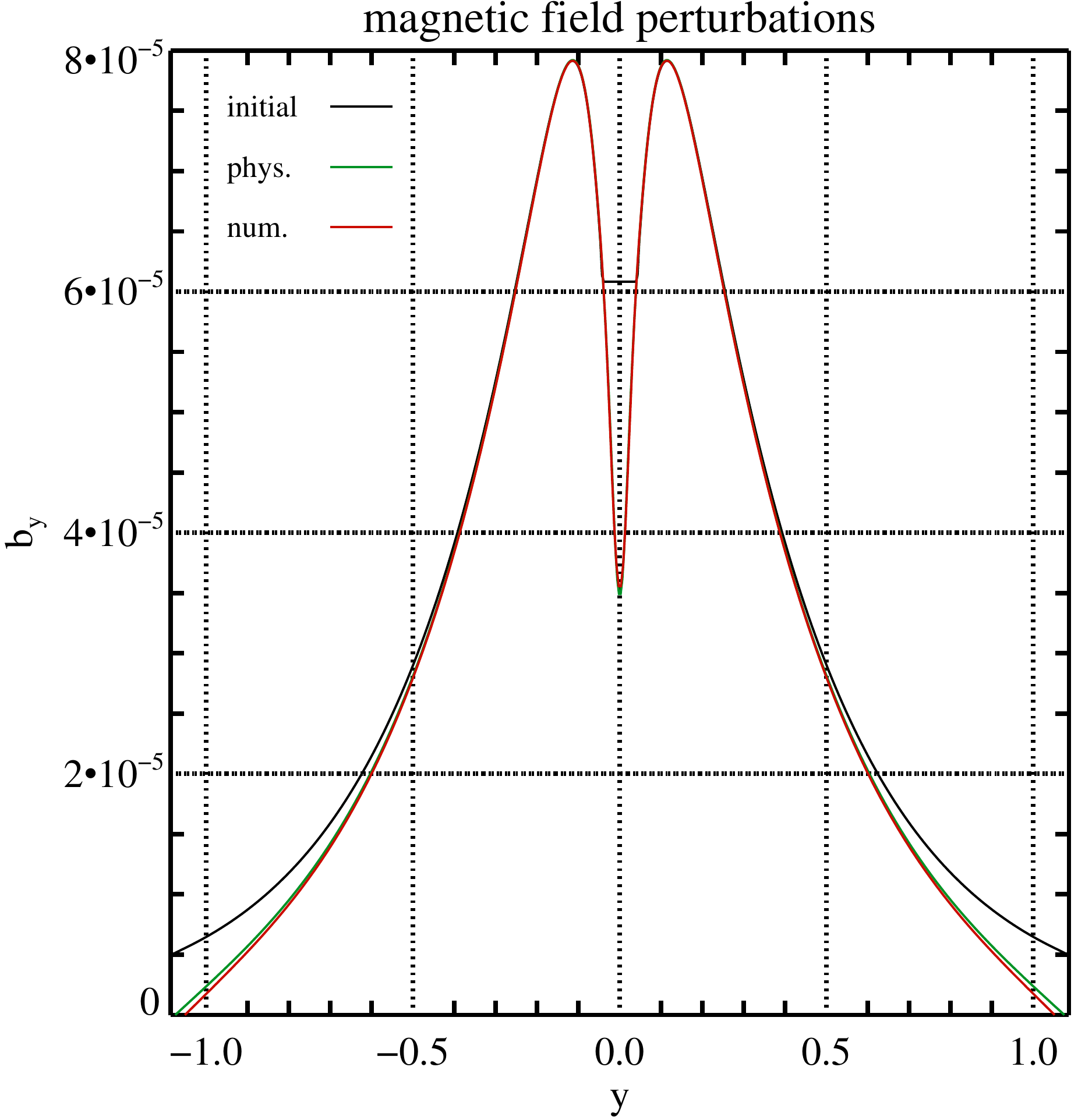} \\
  \includegraphics[width=0.45\textwidth]{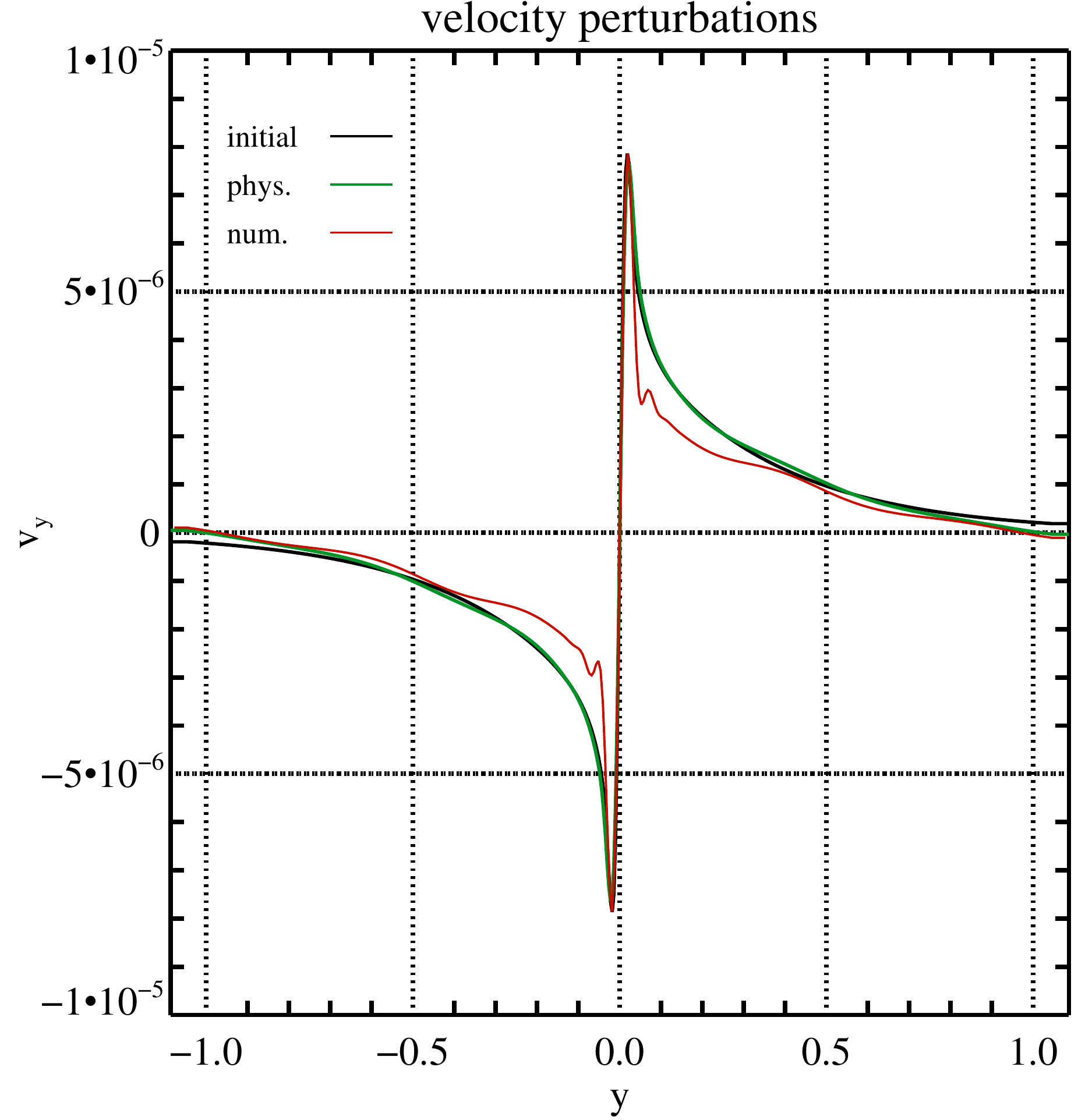}  \phantom{MM}
  \includegraphics[width=0.45\textwidth]{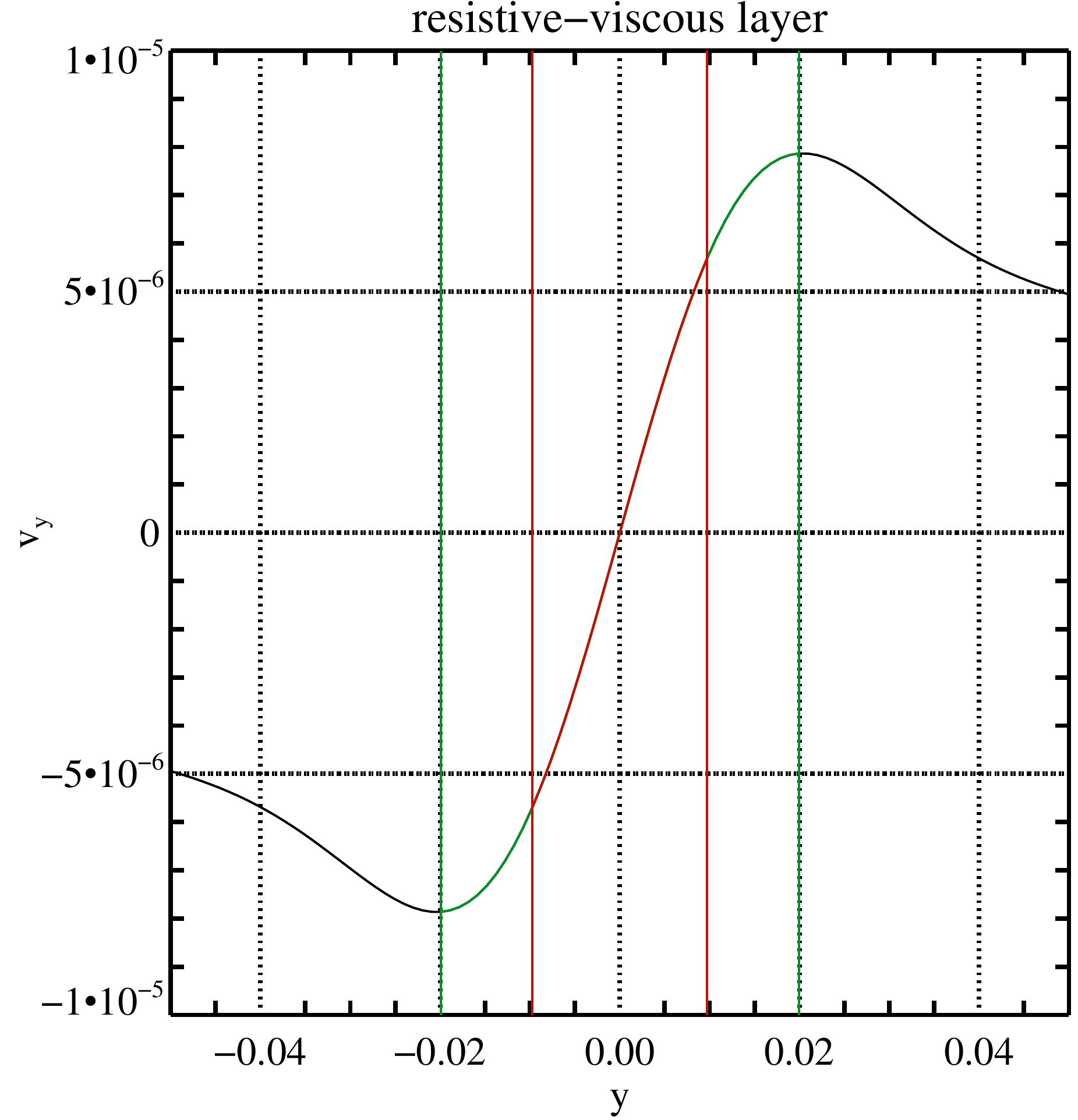}
  \caption{
    {\it Reference simulation} (\tr{\#Rf}, $2048 \times 2048$
    zones, MP9, HLL Riemann solver) of TM driven by physical
    resistivity ($\eta = 10^{-5}$) and \emph{exemplary simulation}
    (\tr{\#Ex}, $384 \times 384$ zones, MP5, LF Riemann solver) of 
    TM driven by numerical resistivity (\ie $\eta = 0$).  The 
    numerical resistivity is negligible for \tr{\#Rf} 
    ($\eta_{\ast} \ll \eta = 10^{-5}$), and has value of
    $\eta_{\ast} = 6.5 \times 10^{-6}$ for \tr{\#Ex} (marked by the 
    third rightmost asterisk in \figref{fig:tm_reso}).
    \emph{Top left:} evolution of the magnetic energy,
    $\bar{E}_{\mathrm{mag},y}$ (Eq.\,\ref{eq:byby}) for simulation
    \tr{\#Rf} (black line) and \tr{\#Ex} (red line)
    together with linear fits to the logarithm of
    $\bar{E}_{\mathrm{mag},y}$ for $t \in [20,40]$ (green and blue
    dashed lines) from which the instability growth rate can be
    measured.
    \emph{Top right:} initial (black) and evolved profiles of the
    magnetic field perturbation $b_{1y}$ at $x=-0.5$ in simulation
    \tr{\#Rf} (at $t=40$; green) and \tr{\#Ex} (at $t=100$; red)
    normalised to the ratio $\max{|b_y(t=0)|} / \max{|b_y(t=100)|}$.
    \emph{Bottom left:} same as upper right panel, but showing the
    velocity perturbation $v_y$ at $x=0$. The two extrema at 
    $y \approx \pm 0.02$ determine the width of the resistive-viscous 
    layer $L_{\rm RV}$.
    \emph{Bottom right:} zoom of the bottom left panel near the
    current sheet showing the evolved velocity perturbation for 
    simulation \tr{\#Rf}. The two green and two red vertical lines, 
    which are located at $y = \pm L_{\mathrm{RV}}$ and
    $y = \pm \epsilon_{\mathrm{RV}}$ (Eq.\,\ref{eq:epsilonr_def}), 
    respectively, bracket the resistive viscous layer and the region 
    where the modulus of the dimensionless parameter $\tilde{s}$  
    (Eq.\ \ref{eq:epsilon_rv_org}) is smaller than one.  
  }
\label{fig:tm_example_4_in_1}
\end{figure*}

The reference simulation \tr{\#Rf} was performed with the HLL
Riemann solver, a MP9 reconstruction scheme, and with a grid of
$2048 \times 2048$ zones (\figref{fig:tm_example_4_in_1}).  We find
that the initially imposed magnetic field and velocity perturbations
do not evolve much with time, except for a growth of their
amplitudes. This indicates that our initial perturbations, which are
based on the TM solution in resistive-non-viscous MHD (in the
\emph{constant $\psi$ approximation}), are very similar to the
eigenfunctions of resistive-viscous TM.

%----------------------------------------------------------------------------
%%
%----------------------------------------------------------------------------

The upper right panel of \figref{fig:tm_example_4_in_1} shows profiles
in $y$ direction of the initial (black) and the evolved (at $t = 40$;
green) magnetic field perturbations at $x = -0.5$, the latter being
normalised to the ratio $\max{|b_y(t=0)|} / \max{|b_y(t=100)|}$.  The
corresponding velocity perturbation at $x = 0$ (bottom panels) exhibit
two pronounced extrema surrounding the magnetic shear layer (marked by
the two vertical green lines in the lower right panel, which is a zoom
of the lower left panel), which are characteristic of TMs.

To measure the TM growth rate, we compute the evolution with time of
the quantity
\begin{equation}
 \bar{E}_{\mathrm{mag},y} \equiv \int_{-L_x}^{L_x} \int_{-a}^{a} 
                               \frac{ b_y^2}{2}\ \mathrm{d}x
                               \mathrm{d}y \, ,
 \label{eq:byby}
\end{equation}
where the integration is performed only up to $|y| \le a$ to reduce a
potential influence of boundary conditions.  After an initial
transient lasting up to 20 time units during which the initial
perturbation adjusts to the analytic solution,
$\bar{E}_{\mathrm{mag},y}(t)$ grows exponentially at a constant
rate. Since $b_y \propto \exp(\gamma t)$,
\begin{equation}
 \frac{1}{2} \ln \left( \bar{E}_{\mathrm{mag},y} \right) =  
 \gamma t + \mathrm{const.},
\label{eq:logbyby}
\end{equation}
where the constant depends on the initial perturbation amplitude and the box size. Using the above equation, we compute
the instability growth rate by means of a simple linear regression.  The black line in the upper left panel of
Fig.\,\ref{fig:tm_example_4_in_1} shows the time evolution of $\bar{E}_{\mathrm{mag},y}$, while the green dashed line is
the linear fit according to Eq.\,(\ref{eq:logbyby}) -note that both lines are almost indistinguishable after the initial
transient time.

To obtain the width of the resistive viscous layer, we plot
$v_y(x=0,y)$ for every simulation at $t=30$ and measure the locations
$L^{+}_{\mathrm{RV}}$ and $L^{-}_{\mathrm{RV}}$ of the two velocity
extrema (see vertical green lines in the bottom right panel of
Fig.\,\ref{fig:tm_example_4_in_1}).  To attribute a measurement error,
we note that the extremum can be located anywhere inside of the
corresponding computational zone of vertical size $\deltx y$.  Thus,
the actual location of the extremum is uncertain up to an error
$\pm \deltx y / 2$, \ie the layer width is
\begin{equation}
L_{\mathrm{RV}} = \frac{L^{+}_{\mathrm{RV}} -
  L^{-}_{\mathrm{RV}}}{2} \pm \frac{\deltx y}{2}.
\end{equation}

The methodology explained above to measure the growth rate $\gamma$
and the width of the resistive-viscous layer $L_{\mathrm{RV}} $ was
applied to all TM simulations discussed below.  To understand the
dependence of these quantities on the different relevant parameters
and to compare with the analytic results, we performed several series
of simulations exploring the parameter space in the neighbourhood of
the {\it reference model}, by varying $\eta$, $\nu$, $b_0$ and
$k$. Details of these simulations can be found in
Appendix~\ref{app:tm}.

The main result extracted from this set of (numerically converged)
simulations is the disagreement between the numerically obtained
growth rates and the analytic ones given by
Eq.~(\ref{eq:gr_visc_main}).  The most likely explanation for the
discrepancy is that the parameters of our TM simulations are outside
of the regime of validity of the analytic results, particularly
because of the difficulty to guarantee $L_{\mathrm RV} \ll a$.
Unfortunately, this means that the analytic expressions\
(\ref{eq:gr_visc_main}) and\ (\ref{eq:epsilon_rv_org_main}) cannot be
used to measure numerical resistivity.  Thus, we decided to use an
empirical approach to the problem.

\begin{figure}%[h]%[t]
  \centering
  \includegraphics[width=0.45\textwidth]{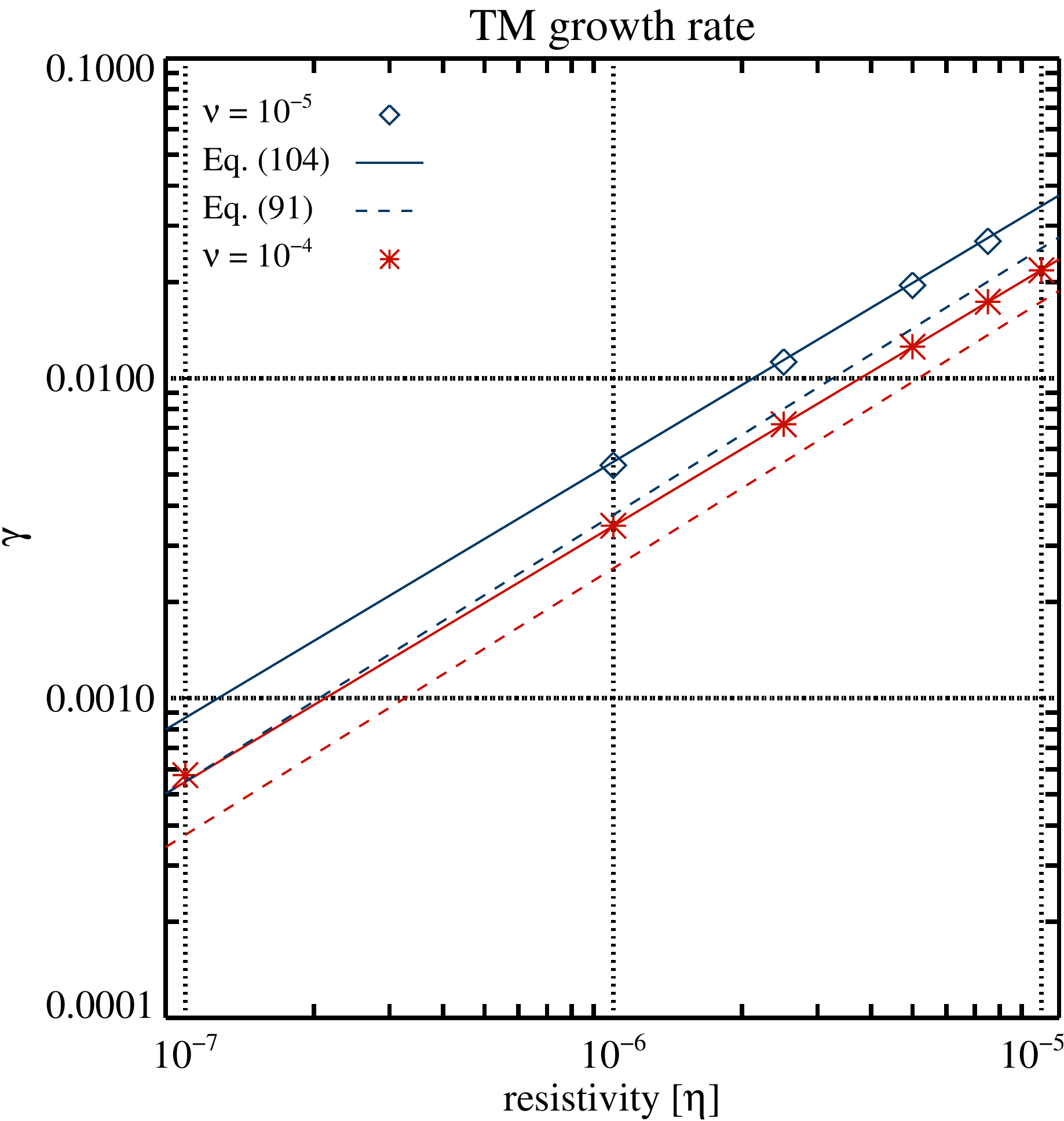} \\*[0.1cm]
  \includegraphics[width=0.45\textwidth]{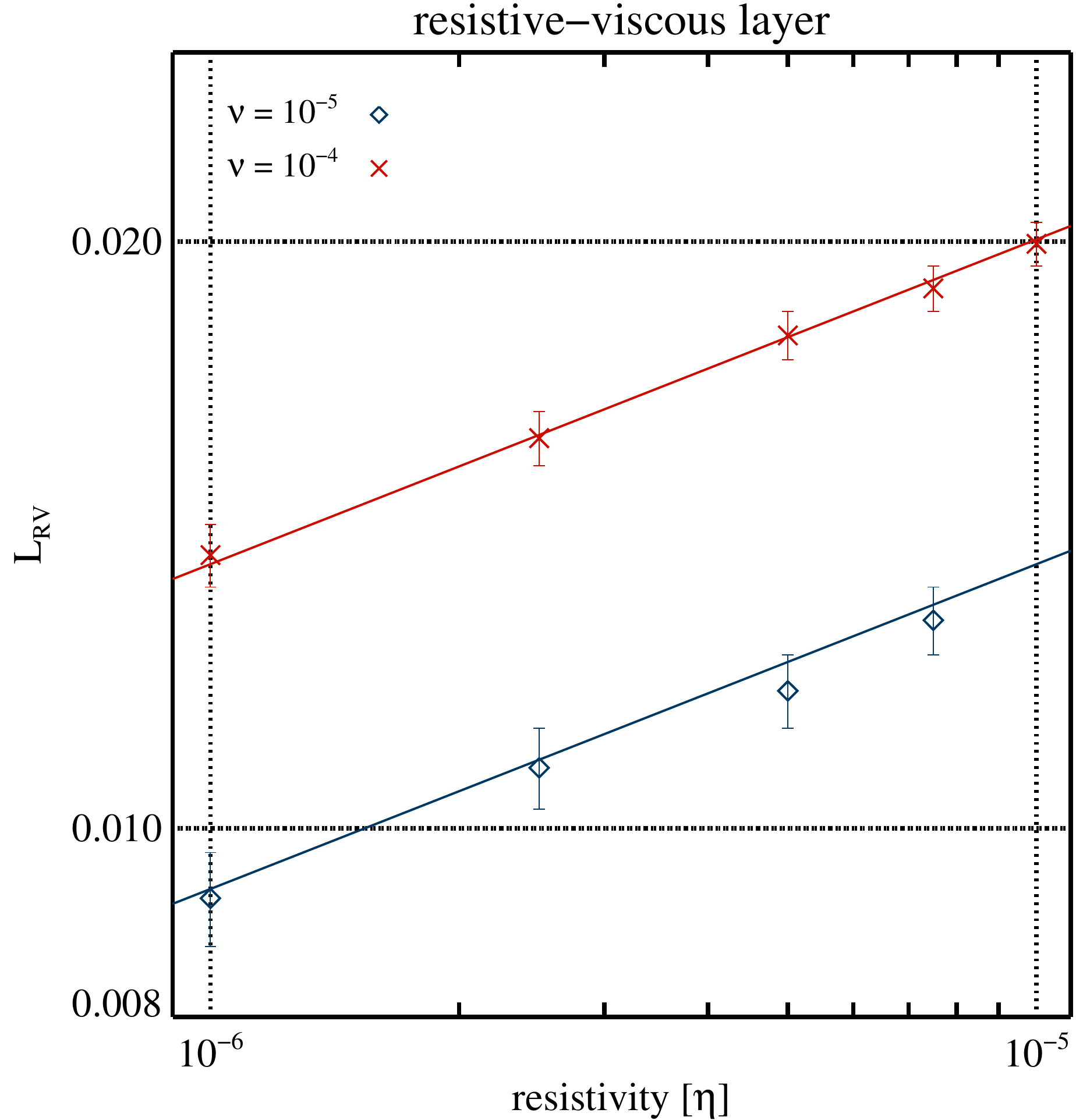}
  \caption{ 
TM growth rate (\emph{top}) and resistive-viscous layer
    width (\emph{bottom}) as a function of resistivity.  Blue diamonds
    and red asterisks denote results of simulations performed with a
    constant viscosity of $\nu = 10^{-5}$ and $\nu = 10^{-4}$,
    respectively. The simulations employed the HLL Riemann solver,
    MP9, and a grid resolution of $2048 \times 2048$ zones. 
    The solid lines are empirical expressions given by 
    \eqsref{eq:gr_calib_main} and (\ref{eq:vl_calib_main}),
    respectively. The dashed lines in the upper panel are the 
    analytical predictions given by Eq.\,(\ref{eq:gr_visc_main}).}
\label{fig:tm_calib_theory}
\end{figure}

Using the insight gained from the theoretical work of FKR63, we
postulate an ansatz for the dependence of both the TM growth rate and
the width of the resistive-viscous layer on the physical parameters,
which we then calibrate using the series of numerical simulations
mentioned above.  The whole procedure is described in detail in
Appendix~\ref{app:tm}.  The empirical expressions resulting for
$k = 3$ and $a = 0.1$ are
\begin{align}
\gamma(k=3,a = 0.1) &=  34.56  \eta^{4/5} \nu^{-1/5} \left(
  \frac{b_0}{\sqrt{\rho_0}} \right)^{2/5} , 
\label{eq:gr_calib_main} \\
  L_{\mathrm{RV}} (k=3,a=0.1)  &=   0.634  (\eta \nu)^{1/6} \left(
  \frac{b_0}{\sqrt{\rho_0}} \right)^{-1/3}.
\label{eq:vl_calib_main}
\end{align}
Figure \ref{fig:tm_calib_theory} shows the TM growth rates
(\emph{upper panel}) and the width of the resistive-viscous layer
(\emph{lower panel}) measured from two series of simulations done with
a viscosity $\nu = 10^{-4}$ (\emph{blue diamonds}) and $\nu = 10^{-5}$
(\emph{red asterisks}) for different values of the resistivity.  Solid
lines represent the empirical expressions given by
Eqs.\,\bref{eq:gr_calib_main} and \bref{eq:vl_calib_main}, while the
analytic results given by Eq.~(\ref{eq:gr_visc_main}) are plotted with
dashed lines in the upper panel of \figref{fig:tm_calib_theory}.  The
discrepancy in the growth rate between analytic and numerical results
is obvious, whereas our empirical expression (\ref{eq:vl_calib_main})
for the width of the resistive-viscous layer is compatible with the
analytical one (Eq.\,\ref{eq:epsilon_rv_org_main}).

%----------------------------------------------------------------------------
%----------------------------------------------------------------------------

%%%%%%%%%%%%%%%%%%%%%%%%%%%%%%%%%%%%%%%%%%%%%%%%%%%%%%%%%%%%%%%%%%%%%%

\subsubsection{Numerical TM}
\label{sec:num_tm}

With the knowledge acquired from the resistive-viscous simulations of
the previous section, we can tackle the problem of estimating the
numerical resistivity of the code.  If we perform a simulation with
$\eta=0$, the development of TM signals the presence of a non-zero
numerical resistivity $\eta_{\ast}$, because TM are not present in
ideal MHD.

For the numerical setup presented in the previous section and for a
viscosity $\nu=10^{-4}$, the TM growth rate should be well described
by Eq.\,(\ref{eq:gr_calib_main}), if $\eta_{\ast} \lesssim 10^{-5}$.
In this case, we can determine $\eta_{\ast}$ using the expression
\begin{equation}
 \eta_{\ast} =  \left( \frac{ \gamma(k=3,a = 0.1) }{34.56} \right)^{5/4}
                \nu^{1/4} \left( \frac{\sqrt{\rho_0}}{b_0} \right)^{1/2},
\label{eg:resi_from_gr_main}
\end{equation}
where we need to measure only the growth rate of the instability,
$\gamma$, for a simulation with $k=3$ and $a=0.1$.

Alternatively, one could measure the resistive-viscous layer width
(Eq.\,\ref{eq:vl_calib_main}) to obtain $\eta_{\ast}$ from the
expression
\begin{equation}
 \eta_{\ast} =  \left( \frac{
   L_{\mathrm{RV}}(k=3,a = 0.1)  }{0.634} \right)^{6} \nu^{-1} \left(
  \frac{b_0}{\sqrt{\rho_0}} \right)^2.
\end{equation}
This method is much less accurate, however, because measuring
$L_{\mathrm{RV}}$ from a simulation is prone to rather large relative
errors (of the order of $0.1$), and because
$ \eta_{\ast} \propto L_{\mathrm{RV}}^6$.

To compute the numerical resistivity from
Eq.~(\ref{eg:resi_from_gr_main}) we need to know the value of the
viscosity $\nu$.  However, for a coarse numerical resolution the value
of the numerical viscosity can be of the same order. Therefore, we
should require $\nu \gg \nu_{\ast}$.  Expecting that the numerical
resistivity and viscosity are of the same order, we had to choose a
value of $\nu$ that is larger than the typical values of both
numerical resistivity and numerical viscosity. On the other hand,
$\nu$ must not be too large because the growth rate of the instability
decreases with increasing $\nu$, \ie more expensive simulations are
required.  The size of the resistive-viscous layer also grows with
$\nu$ and may become comparable to $a$, thus violating the condition
$L_{RV} \ll a$, \ie Eq.\,(\ref{eq:gr_calib_main}) no longer holds.  As
a compromise, we chose a value of $\nu=10^{-4}$ and performed all
simulations with sufficiently high resolution to ensure that
$P_{\mathrm m} \gg 1$.

Eq.\,(\ref{eq:gr_calib_main}) was obtained from numerical simulations
in which we removed the background field from the resistive term of
the MHD equations (see Eq.~\ref{eq:trick}) to prevent diffusion of the
background field.  The simulations to be discussed in the remainder of
this section did not require this measure, because they were performed
\emph{without} physical resistivity.  In spite of this difference, we
can still apply the calibration obtained in the former series of
simulations, because we find that the results of both series of
simulations are consistent (TM develop in both cases, but the
background magnetic field dose not diffuse).  Hence, numerical
resistivity seems to act independently on large scales (diffusion of
the background field across the box) and small scales (development of
TM). This finding confirms our ansatz, which postulates that the value
of numerical resistivity differs for phenomena occurring at different
length scales, because $\eta_{\ast}$ depends on the typical length
($\LL$) and velocity ($\vv$) of the flow.
 
To determine the dependence of the numerical resistivity on the three
Riemann solvers (LF, HLL, HLLD), we performed simulations with the MP5
reconstruction scheme, the RK3 time integrator, a CFL factor of $0.7$,
and grids of $128 \times 128 $ to $1024 \times 1024$ zones. The default
physical parameters were $a = 0.1$, $k =3$, $b_0 = \rho_0 = 1$,
$\nu = 10^{-4}$, and $\eta=0$.  

%----------------------------------------------------------------------------
\begin{figure}%%[h]
  \centering
 \includegraphics[width=0.45\textwidth]{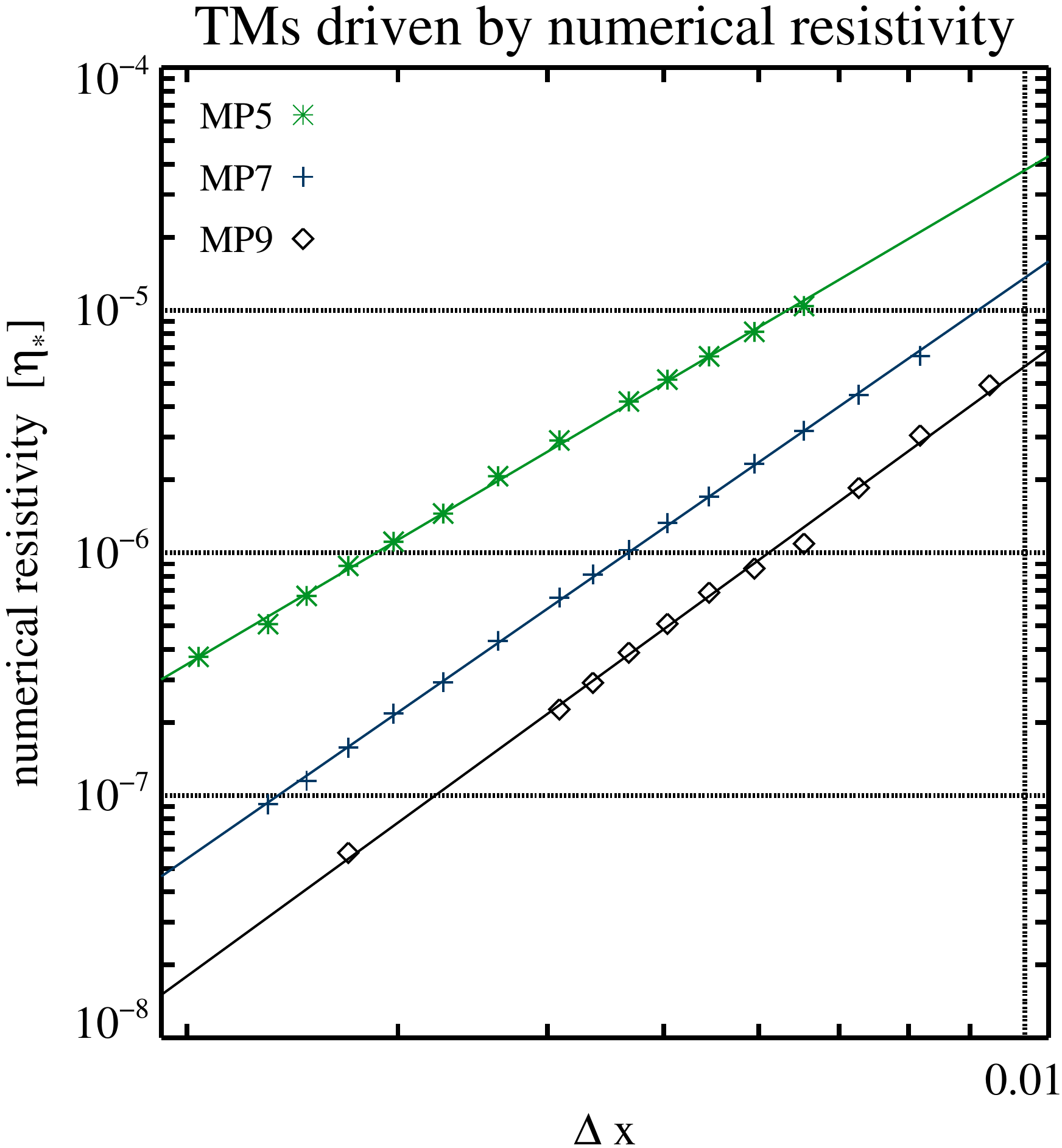}
 \caption{Numerical resistivity of TM simulations performed with $224$
   to $1024$ zones (per dimension), the LF Riemann solver, a viscosity
   $\nu = 10^{-4}$, no (physical) resistivity, and the MP5 (green
   asterisks), MP7 (blue plus signs), and MP9 reconstruction scheme
   (black diamonds).  The straight lines are linear fits to the data.
 }
  \label{fig:tm_reso}
\end{figure}%%
%----------------------------------------------------------------------------

We find that TM are instigated by numerical resistivity for the LF
solver. In the simulations performed with the HLL and HLLD Riemann
solvers no TM are observed, \ie the numerical resistivity resulting
from these solvers, although undetermined, must be much less than that
of the LF solver. For the latter solver, as expected, the higher the
grid resolution the smaller the instability growth rate, \ie the lower
the numerical resistivity (see Fig.\ \ref{fig:tm_reso}), since
$\gamma \propto \eta^{4/5}$ (Eq.\ \ref{eq:gr_calib_main}).  Coarsening
the grid resolution, the numerical resistivity eventually becomes so
high that the width of the resistive-viscous layer is so large that
Eq.\,(\ref{eq:gr_calib_main}) is invalid, and we can no longer
precisely measure the magnitude of the numerical resistivity.  The
resolution limit depends on the order of the reconstruction scheme,
being $320$, $256$, and $224$ zones per dimension for the MP5, MP7,
and MP9 scheme, respectively.

The results of an exemplary simulation (\tr{\#Ex}) without
resistivity obtained with MP5 reconstruction scheme on a grid of
$384 \times 384$ zones are shown in \figref{fig:tm_example_4_in_1}.
Like in the reference model \tr{\#Rf} (with $\eta = 10^{-5}$; black
dashed-dotted line in upper left panel), a TM grows exponentially with
time in model \tr{\#Ex} (red dashed-dotted line in the panel), this
time being driven by numerical resistivity (in this simulation
$\eta_{\ast}= 6.5 \times 10^{-6}$ marked with the third rightmost
asterisk in \figref{fig:tm_reso}).  The TM induced growth of the
magnetic field (upper right panel of \figref{fig:tm_example_4_in_1})
and velocity (bottom left panel) perturbations in model \tr{\#Ex}
(without resistivity; red lines) are similar to those in model
\tr{\#Rf} (with resistivity; green lines).  This comparison clearly
demonstrates that the behaviour of the numerical resistivity closely
resembles that of (real) physical resistivity.

We anticipate  that the main contribution to the numerical resistivity
comes from 
the $y$-direction.  All variables exhibit a much stronger variation in
$y$-direction than in $x$-direction.  Hence, the characteristic length
scales in our multi-D ansatz for the
numerical resistivity (analogous one to ansatz~(\ref{eq:visc:2D}) for
numerical shear viscosity), are much larger in
$x$-direction than in $y$-direction, $\LL_x \gg \LL_y$.  More
specifically, we can preliminarily 
estimate that $\LL_x \propto k^{-1} \propto L_x$ and
$\LL_y \lesssim a$.  Consequently, the total numerical errors coming
from the $x$-direction will be
negligible compared to the ones due to the discretisation in $y$,
allowing us to use the simpler one-dimensional ansatz in the following. 
Therefore, in the further discussion of TMs, we will refer to the 1D
ansatz for numerical resistivity (Eq.\ \ref{eq:eta*}) for the sake of
simplicity. Furthermore, we will use $\Delta x$ instead of $\Delta y$,
since they are equal in our simulations, having in mind, however that
the main contribution comes form errors proportional to
$(\Delta y / \LL_y )^r$. 
A similar situation occurred in the 2D simulation of sound waves (series
\tr{\#LS6}--\tr{\#LS11} from \tabref{tab:2d_waves}) in which 
$ ( \Delta x / \LL_x )^r \gg  (\Delta y  / \LL_y )^r$ and therefore
the contribution of the $y$ sweep to the numerical dissipation was
negligible and one could equally well use 1D ansatzes for numerical
dissipation. 

The dependence of the numerical resistivity (which is determined from
the measured growth rate of the instability using
Eq.~\ref{eg:resi_from_gr_main}) on the grid resolution is shown in
Fig.\,\ref{fig:tm_reso}. 
The results are fitted with the functions
\begin{align}
 \ln (  \gamma ) &=  \alpha_5 \ln ( \deltx x ) + c_5, \nonumber \\
 \ln (  \gamma ) &=  \alpha_7 \ln ( \deltx x ) + c_7, \nonumber \\
 \ln (  \gamma ) &=  \alpha_9 \ln ( \deltx x ) + c_9,
\label{eq:ln_gamma_fit}
\end{align}
where $\alpha_i$ and $c_i$ are the coefficients of the MP
reconstruction scheme of $i-$th order. Their values are (for a CFL
factor in the range $0.1$ to $0.7$)
\begin{align}
 \alpha_5 &= 2.35  \pm 0.02,  & c_5= 9.85  \pm  0.13 , \nonumber  \\
 \alpha_7 &= 2.740 \pm 0.015, & c_7= 11.06 \pm  0.09 ,  \nonumber \\
 \alpha_9 &= 2.88  \pm 0.05,  & c_9= 11.1  \pm  0.3 .
\label{eq:ln_gamma_fit_results}
\end{align}

If the numerical resistivity scales as
$\eta_{\ast} \propto \vv \LL \, (\deltx x /\LL)^{r}$ as in
Eq.\,(\ref{eq:eta*}), one would naively expect that the growth rate
scales as $\gamma \propto (\Delta x)^{4 r /5}$, \ie the expected
theoretical values should be $\alpha_5 = 4$, $\alpha_7 = 5.6$, and
$\alpha_9 = 7.2$, which do not agree with our results.  As we explain
below, this argumentation is wrong, however, because it fails to
account for an (implicit) dependence of the quantities $\vv$ and $\LL$
on $\eta_{\ast}$.

To explain the apparent considerable reduction of the convergence
order $r$ of the MP reconstruction schemes in
(\ref{eq:ln_gamma_fit_results}), we need to have a careful look at the
ansatz (\ref{eq:eta*}) for the numerical resistivity (neglecting the
contribution of time integration errors)
\begin{equation}
  \eta_{\ast} = \nn_{\eta}^{\deltx x} \timex \vv \timex \LL \timex
                \bigg( \frac{\deltx x}{\LL}\bigg)^{r},
\label{eq:eta*tm}
\end{equation}
where $\vv$ and $\LL$ are the system's characteristic speed and
length, respectively.

If we were to assume $\LL \propto a$ (which is constant), we would
obtain $r_i = (5/4) \alpha_i$.  The conceptual mistake we have made
here is that $a$ is the correct choice for the characteristic length
of the background magnetic field diffusion problem, but not for a TM
whose length scale is much smaller than the shear width.  It turns out
(as we demonstrate below) that the characteristic length of the system
is proportional to the width of the resistive-viscous layer, \ie
$\LL \propto L_{\mathrm{RV}}$.  This seems logical because the current
sheet can be described very well neglecting Ohmic dissipation
everywhere outside the narrow resistive-viscous layer whose width is
$L_{\mathrm{RV}}$ rather than $a$.

The value of $\LL$ is somewhat arbitrary, because the boundary of the
resistive-viscous layer is (physically) not sharp.  We defined its
width to be set by the characteristic velocity peaks (see
\figref{fig:tm_theory}), which is a useful convention for our purpose.
In fact, there exists a transition region (marked in shaded yellow in the
figure), where the ideal MHD equations can still approximately
  be applied, although one is already in the non-ideal regime.

For our applications, we found a useful definition based on the fact
that resistive and viscous effects are largest in the vicinity of
steep gradients of the MHD variables.  The (in relative terms) most
important gradient is that of the $y$-component of the velocity, which
is very large between the two extrema close to the current sheet (see
bottom left panel of \figref{fig:tm_example_4_in_1}).  Taking into
account that $L_{\mathrm{RV}}$ is the half distance between the two
extrema, which approximately corresponds to $1/4$ of a wavelength of a
sine function, we propose to use as the proper length scale
\begin{equation}
  \LL  =  4 L_{\mathrm{RV}} .
  \label{eq:4m}
\end{equation}
This choice is consistent with identifying $\LL$ with the wavelength
in the wave-damping tests.  It further suggests that a similar
reasoning based on local extrema may lead to the appropriate value in
other systems, too.  Combining Eqs.\,(\ref{eq:4m}) and
(\ref{eq:eta*tm}), we obtain
\begin{equation}
\eta_{\ast} = \nn_{\eta}^{\deltx x}  \timex \vv \timex  4 L_{{\mathrm{RV}}}  \timex
\bigg( \frac{\deltx x}{ 4 L_{{\mathrm{RV}}}}\bigg)^{r}.
\label{eq:eta*tm_with_length}
\end{equation}
On the other hand, from Eq.\,(\ref{eq:vl_calib_main}), we have
\begin{equation}
  L_{{\mathrm{RV}}}(k=3,a=0.1) =   0.634  \eta_{\ast}^{1/6} \nu^{1/6} \left(
  \frac{b_0}{\sqrt{\rho_0}} \right)^{-1/3}.
\label{eq:m_eps*}
\end{equation}
Note the explicit dependence of Eq.~(\ref{eq:eta*tm_with_length}) on
$L_{{\mathrm RV}}$ and of Eq.~(\ref{eq:m_eps*}) on $\eta_{\ast}$. This
dependence can be easily removed obtaining the expressions
\begin{align}
 \eta_{\ast} =& \left [2.536^{(1-r)}  \nn_{\eta}^{\deltx x} \vv \nu^{(1-r)/6} 
                \left( \frac{b_0}{\sqrt{\rho_0}} \right)^{-(1-r)/3}
                (\Delta x)^{ r} \right ]^{6/(5+r)} 
\label{eq:eta*_} \\
%L_{{\mathrm{RV}}} =&\left [ 0.0649 \,(4)^{(1-r)} \nu %%} 
L_{{\mathrm{RV}}} =&\left [ \frac{0.2596}{4^r} \nu 
                   \left ( \frac{b_0}{\sqrt{\rho_0}} \right)^{-2} 
                   \nn_{\eta}^{\deltx x} \vv  (\Delta x)^{ r} \right ]^{1/(5+r)} \,,
\label{eq:eps_vs_dx}
\end{align}
which are valid only for $a = 0.1$ and $k=3$, and give the true
dependence of $\eta_\ast$ and $L_{{\mathrm{RV}}}$ on the grid
resolution. Consequently, the TM growth rate is expected to depend on
$\Delta x$ with an exponent $\alpha_i = 24 r/ [5(5+r)]$, which allows
us to compute the order of convergence from the numerical values
$\alpha_i$ as
\begin{equation}
r= \frac{25 \alpha_i}{24-5 \alpha_i}. %,
\label{eq:r_ast}
\end{equation}
Similarly, Eq.\,(\ref{eq:eta*_}) can be used to compute
$\nn_{\eta}^{\deltx x}$ resorting to the coefficient $c_i$ from the
fit to the growth rate and identifying $\vv$ as the magnetosonic
speed, \ie $\vv = c_{ms} \approx 1.63$ in this case ($\theta=\pi/2$ in
Eq.\,\ref{eq:ms_full}).

%----------------------------------------------------------------------------
\begin{table}
\centering
    \begin{tabular}{lcc}
      \tableline
      Reconstruction & $ \nn_{\eta}^{\deltx x} $ & $r$    \\ 
      \tableline
      MP5 & $ 16  \pm  5$ &$  4.81 \pm 0.09$ \\
      MP7 & $ 142  \pm  33$ & $ 6.65 \pm 0.08 $\\ % CHECK THIS ERROR!
      MP9 & $ \ \ 170  \pm  220$ & $ 7.6 \pm 0.6$  \\
      \tableline
    \end{tabular}
    \caption{Resistivity coefficient $\nn_{\eta}^{\deltx x}$ 
      (for the definition see Eq.\,(\ref{eq:eta*})) and 
      reconstruction scheme order $r$ (Eq.\,(\ref{eq:r_ast}))
      determined from TM simulations performed  with 
      the LF Riemann solver (see also Fig.\,\ref{fig:tm_ms_comparison}).
    }
\label{tab:tm_reco}
\end{table}
%----------------------------------------------------------------------------

In Table \ref{tab:tm_reco}, we list the values of $\nn_{\eta}^{\deltx
  x}$ and $r$ computed with the procedure outlined above.  The MP5 and
MP7 schemes are almost 5th and 7th order accurate, whereas the MP9
scheme performs below the theoretical expectation.  In other words,
the higher the order of the reconstruction scheme, the higher the
reduction of the convergence order.  A possible explanation of
  this fact is the following.  The function $\velo_y$ is proportional
  to $y^{-1}$
  for $|y| \ll a$, i.e., outside of the resistive-viscous layer 
%$y \rightarrow 0$ 
  (see the discussion in Appendix \ref{app:tm}, in particular
  Eq.\,\ref{eq:ind_out} and \figref{fig:tm_phi}).
  For this reason all the derivatives of $\velo_y$ in the
  $y-$direction diverge. Thus, the neglected higher order terms of the
  Taylor expansion in the reconstruction of $\velo_y $ for
  $y \approx 0$ can actually be dominant.
  Taking this into consideration, it is rather more surprising that the MP5
  and MP7 schemes almost achieve their theoretical order of accuracy
  than the fact that the MP9 scheme performed below the
  theoretical expectation.

%----------------------------------------------------------------------------
\begin{figure}%%[h]
\centering
\includegraphics[width=0.45\textwidth]{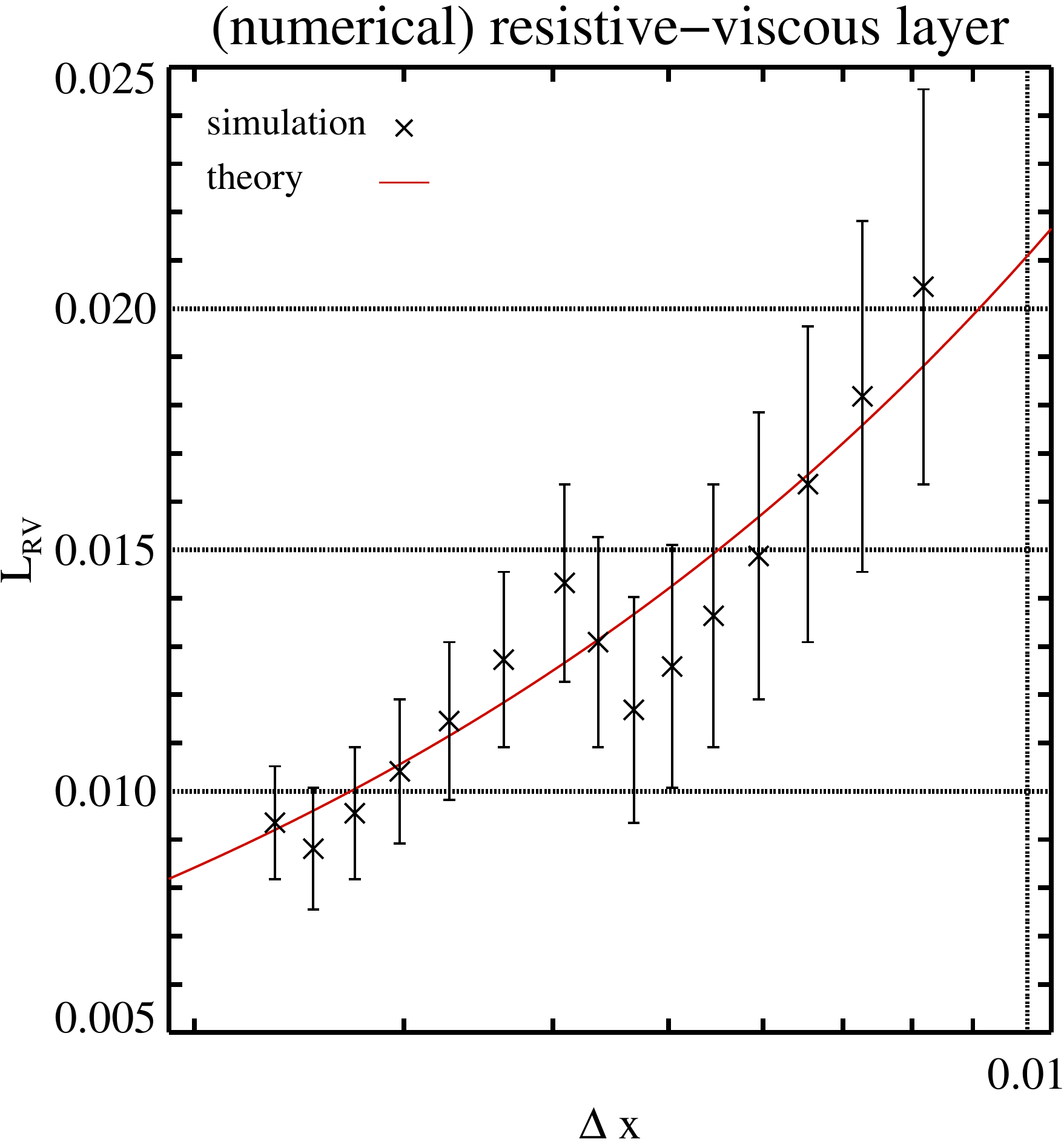}
\caption{Width of the resistive-viscous layer as a function of
  $\Delta x$ (which is proportional to the value of the numerical
  resistivity) in simulations of TM driven by numerical resistivity.
  Black crosses depict measured values obtained in simulations with
  MP7.  The red curve shows the expected layer width
  (Eq.\,\ref{eq:eps_vs_dx}), given the parameters for the numerical
  resistivity determined from the measured growth rate.  }
\label{fig:tm_reso_layer}
\end{figure}%%
%----------------------------------------------------------------------------

According to \figref{fig:tm_reso_layer}, the values of
$L_{{\mathrm RV}}$ measured directly from the numerical simulations
(see previous section) agree well with those computed with
Eq.~(\ref{eq:eps_vs_dx}), where the values of $r$ and
$\nn_{\eta}^{\deltx x}$ needed in this equation are extracted from the
growth rate using Eqs.\,(\ref{eq:ln_gamma_fit_results}) and
(\ref{eq:r_ast}).  This result shows that our assumptions are correct,
which is far from being obvious, because we assumed that (i) numerical
errors can be called ``numerical resistivity'', (ii) this numerical
resistivity can be treated as normal physical resistivity, and (iii)
the same equations can be used to determine its magnitude or predict
its influence on the system.  Moreover, we also had to make use of
ansatz (\ref{eq:eta*}) for the numerical resistivity.
 
\begin{figure}%%%[h]
  \centering
  \includegraphics[width=0.45\textwidth]{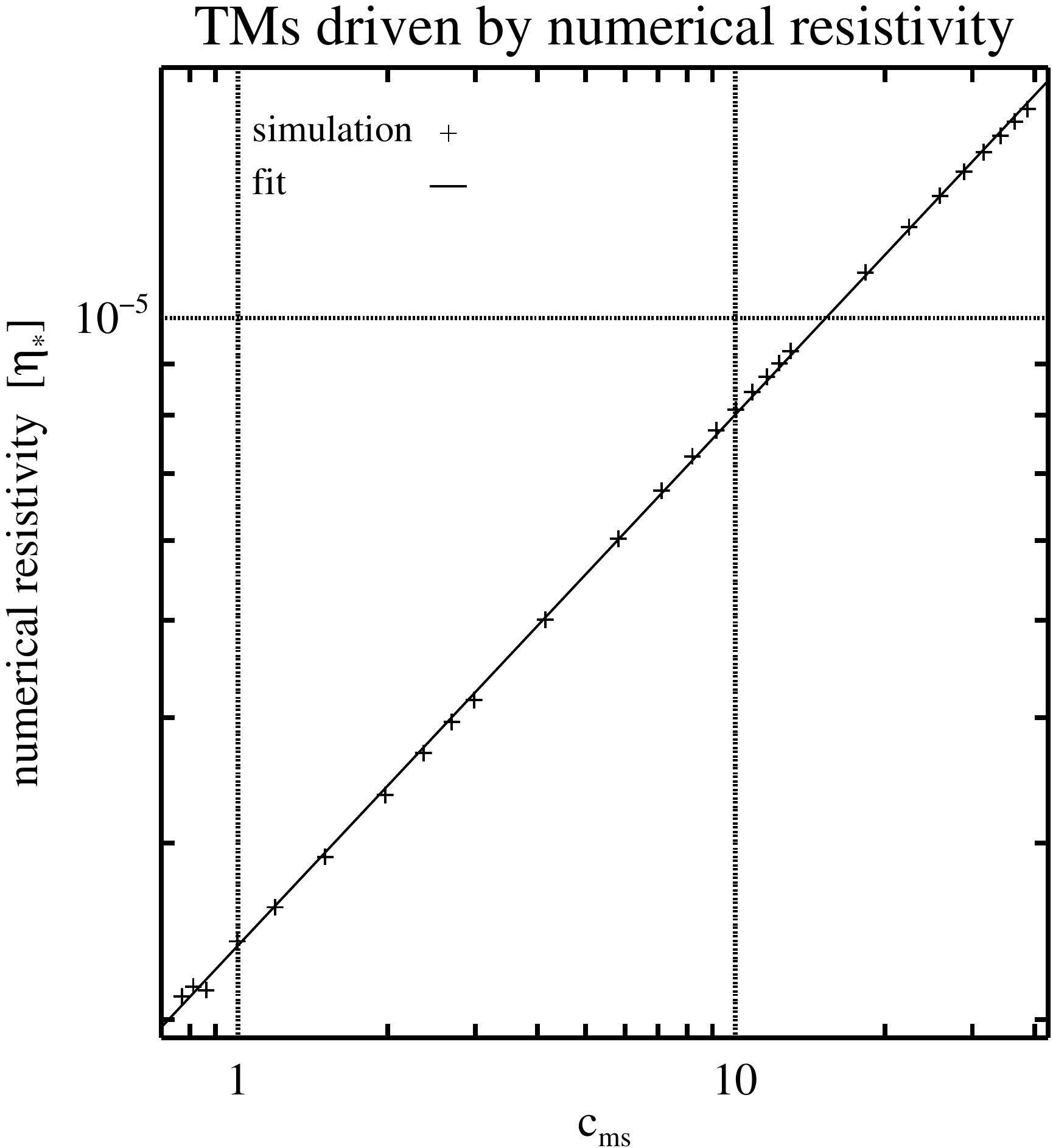}\\*[0.1cm]
  \includegraphics[width=0.45\textwidth]{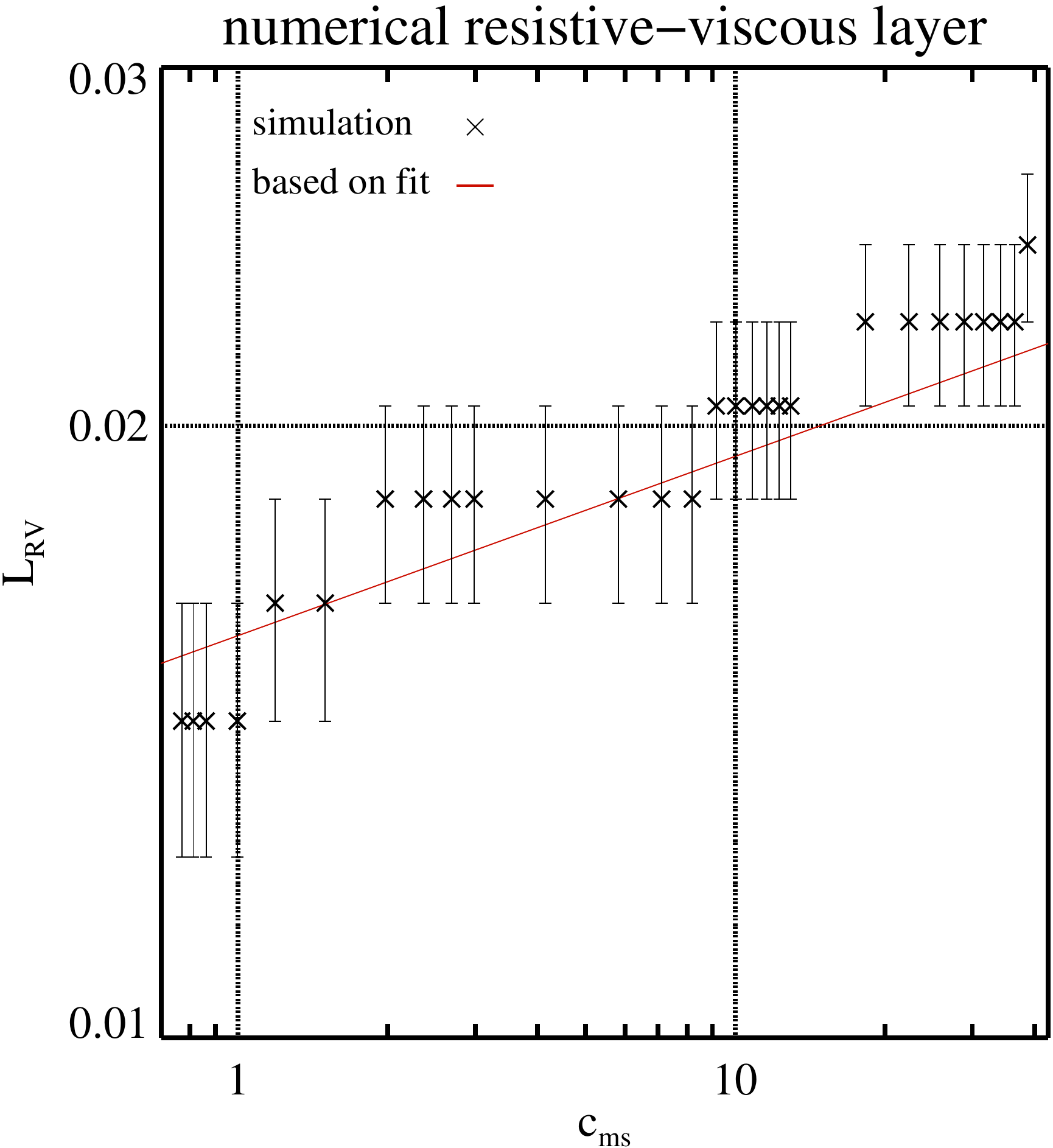}
  \caption{\emph{Upper panel:} Numerical resistivity in TM simulations
    performed with MP5 on a grid of $512 \times 512$ zones as a 
    function of the fast magnetosonic speed $\cmag$ 
    (Eq.~\ref{eq:ms_best}), which was changed varying $p_0$ but
    keeping $b_0$ and $\rho_0$ constant. The black curve is 
    a fit to the simulation data.  
    \emph{Bottom panel:} Width of the resistive-viscous shear layer
    (black crosses) as a function of $\cmag$ for the simulations shown
    in the upper panel.  The red curve gives the predicted width of
    the resistive-viscous layer based on the fit in the upper panel.
  }
  \label{fig:tm_pressure}
\end{figure}

To test whether the magnetosonic speed is indeed the characteristic
velocity of our TM setup, as expected from the arguments given in the
discussion of the wave damping simulations, we ran several simulations
with the same setup, but varying the fast magnetosonic speed from
$\cmag \approx 1$ to $\approx 39$. This was achieved by changing the
background pressure from $0.01$ to $900$, keeping $b_0 = \rho_0 = 1$.
The upper panel of Fig.\ \ref{fig:tm_pressure} shows that the numerical
resistivity increases with $\cmag$.

Different from the wave damping tests, it is not straightforward,
however, to compute the fast magnetosonic speed, because in TM
simulations the perturbed fluid makes a ``U-turn'' in the vicinity of
the magnetic shear layer (\ie for $| y / a| \ll 1$).  Therefore,
determining the ``correct'' values of $\theta$ (which changes from $0$
to $\pi/2$) and the background magnetic field strength (which changes
from $1$ for $| y / a| \gg 1$ to $0$ for $| y / a| \approx 0$) is very
error-prone.  That is why we introduced ansatzes (\ref{eq:nu*}),
(\ref{eq:xi*}), and (\ref{eq:eta*}) to have a simple way of estimating
the code's numerical dissipation. Consequently, we took the maximum
possible magnetosonic speed (obtained for $\cos \theta = 0$;
Eq.\,\ref{eq:ms_full}) in our previous analysis, \ie
\begin{equation}
  c_{\mathrm{ms}} = \sqrt{\ca^2 + \cs^2}.
\label{eq:ms_simple}
\end{equation}
and put $\ca = 1$, which is well motivated by practical purposes (\ie
to keep the ansatzes as simple as possible).  However, in the current
analysis (simulations presented in \figref{fig:tm_pressure}), we
obtained a better fit with respect to $c_{\rm ms}$ with $\ca =0.76$,
which corresponds to the \alf speed at a distance $y = a$, \ie
\begin{equation}
  c_{\mathrm{ms}} = %\sqrt{ ( \Gamma p_0 + (0.72 b_0) ^2)/\rho_0 } = 
\sqrt{ 5/3 p_0  + 0.76^2 } .
\label{eq:ms_best}
\end{equation}
We note that the precise choice of $\ca$ is irrelevant in the high
plasma $\beta$ regime, while it only slightly affects the quality of
the fits in the low plasma $\beta$ regime (where the parameter
$\beta \equiv b_0^2 /(2p_0)$) .

%----------------------------------------------------------------------------
%%
%----------------------------------------------------------------------------

From the measured growth-rates $\gamma$, we determined the numerical
resistivity in each simulation ($\eta_\ast \propto \gamma^{5/4}$), and
fitted the results with
\begin{equation}
  \ln (\eta_{\ast}) = s \ln(\cmag) + d
\label{eq:tmp1}
\end{equation}
obtaining
\begin{align}
  s&= 0.524 \pm 0.002,
\nonumber \\
  d&= -12.9 \pm 0.5.
\label{eq:tmp3}
\end{align}
From Eq.~(\ref{eq:eta*_}) and Eq.~(\ref{eq:gr_calib_main}), we find
that
\begin{equation}
 \eta_{\ast} \propto \vv^{6/(5+r)} \,,
\label{eq:tmp2}
\end{equation}
and putting $\vv = c_{\mathrm{ms}}$ in Eq.\ (\ref{eq:tmp2}), we
finally obtain
\begin{equation}
  s = \frac{6}{5 + r}. \label{eq:bumba7}
\end{equation}

For MP5 reconstruction ($r_{\rm th}=5$), the expected value is
$s=0.6$, which is close to the measured one.  Using $s$ from
Eq.\,(\ref{eq:tmp3}), we determine the reconstruction scheme order to
be $r = 6.45 \pm 0.05$, which is neither equal to $5$ within the
errors nor consistent with the value $r= 4.81 \pm 0.09$ from
\tabref{tab:tm_reco}.  This discrepancy should not concern us,
however, because we included only statistical errors in the
measurement errors from the linear fit neglecting other errors, \eg
those originating from estimating the fast magnetosonic speed (which
changes from zone to zone in the simulation). This implies that this
way of determining the order of the reconstruction scheme is much less
reliable than from the resolution studies.

In the bottom panel of \figref{fig:tm_pressure}, we show the measured
width of the resistive-viscous layer and its value (red curve)
predicted from the fit Eq.~(\ref{eq:tmp1}). Using the values of $s$
and $d$ in Eq.~(\ref{eq:tmp3}) we determined $r$ and
$\nn_{\eta}^{\deltx x}$, which we then inserted into
Eq.~(\ref{eq:eps_vs_dx}).  This methodology demonstrates that our
model is self-consistent, since the width of the resistive-viscous
layer does depend on $c_{\rm ms}$ as expected.

At the beginning of this section we made the assumption that the
numerical resistivity is not changing the background magnetic profile,
but affects only the flow in the resistive-viscous layer, where TM
grow.  All consistency checks performed in this section seem to
indicate that our assumption is correct. As a confirmation, we checked
that in none of the simulations there has been a significant
modification of the background profile.  This finding differs from the
one obtained in the simulations with physical resistivity but without
removing the background field from the induction equation
(\ref{eq:trick}).  In those simulations, the background field started
to diffuse during the simulations (which was the reason why we
modified the induction equation (Eq.\ \ref{eq:trick}) in the first
place).

\begin{figure}%%[h]
  \centering
  \includegraphics[width=0.45\textwidth]{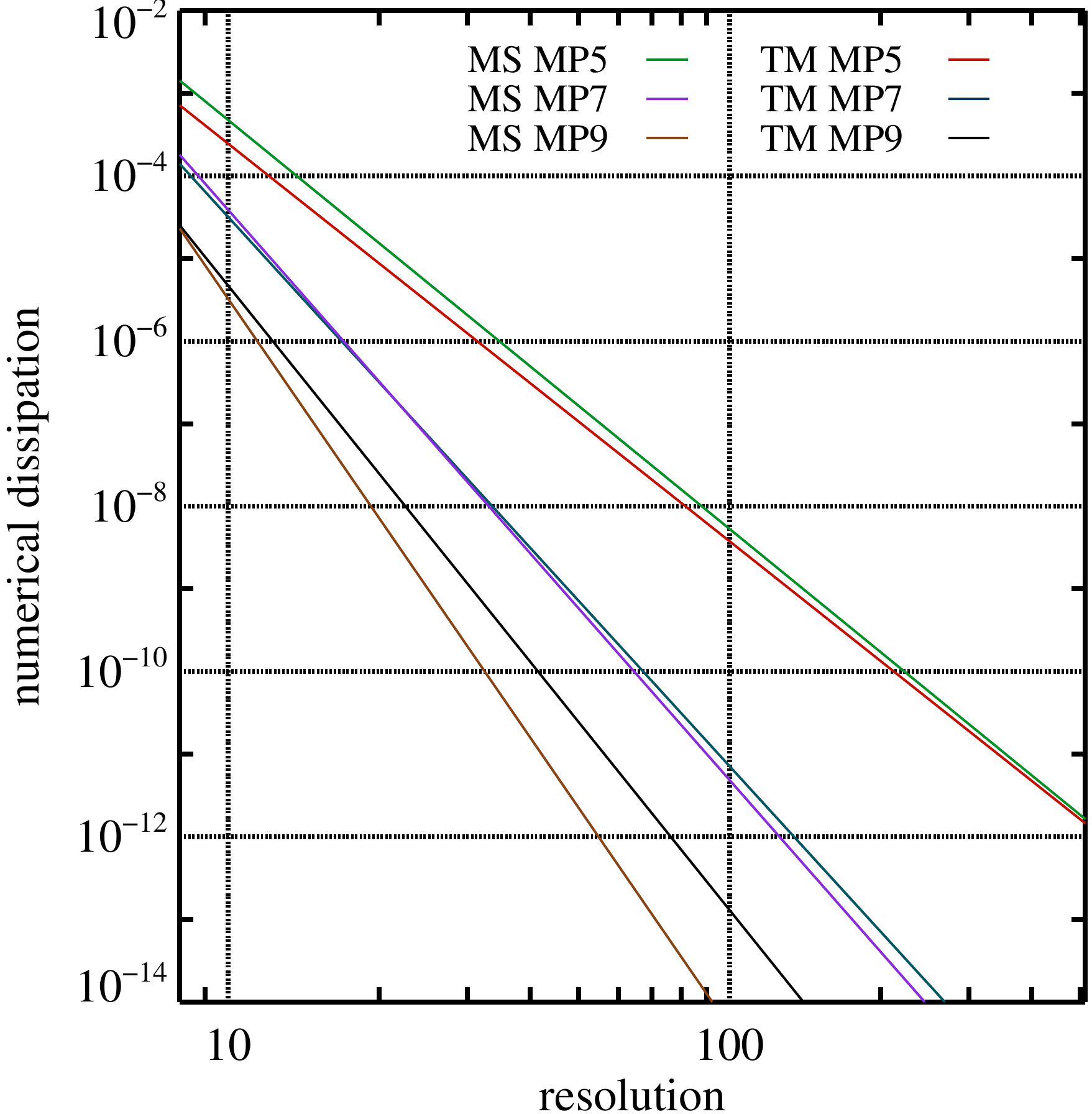}
  \caption{
    Comparison of the numerical dissipation of \textsc{Aenus} when
    simulating magnetosonic waves (MS) and tearing modes (TM) with
    different reconstruction schemes (MP5, MP7, and MP9) and RK3.  The
    numerical dissipation is given by
    $(4/3) \nu_{\ast} + \xi_{\ast} + \eta_{\ast}/(1+\cs^2/\ca^2)$ in
    the former case and by $\eta_{\ast}$ in the latter case. For the
    wave simulations, we found that
    $\eta_{\ast}/(1+\cs^2/\ca^2) \ll (4/3) \nu_{\ast} + \xi_{\ast} $
    (see Sec.\,\ref{sec:ms_waves}). For the wave problem both the HLL
    and the LF Riemann solver result in a very similar amount of
    dissipation, whereas for the TM simulations
    $\eta_{\ast}(\mathrm{HLL}) \ll \eta_{\ast}(\mathrm{LF})$. The CFL
    factor was chosen such that time integration errors were
    negligible.  The results are renormalised by setting the
    characteristic velocity and length equal to one, \ie
    $\vv = \LL = 1$.}
  \label{fig:tm_ms_comparison}
\end{figure}

Finally, in Fig.\,\ref{fig:tm_ms_comparison}, we present a
  comparison of the expected numerical dissipation in simulations of
  TM and magnetosonic wave damping based on our ansatz (see
  Eqs.\,\ref{eq:nu*}, \ref{eq:xi*}, and \ref{eq:eta*}) and the
  estimators from Tables \ref{tab:waves} and \ref{tab:tm_reco}. For
  the simulations, we set the characteristic velocities and lengths
  equal to one, \ie $\vv = \LL = 1$. The box length is set to 1 too,
  hence ``resolution'' in the abscissa of
  Fig.\,\ref{fig:tm_ms_comparison} refers to the number of zones per
  characteristic length.  As we can see, the expected numerical
  dissipation based on calibration with the help of both types of
  simulations (MS waves and TM) is similar.  This is an indication
  that our approach is presumably universal, and it makes us confident
  that it can be used to estimate dissipation coefficients for other
  flows.

%%%%%%%%%%%%%%%%%% Magnetosonic waves vs TM Comparison %%%%%%%%%%%%%%%%%%%%%%%
%%
%%%%%%%%%%%%%%%%%%%%%%%%%%%%%%%%%%%%%%%%%%%%%%%%%%%%%%%%%%%%%%%%%%%%%%%%%%%%%%%

%%%%%%%%%%%%%%%%%%%%%%%%%%%%%%%%%%%%%%%%%%%%%%%%%%%%%%%%%%%
%%%%%%%%%%%%%%%%%%%%%%%%%%%%%%%%%%%%%%%%%%%%%%%%%%%%%%%%%%%
\section{CASE STUDY: MAGNETOROTATIONAL INSTABILITY}
\label{sec:case}

In the previous section, we present a methodology which allows us to
estimate the numerical viscosity (bulk and shear) and resistivity of a
code. It also serves to determine the characteristic velocity, $\vv$,
relevant for the numerical dissipation coefficients. We applied this
methodology to the \textsc{AENUS} code for different numerical
schemes. Using the ansatzes Eqs.~(\ref{eq:nu*}), (\ref{eq:xi*}) and
(\ref{eq:eta*}) and the numerical dissipation coefficients given by
Tables \ref{tab:resvis} (for shear and bulk viscosity coefficients)
and \ref{tab:tm_reco}  (for resistivity coefficients), it is
possible to determine the numerical resolution, $\Delta x$, needed to
perform a numerical simulation with a numerical viscosity and
resistivity lower than a given threshold. This allows us to estimate
the computational resources needed for a particular application and
helps us to choose the numerical scheme that minimises the
computational cost. To show the feasibility and the usefulness of this
analysis, we present here the estimates for a particular application of
\textsc{AENUS} to simulations of the MRI. 

The MRI \citep[][]{Velikhov__1959__SovPhys__MRI,
  Chandrasekhar__1960__PNAS__MRI} is an instability which can develop
in a differentially rotating fluid in the presence of a magnetic
field. In the case of an homogeneous vertical magnetic field the MRI
develops non-linear channel flows which are then disrupted by
parasitic instabilities \citep{Goodman_Xu, Pessah_Goodman} into a
turbulent flow. The instability has been proposed as the main driver
of accretion in accretion disks \citep{Balbus_Hawley__1991__ApJ__MRI}
and may play an important role in the amplification of magnetic field
during the collapse of stellar cores \citep{Akiyama:2003}. There are a
number of MHD simulations devoted to the study of the MRI in local box
simulations \citep[see e.g.][]{Obergaulinger_et_al_2009,
  Rembiasz:2016a, Rembiasz:2016b} in which we can aim to resolve
numerically the magnetised flow with minimal numerical viscosity and
resistivity. The main difficulty of those simulations is that the
characteristic length scale relevant for the growth of the MRI channel
modes and its termination due to parasitic instabilities
(Kelvin-Helmholtz vortices) is very small ($\LL \sim 100$~m) compared
to the size of the system ($\sim 10$~km) when realistic conditions are
considered \citep[see][for a
discusion]{Obergaulinger_et_al_2009}. Here we show, as an example, how
one can estimate the resolution requirement and computational cost for
the numerical simulations presented in \cite{Rembiasz:2016b}.

In the typical setup of \cite{Rembiasz:2016b} the box size is
$2 {\rm \,km }\times 2 {\rm \,km }\times 0.666 {\rm \,km }$ or
$2 {\rm \,km } \times 8 {\rm \,km } \times 2 {\rm \,km }$ (radial,
azimuthal and vertical extension) and channel modes develop with a
size of $\lambda_{\rm MRI} = 0.666$\,km in the vertical direction (one
or three channel modes fit in the vertical direction, depending on the
box size). The characteristic length is set by the width of the
channel mode, \ie $\LL=\lambda_{\rm MRI}$. The characteristic speed
may be chosen among the fast magnetosonic speed
($c_{\rm ms} = 3\times 10^9$~cm~s$^{-1}$), the \alf speed,
($c_{\rm A} = 7.8\times10^{7}$~cm~s$^{-1}$), or the flow velocity
($v_{\rm flow} = 2.4 \times 10^{9}$~cm~s$^{-1}$). To make a
conservative estimate we take the largest of the three, \ie
$\vv=3\times 10^{9}$~cm~s$^{-1}$.  The goal of
\cite{Rembiasz:2016b} was to run the simulations with numerical
viscosities and resistivities below a value of
$7.48\times 10^8$~cm$^2$~s$^{-1}$  
\citep[according to estimates of][(physical) viscosity due to
  neutrinos can vary from $10^9\tto 10^{12}$~cm$^2$~s$^{-1}$ inside of
  a proto-neutron star]{Guilet_et_al_2015}%
.  They estimated that in that regime
the Reynolds numbers are above $\sim100$, which would be sufficient to
have convergent results for the growth of the channel flows up to the
termination due to parasitic instabilities. The post-termination
evolution of the generated turbulent flow would have probably required
more stringent resolutions, but this was not the aim of
\cite{Rembiasz:2016b}.

\begin{figure}%%[h]
\centering
  \includegraphics[width=0.45\textwidth]{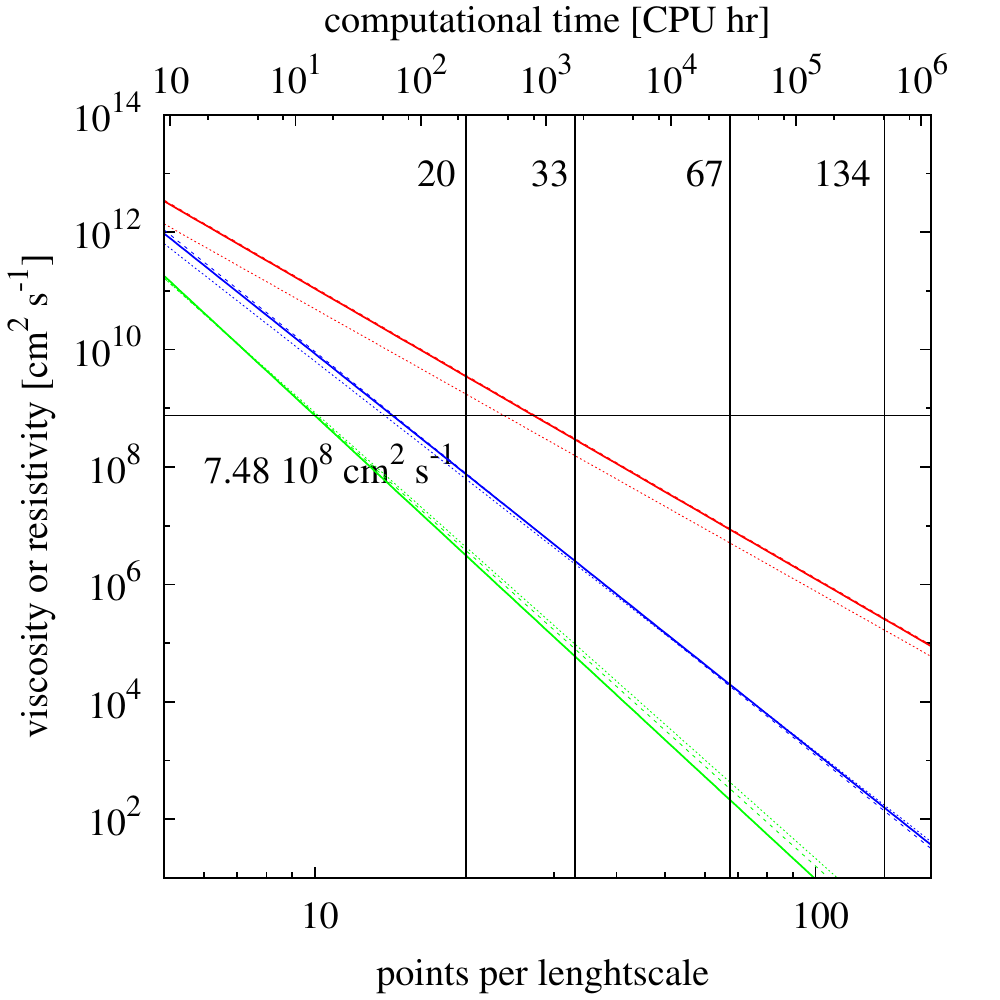}
  \caption{Estimation of the numerical shear viscosity (solid lines),
    bulk viscosity (dashed lines), and resistivity (dotted lines) for
    the MRI simulations of \cite{Rembiasz:2016b} for different 
    numbers of grid points per length scale ($\Delta x / \LL$; lower
    abscissa). Results are shown for three different reconstruction
    schemes: MP5 (red), MP7 (blue), and MP9 (green). Vertical black
    lines show the resolution used in that work. The horizontal black
    line shows the viscosity and resistivity goal, below which the
    Reynolds numbers exceed a value of $\sim 100$. The upper abscissa
    gives an estimate of the corresponding computational cost of a
    $12$~ms simulation with \textsc{AENUS} for the smallest box size
    ($2 {\rm \,km } \times2 {\rm \,km } \times 0.666 {\rm \,km }$). 
}
\label{fig:case}
\end{figure}

Figure~\ref{fig:case} shows the estimated numerical viscosities and
resistivities of the \textsc{AENUS} code according to the measurements
of the present work rescaled to the simulations of
\cite{Rembiasz:2016b} for an HLLD solver and three different
reconstruction schemes (MP5, MP7 and MP9).  According to the figure,
it is sufficient to perform simulations with  $10$, $14$, and $28$
zones per $\LL$ to reach the accuracy goal for MP9, MP7, and MP5,
respectively. This result demonstrates the advantage of using
ultra-high order schemes (MP9) in simulations of smooth flows

Once the numerical resolution is known, it is possible to estimate the
computational cost of a given simulation. As an example, we consider
our smallest simulation box
($2 {\rm \,km } \times2 {\rm \,km } \times 0.666 {\rm \,km } $), a
typical simulation period of $12$~ms duration, and a CFL factor of
$0.7$. \textsc{AENUS} performs on the SuperMUC supercomputer
(www.lrz.de) typically at $0.15$~ms/iteration/zone.  The upper
abscissa of Fig.~\ref{fig:case} shows the resulting total CPU time for
this setup. To reach the accuracy goal, we estimate that we need at
least $15$, $60$, and $842$~CPU~hrs for MP9, MP7, and MP5,
respectively. Even if MP9 suffers of a slight computational overhead
in comparison to MP7 and MP5, due to the larger size of the numerical
stencil, our analysis shows that it pays off to perform simulations
with MP9.  Indeed, \cite{Rembiasz:2016b} performed their simulations
with resolutions ranging from $20$ to $134$ zones per
$\lambda_{\rm MRI}$ and showed convergence of the results.

%%%%%%%%%%%%%%%%%%%%%%%%%%%%%%%%%%%%%%%%%%%%%%%%%%%%%%%%%%%
%%%%%%%%%%%%%%%%%%%%%%%%%%%%%%%%%%%%%%%%%%%%%%%%%%%%%%%%%%%
\section{SUMMARY AND CONCLUSIONS}
\label{sec:summary}

We have presented a reliable methodology to measure the numerical
  shear and bulk viscosity, and the resistivity of Eulerian finite
  volume MHD codes by means of a simple ansatz for each of these
  numerical effects, which are inevitably present in any such code.
  We have postulated that the amount of numerical dissipation depends
  on the characteristic length and velocity of the system under
  consideration, the numerical resolution, and some free parameters
  which have to be calibrated depending on the numerical scheme in
  use. Hence, our ansatz for each of the three numerical effects
  consists of two additive terms describing the contribution of
  spatial and temporal discretisation errors, which both depend on the
  characteristic length and velocity of the system, and on the grid
  resolution and the size of the time step, respectively.

We performed the parameter calibration by means of a set of test
  simulations using the \textsc{Aenus} code.  However, because the
  procedure is not restricted to this code, we provide potential users
  of our methodology with the detailed results of our test suite at
% changes
 \url{http://www.uv.es/camap/tmweb/Web_tm.html}.
%  \url{http://puszek.cats}.
% changes
 These data should help to measure the
  dissipation coefficients of other Eulerian finite volume MHD codes.

Firstly, we have considered three wave damping tests from which one
  can directly extract a linear combination of the numerical
  resistivity, and the numerical shear and bulk viscosities.  These
  simulations allowed us to estimate the latter two quantities
  accurately. However, we failed to obtain a physically sound value of
  the numerical resistivity, because it is much smaller than that of
  the numerical viscosity of \textsc{Aenus}, \ie our estimate of the
  numerical resistivity is dominated by systematic errors.

Nevertheless, the wave damping simulations confirm the
   appropriateness  of our ansatz for the numerical
  shear and bulk viscosity, and the resistivity.  In almost all
  simulations performed by us, the spatial reconstruction schemes and
  the RK time integrators have the theoretically expected order of
  accuracy. We also find that in simulations of sound waves, \alfc
  waves, and fast magnetosonic waves the characteristic length and
  velocity of the system are always the wavelength of the wave and the
  fast magnetosonic speed (which reduces to the sound speed in the
  case of sound waves), respectively.  Because the value of the
  numerical resistivity is substantially lower than that of the
  numerical viscosity in the wave damping tests (see above), the
  numerical magnetic Prandtl number
  $\Pmast \equiv \nu_{\ast} / \eta_{\ast}$ is not close to $1$, as it
  is commonly suspected among practitioners in the field.

  Secondly, we have performed TM simulations since the wave damping ones were hardly useful to asses the value of the
  numerical resistivity. By measuring the growth rate of the TM instability and fixing the value of the physical
  viscosity, we have been able to estimate the numerical resistivity of \textsc{Aenus}.  Moreover, from the estimated
  value of the numerical resistivity, we could correctly predict the expected width of the resistive viscous layer. This
  indicates that our method is (self-)consistent.  A cautionary note must be added here. In order to obtain reliable
  estimates of the numerical resistivity in TM simulations it is necessary to employ spatial reconstruction schemes of
  order $r\ge 5$.  Extensive numerical experience \citep{Rembiasz} shows that the TM setup employed here becomes
  numerically unstable for lower order reconstruction schemes in the resolution range considered by us, i.e.\, it was
  difficult to maintain a magnetohydrodynamical equilibrium of the background flow for tests in which less than a
    fifth order spatial reconstruction scheme was employed.

Comparing the expected numerical dissipation in simulations of
  magnetosonic wave damping and TM, we find that the expected
  numerical dissipation based on a calibration with the help of both
  types of simulations is similar.  This indicates that our approach
  is supposedly universal, which gives us confidence that it is also
  applicable to estimate the dissipation coefficients for different
  flow regimes and numerical setups.
For illustration of our approach in an astrophysical context, we have estimated in
Sec.\,\ref{sec:case} the numerical viscosities and resistivity for the
MRI models of \cite{Rembiasz:2016b}, which in turn have allowed us to
obtain a reliable estimate of the computational needs in terms of
numerical resolution and computational time for these particular
simulations.

We have found that the high orders of convergence of the MP
reconstruction schemes obtained in the wave damping tests are retained
in the 2D simulations  of sound waves and TMs.  This is a remarkable result, because our
approach is based on several (one per dimension) independent
one-dimensional reconstruction steps instead of a genuinely
multi-D reconstruction algorithm as proposed by, \eg
\citet{Colella_et_al_2011,McCorquodale_Colella_2011,Zhang_et_al_2011,Buchmueller_Helzel_2014}.  
In general, such a simplification may introduce additional numerical errors, thus degrading the order of
accuracy of multi-D simulations to second order, \ie well
below that of 1D ones.

However, our 2D simulations do not suffer from this degradation. This may be the result of our
somewhat simple tests, i.e.\ in the TM case, the derivatives in the $y$-direction dominate
over the ones in the $x$-direction, and in 2D simulations of sound waves, we consider small sinusoidal perturbations of
the background. One should not expect this behaviour to hold in a general case of a non-linear multi-D problem. 
Exploring this topic is beyond the  scope of this work though.

It is important to note that in our ansatz there is a single
length scale $\LL$ for which the dissipation coefficients are
estimated. If applied to systems in which dissipation occurs at
multiple length scales (\eg turbulence), then the interpretation
of our ansatz is a measurement of the dissipation coefficients at each
length scale, which may give rise to different values of
  these coefficients. Indeed, we observe this effect in our TM
simulations, if the background magnetic field is not eliminated
from the induction equation. If not eliminated,
we have dissipation occurring both in the resistive-viscous layer (of
size $L_{\rm RV}$) and within the background shear profile (of size
$a >> L_{\rm RV}$) at the same time.  This scale-dependent definition
of the dissipation coefficients is similar to that used in large eddy
simulations of anisotropic weakly compressible turbulence \citep[see
e.g.][]{Fureby:1999,Zhou:2014,Radice:2015}.  Alternatively, we could
have formulated our ansatz equivalently in terms of scale-independent hyperviscosity and
hyperresistivity coefficients. This formulation has the disadvantage of not having physical counterparts for the purely numerical
  hyperdissipation coefficients but, on the other hand, it allows to interpret the high-order derivative terms appearing in
the Taylor expansion of the space and time derivatives (see
Eq.\,\ref{eq:scalargen}).

Using our ansatz may not always be
straightforward, since it requires identifying the relevant
characteristic velocity $\VV$ and length $\LL$, of the system.
  As for the former, we have shown that in
all our tests (wave damping and TM) without background flow, the
characteristic velocity is equal to the fast magnetosonic speed.  In
the wave-damping tests with a non-zero background velocity (see
Sec.\,\ref{sec:background_velocity}), we have found that the
characteristic velocity depends on the direction of the wave
propagation relative to the
direction of the background flow. However, one can easily 
obtain an estimate for the upper
limit of the characteristic velocity, \ie for the sum of the moduli of
the background velocity and the fast magnetosonic speed.  The
determination of the characteristic length, $\LL$, of the system can
be tricky and requires a good understanding of the simulated
system. For example, in the case of the TM simulations, the first
``natural candidate'' for $\LL$ seems to be the width of the magnetic
shear layer, $a$, (Eq.~\ref{eq:b0_tan}), but it turns out to be
the width of the resistive-viscous layer, $L_{\rm RV}$
(Eq.~\ref{eq:vl_calib_main}).

Even though all our test were performed in boxes with uniform
  grid resolutions, our results can  be translated to
adaptive mesh
  refinements (AMR) codes that refine the grid where the flow
    develops small-scale structures. A natural way of using our
    ansatzes in AMR simulations would be to apply it at each
    refinement level separately, i.e.~to compute the numerical
    dissipation coefficients (\ref{eq:eta*}) based on the refined grid
    width.  Numerical viscosity and resistivity then quite obviously
    are position-dependent.  This is, however, not a unique feature of
    AMR simulations because, in general,  numerical viscosity and
    resistivity are local quantities because they depend on the
    characteristic velocity and length scale that may vary strongly
    throughout the simulation domain.

Finally, we note that in the (astro-)physics literature hydrodynamic
and magnetohydrodynamic simulations are often categorised by the
numerical Reynolds numbers
\begin{equation}
R_{{\rm e}\ast}      = \frac{ V \timex L }{\nu_\ast} \qquad , \qquad
R_{{\rm m}\ast} = \frac{ V \timex L }{\eta_\ast},
\end{equation}
where $L$ and $V$ are the \emph{typical length} and \emph{typical
  velocity} of the system, respectively. However, often both
quantities are subjectively chosen. In other words, the assumed
typical length and velocity of the system are not the values obtained
after a thorough calibration analysis as we have conducted it
here. Indeed, in general, $L \neq \LL$ and $V \neq \vv$, \ie the
typical values (chosen by the respective author) are not equal to the
characteristic values, which are uniquely set by the physics and
numerics of a given simulation.  Hence, the Reynolds numbers commonly
estimated can vary by a few orders of magnitude for the same physical
system and numerical setup across the literature.  For this reason,
the typical values of the numerical Reynolds number seldomly are a
useful quantity to cross-compare different numerical models.  However,
once the numerical viscosity and resistivity are measured (as we
propose it in this paper), one can easily express simulation results
in terms of numerical hydrodynamic and magnetohydrodynamic Reynolds
numbers, in as much as the proper characteristic length and velocity
of the system have been identified. Because
this is not a common practice in the community, we conclude that the
use of numerical Reynolds numbers more often obscures than reveals
the true nature of numerical viscosity and resistivity.

\acknowledgments
TR acknowledges support from The International Max Planck Research
School on Astrophysics at the Ludwig Maximilians University Munich.
 EM \& TR acknowledge support from the Max-Planck-Princeton Center for
Plasma Physics as well as  from the Deutsche
Forschungsgemeinschaft through the grant EXC 153 "Origin and Structure of
the Universe". MA, MO, PCD, and TR acknowledge support from the
European Research Council (grant CAMAP-259276). We also acknowledge
support from grants AYA2015-66899-C2-1-P, AYA2013-40979-P, and
PROMETEOII/2014-069.  The computations were performed at the
Max Planck Computing and Data Facility (MPCDF)
and at the Servei d'Inform\`atica of the University of Valencia.
We thank J.~Guilet for his useful comments on the manuscript.
We also thank the anonymous referee whose valuable comments and suggestions
allowed us to improve the quality of this manuscript.

 \software{Aenus \citep{Obergaulinger__2008__PhD__RMHD}}

\appendix

\section{TEARING MODE GROWTH RATE}
\label{app:tm}

In Appedix \ref{app:tm_ana}, we present an analytical derivation of
the TM growth rate in resistive MHD, and we briefly discuss a
generalisation of these results to resistive-viscous MHD, which
was already done by FKR63.

In Appendix \ref{app:tm_semi}, we postulate empirical equations for
the TM growth rate in resistive-viscous MHD and calibrate these rates
with the help of numerical simulations.

\subsection{Analytical Approach}
\label{app:tm_ana}

In their analytical study of the TM instability, FKR63 considered the
effects of various physical factors, like a position dependent
background density, temperature, and resistivity.  We will restrict
ourselves to a much simpler system, yet demonstrating the key features
of the TM instability.  Our presentation is greatly inspired by the
excellent discussion of this topic in \citet{gkp}.  We also recommend
\cite{Schnack} for a concise introduction to the TM instability.

\subsubsection{General Case}

We consider perturbations of the system described in
Sec.\,\ref{sec:tm_theory} whose wavelength in $x$ direction is
comparable to the shear width, \ie $k \lesssim a^{-1}$
(Eq.\,\ref{eq:k_delta_main}).  Note that one only needs to solve
Eqs.\,(\ref{eq:dt_b}) and (\ref{eq:dt_v}) for $ \velo_y $ and
$b_{1y}$, respectively, while the other perturbation components can be
easily determined from conditions (\ref{eq:div_v}) and
(\ref{eq:div_b}).  For the WKB ansatz (Eqs.\,\ref{eq:wkb1} and
\ref{eq:wkb2}) to be justified, condition (\ref{eq:timescales_main})
must be satisfied.

Inserting the WKB ansatz into Eqs.\,(\ref{eq:dt_b}) and
(\ref{eq:dt_v}) and taking the curl of the latter equation to
eliminate the pressure, the $y$ component of the induction equation
and the derivative $\partial_x$ of the $z$ component of the equation
of motion read
\begin{align}
  \gamma b_{1y} =& [\nabla \times ({\bf v} \times {\bf b_0})]_y + \eta
  (- k^2 + \partial^2_y ) b_{1y},   \\
  \gamma \rho_0 (- k^2  +  \partial^2_y ) \velo_y   =&  \partial_x \{ \nabla
  \times [  (\nabla \times {\bf b_1})
  \times {\bf b_0} +  (\nabla \times {\bf b_0}) \times {\bf b_1}
  % \nonumber \\ & 
+ \rho_0 \nu   \nabla^2 {\bf v} ] \}_z. 
\end{align}
After some algebra we arrive at
\begin{align}
  \gamma b_{1y} = & i k \velo_y b_{0x} + \eta (- k^2 + \partial^2_y ) b_{1y},    
\label{eq:flux_freezing}   \\
  \gamma \rho_0 (  - k^2 + \partial^2_y ) \velo_y   
   =& \rho_0 \nu (k^4 -2 k^2 \partial_x^2 + \partial_x^4) \velo_y 
    + i k [ -b_{1y} \partial_y^2 b_{0x} + b_{0x} (-k^2 + \partial_y^2) b_{1y}]. 
\label{eq:force_balance}
\end{align}
Because this system cannot be integrated analytically, FKR63 used the
BLA method (see Sect.\,\ref{sec:tm_theory}). They divided the domain
into two regions, an outer one ($-L_y \le y < - y_{\epsilon}$ and
$ y_{\epsilon} < y \le L_y$, where $y_{\epsilon}$ is a small positive
constant such that $ y_{\epsilon} \ll a$) in which dissipative effects
can be neglected, and an inner layer
($ - y_{\epsilon} < y < y_{\epsilon}$) in which resistivity (and
viscosity) are important (see \figref{fig:tm_theory}).  The complete
solution is determined by an interplay between both regions:
resistivity acts in the inner region, but the rate at which magnetic
field lines are reconnected also depends on the rate at which the
field can be advected into and out of the inner region. Hence, the
growth rate can be quite different for field profiles that are the
same close to $y = 0$, but differ elsewhere.

\paragraph{Outer Layer}
Condition (\ref{eq:timescales_main}) implies that there are negligible
field gradients in the outer regions ($|y|> y_{\epsilon}$), \ie
resistive processes are slow there compared to the growth of  TMs.  On
the other hand, $\gamma \gg a^{-2} \eta \sim k^2 \eta$ (the second
relation follows from assumption \ref{eq:k_delta_main}). Hence, to solve
Eqs.\,(\ref{eq:flux_freezing}) and (\ref{eq:force_balance}) in the
outer region the resistivity term in the induction equation
(\ref{eq:flux_freezing}) can be neglected, \ie
\begin{equation}
 i k \velo_y b_{0x} =  \gamma b_{1y}\ . 
\label{eq:ind_out_0}
\end{equation}
Furthermore, from the second part of condition
(\ref{eq:timescales_main}), we have $\gamma \ll \ca / L_y \sim \ca k$.
This inequality together with Eq.\,(\ref{eq:ind_out_0}) imply that
terms proportional to velocity (gradients) in Eq.\
(\ref{eq:force_balance}) are negligible, \ie
$| \gamma \rho_0 k^2 \velo_y| \ll |i k^3 b_{0x} b_{1y}|$. Hence, TMs
evolve so slowly that the plasma inertia (terms containing
$\rho_0 \velo_y$ in Eq.\,\ref{eq:force_balance}) can be neglected on
the ideal MHD time scale, and Eqs.\,(\ref{eq:flux_freezing}) and
(\ref{eq:force_balance}) simplify to
\begin{align}
  i k \velo_y  &=  \frac{\gamma b_{1y}}{ b_{0x}},  \label{eq:ind_out}  \\
  i k [ -b_{1y} \partial_y^2  b_{0x} + b_{0x} ( -k^2 + 
  \partial_y^2) b_{1y}] &= 0.  \label{eq:force_balance_out}
\end{align}
in the outer layers.  So far, we have not yet made any assumption
concerning the background magnetic field.  For
$b_{0x}(y) = b_0 \tanh( y / a)$, the solution of
Eq.\,(\ref{eq:force_balance_out}) reads
\begin{equation}
  b_{1y}(y) = b_1 \left(\frac{1 + \tanh (y/a)}{1 - \tanh(y/a)}\right)^{ a\, k /2 }
             \frac{\tanh(y/a)-ak}{ \Gamma(2 - ak)}, 
\label{eq:b1y_out}
\end{equation}
where $b_1$ is a constant (initial perturbation amplitude) and
$\Gamma$ denotes the \emph{Euler gamma function}.  The velocity
perturbations can be easily determined combining
Eqs.\,(\ref{eq:ind_out}) and (\ref{eq:b1y_out}):
\begin{equation}
  \velo_{1y}(y) = \frac{\gamma b_1}{ik b_0} 
                  \left(\frac{ 1 + \tanh (y/a)}{1 - \tanh(y/a)}\right)^{ a\, k /2 }
                  \frac{\tanh(y/a)-ak}{\Gamma(2 - ak) \tanh(y/a)}. 
\label{eq:v1y_out}
\end{equation}
Note that Eq.\,(\ref{eq:ind_out}) has a singularity for
$| y / a| \rrr 0$, \ie for $|b_{0x}| \rrr 0$, which is removed (\ie
smearded out) by resistivity in the inner region.

\paragraph{Inner Layer}
Resistive (and viscous) terms can no longer be neglected in the inner
layer ($|y|< y_{\epsilon}$), and we have to solve
Eqs.\,(\ref{eq:flux_freezing}) and (\ref{eq:force_balance})
simultaneously.  Because in the inner region $| y / a| \ll 1$, we can
approximate the background magnetic field (Eq.\,\ref{eq:b0_tan}) as
$ b_{0x}(y) \approx b_0 y / a $ (Eq.\,\ref{eq:b_taylor}).  In the
inner region, perturbations in both velocity and magnetic field vary
more strongly in the $y$ direction than in the $x$ direction, \ie
$|k^2 \velo_{y}| \ll |\partial_y^2 \velo_{y}|$ and
$|k^2 b_{1y}| \ll |\partial_y^2 b_{1y}|$.%
\footnote{As an example, we consider velocity perturbations. Assuming
  that $b_{1y}$ is constant and using approximation
  (\ref{eq:b_taylor}), Eq.\,(\ref{eq:ind_out}) implies
  $|\partial_y^2 \velo_{y}| \sim | \velo_{y} / y^2 |$. Because
  $|y| \ll a$ in the inner layer, we have from
  Eq.\,(\ref{eq:k_delta_main}) $a^{-1} \sim k$, and hence
  $| \velo_{y}/ y^{2} | \gg | \velo_{y} k^2 |$. Although
  Eq.\,(\ref{eq:ind_out}) holds only in the outer layer, it should
  still be roughly applicable near the edge of the inner region. Note
  that $b_{1y} = \mathrm{const.}$ was only assumed to simplify the
  calculations, \ie relaxing this assumption does not change the
  estimate.}
Therefore, we can neglect the terms proportional to $k^2$ in
Eqs.\,(\ref{eq:flux_freezing}) and (\ref{eq:force_balance}), and we
obtain
\begin{align}
  \gamma b_{1y} &= i k \velo_y b_0 a^{-1} y  + \eta \partial^2_y b_{1y},    
\label{eq:ind_in} \\
  \gamma \rho_0   \partial^2_y  \velo_y   &= \rho_0 \nu (-2k^2
  \partial_y^2 +   \partial_y^4)  \velo_y + i k  b_0 a^{-1} y
  \partial_y^2 b_{1y} , 
\label{eq:force_balance_in}
\end{align}
where we used also Eq.\,(\ref{eq:b_taylor}).  Combining both equations into
a single one by eliminating terms that contain $b_{1y}$, we arrive at
a sixth-order ordinary differential equation (ODE) for $\velo_y$ that
reads
\begin{gather}
 \velo_{y}^{(6)} [\nu \eta  y^2] + \velo_{y}^{(5)} [-2 \nu \eta  y] +  
 \velo_{y}^{(4)} [-(\gamma(\eta + \nu) + 2\nu \eta k^2) y^2 + 2\nu \eta] + 
\nonumber  \\
 \velo_{y}^{(3)} [2 \eta (\gamma + 2 \nu k^2) y] + 
 \velo_{y}^{(2)} [k^2 a ^{-2} c_{\rm A}^2 y^4 + \gamma(\gamma + 2\nu k^2)
                 y^2 - 2 \eta (\gamma + 2\nu k^2)  ] + 
 \velo_{y}^{(1)} [2 k^2 a^{-2} c_{\rm A}^2 y^3] =0 ,
\label{eq:sixth}
\end{gather}
where $c_{\rm A}^2 \equiv b_0^2/\rho_0$ and
$\velo_{y}^{(n)} \equiv \partial_y^n \velo_y$.  This equation cannot
be integrated analytically, \ie some further approximations are
necessary.

In the inner layer, but sufficiently far away from $y=0$, velocity
perturbations will have a solution of type $\velo_{y} \propto y^{-1}$,
\ie the terms with the highest order derivatives of $\velo_y$ dominate
the solution of (\ref{eq:sixth}).  Thus, we proceed neglecting lower
order derivatives in (\ref{eq:sixth}).  Because the two terms with the
highest order derivatives ($\partial_y^6 \velo_y$ and
$\partial_y^5 \velo_y$) depend on viscosity, we consider
Eq.\,(\ref{eq:sixth}) for two limiting cases, namely in the viscous
and the inviscid regime.

\subsubsection{Inviscid  Case} 

In the inviscid case ($\nu = 0$), Eq.\,(\ref{eq:sixth}) reduces to a
fourth order differential equation that is still too complicated to be
solved analytically.  Therefore, we follow FKR63 and use their
\emph{constant $\psi$ approximation}.%
\footnote{FKR63 used dimensionless quantities, and in particular a
  dimensionless variable $\psi \sim b_{1y}$. This explains the name
  constant $\psi$ approximation, which is commonly used in the
  literature (see, \eg \cite{gkp} and \cite{Schnack}).  }

FKR63 noted that the function $\velo_y$ becomes singular in ideal MHD
because $\velo_y \propto y^{-1}$ for $y \rightarrow 0$ (see
Eq.\,\ref{eq:ind_out}), \ie $\velo_y$ varies strongly in the limit
of small resistivity.  Resistivity regularises this ill-behaved
solution.  The function $b_{1y}$ varies less for $|y/a| \approx 0$ and
can be approximated by a constant $b_{1y}(y) \approx b_{1y} (0)$.
With this approximation, Eqs.\,(\ref{eq:ind_in}) and
(\ref{eq:force_balance_in}) reduce to
\begin{align}
  \gamma  \eta \rho_0 \partial_y^2 \velo_y -k^2 a^{-2} b_0^2 \velo_y y^2
  &= ik \gamma  b_0 a^{-1} y b_{1y}(0),  
\label{eq:upper3} \\
 \partial_y^2 b_{1y} &= \frac{\gamma \rho_0}{ik b_0 a^{-1} y} \partial_y^2 \velo_y.
\label{eq:lower3}
\end{align}
We solve this system of equations by first integrating
Eq.\,(\ref{eq:upper3}) to obtain $\velo_y$, which we insert into
Eq.\,(\ref{eq:lower3}) to get $b_{1y}$.

To express Eqs.\,(\ref{eq:upper3}) and (\ref{eq:lower3}) in
dimensionless form, we introduce the dimensionless variables
\begin{align}
 s &= y \left( \frac{k^2 b_0^2}{a^2 \gamma \eta
               \rho_0}\right)^{\frac{1}{4}} 
   \equiv \frac{y}{\epsilon_{\mathrm{R}}}, 
\label{eq:s_def} \\
 \Phi &= i \velo_y \left( \frac{b_0^2 \rho_0 \eta k^2}{a^2
                          b_{1y}^{4}(0) \gamma^3} \right)^{\frac{1}{4}}, 
\label{eq:Phi} \\
 \psi &= \frac{b_{1y} }{ b_{1y}(0)},  
\\
 \lambda &= \gamma \left( \frac{\rho_0 a^2}{k^2 \eta b_0^2}
                   \right)^{\frac{1}{3}}  
\end{align}
with the length scale
\begin{equation}
\epsilon_{\mathrm{R}} \equiv  
  \left( \frac{\gamma \eta \rho_0 a^2}{k^2 b_0^2} \right)^{\frac{1}{4}}. 
\label{eq:epsilonr_def}
\end{equation}

The physical interpretation of $\epsilon_{\rm R}$ is that resistive
effects are important in the centre ($y=0$) up to
$|y| \sim \epsilon_{\rm R}$. In these new variables, Eqs.\,(\ref{eq:upper3})
and (\ref{eq:lower3}) read
\begin{align}
 \frac{ {\der}^2 \Phi }{ {\der} s^2} - s^2 \Phi  &= -s, 
\label{eq:upper4} \\
\frac{ {\der}^2 \psi}{ {\der} s^2} &= 
  - \lambda^{3/2} \frac{1}{s}\frac{ {\der}^2 \Phi}{{\der} s^2}.
\end{align}

%----------------------------------------------------------------------------
\begin{figure}
  \centering
  \includegraphics[width=0.4\textwidth]{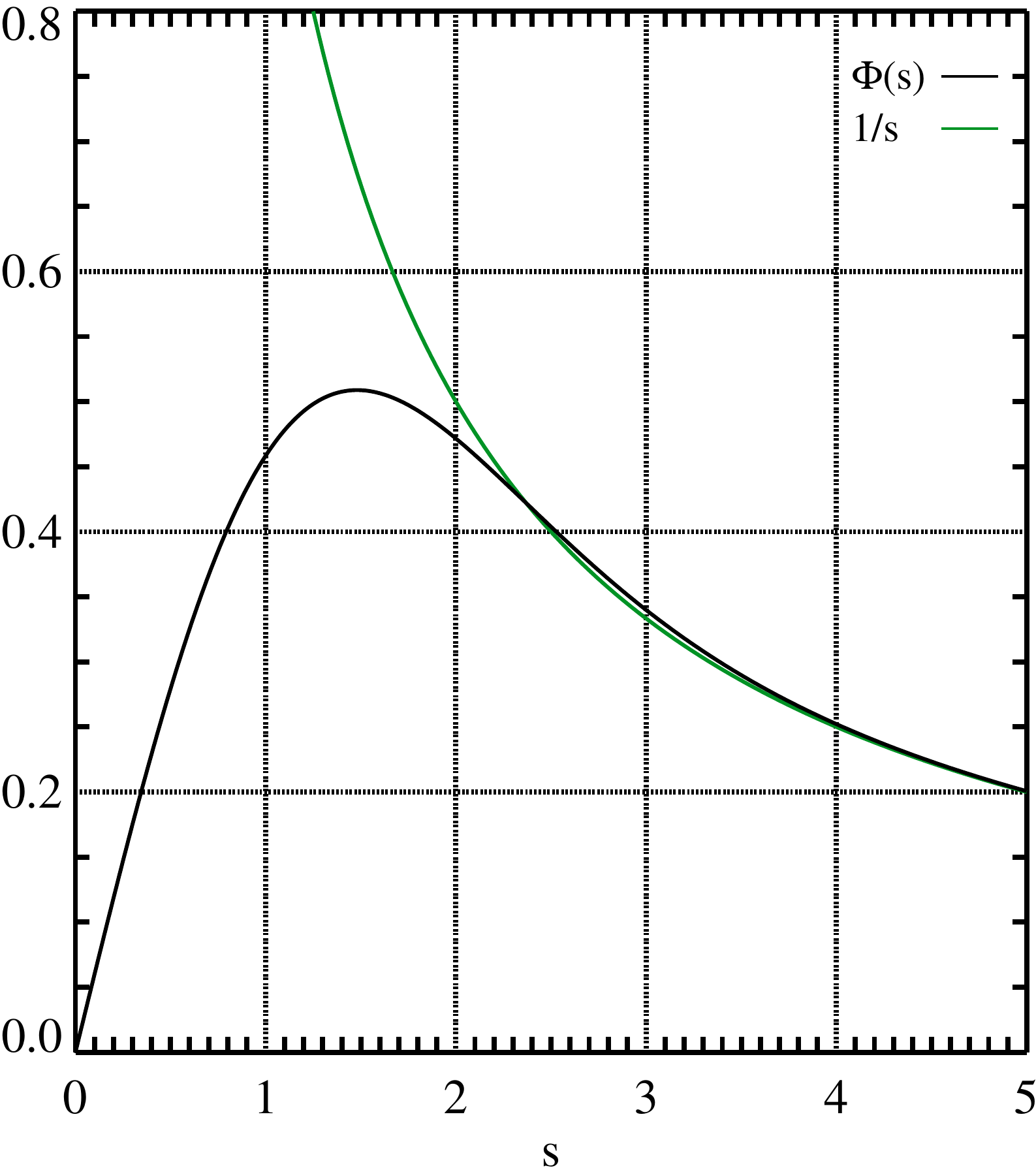}
  \caption{Graphical illustration of the functions $\Phi(s)$ (black;
    defined in Eq.\,(\ref{eq:Phi_integral}) and $1/s$ (green).  The
    velocity perturbations $\velo_y$ are exactly proportional to the
    former function in the inner region and approximately proportional
    to the latter one in the outer region for $| y / a| \ll 1$.  For
    $s \gtrsim 2.5$ both functions are very similar.}
  \label{fig:tm_phi}
\end{figure}%%
%----------------------------------------------------------------------------

The solution of Eq.\,(\ref{eq:upper4}) can be written as an integral
over an auxiliary variable $u$ \citep{gkp}:
\begin{equation}
  \Phi = \frac{s}{2} \int^{1}_{0} (1 - u^2)^{-1/4} e^{-s^2 u/2} \der u.
\label{eq:Phi_integral}
\end{equation}
The function $\Phi$ (black line in Fig.\,\ref{fig:tm_phi}) is positive
for $s > 0$, and has a global maximum at $s_{\rm max} \approx 1.48$.
Moreover, $\Phi(s) \approx 1.06/\sqrt{\pi}\, s$ for $s \ll 1$, and
$\Phi(s) \approx 1/s$ for $s \gg 1$.

To obtain the final form of the velocity perturbations, $\velo_y$, in
the inner layer (given by Eqs.\,\ref{eq:Phi} and
\ref{eq:Phi_integral}), we need to determine the TM growth rate
$\gamma$.  It can be calculated by matching the inner and outer
solutions of $\velo_y$ (the former given by Eq.\,(\ref{eq:v1y_out}) at
a certain point $y_{\mathrm{m}}$ in the region (marked in yellow in
\figref{fig:tm_theory}), where both solutions are valid and overlap,
\ie where they give the same predictions for both the velocity and the
magnetic field perturbations.  The value of $y_\mathrm{m}$ must be
large enough, so that $\Phi(s)$ can be approximated as
$\Phi(s) \approx s^{-1}$, \ie $s \gg 1$, yet small enough that the
outer ideal MHD solution behaves like $\velo_y \propto y^{-1}$ (for
$y \rrr 0$).  Moreover, the inner resistive solution was found for
such small values of $y$ that $b_{0x}(y)$ can be approximated by
Eq.\,(\ref{eq:b_taylor}).  Hence, $y_\mathrm{m} \ll a$ must hold.
Recalling that $s = 1$ for $y = \epsilon_{\mathrm{R}}$, we can combine
the above conditions into
\begin{equation}
  \epsilon_{\mathrm{R}} \ll y_\mathrm{m} \ll a .
\label{eq:scales}
\end{equation}
The remaining part of the matching procedure is conceptually rather
straightforward.  Comparing Eqs.\,(\ref{eq:Phi_integral}) and
(\ref{eq:v1y_out}) at $y_{\mathrm{m}}$, we can determine the tearing
mode growth rate $\gamma$. We omit the details of these calculations%
\footnote{We require that the velocity perturbations $\velo_y(y)$
  (\ref{eq:v1y_out}) in the outer layer and
  $\velo_y(y) = \velo_y(\Phi(s))$ (\ref{eq:Phi_integral}) in the inner
  layer are equal in the vicinity of $y_{\mathrm{m}}$, and that the
  same holds for their first derivatives. FKR63 and \citet{gkp}
  matched instead the magnetic field perturbations.}
and only give here the final expression for the TM growth rate in
resistive (-inviscid) MHD:
\begin{equation}
 \gamma = \left(\frac{2}{2.12}\right)^{4/5} \eta^{3/5}  
          \left(\frac{  b_0 k}{\sqrt{\rho_0} } \right)^{2/5} 
          a^{-6/5} \left(\frac{1}{a\, k} - a\, k\right)^{4/5}.
\label{eq:gr_ideal}
\end{equation}
For $a\, k > 1$, the growth rate is complex (because of the last
factor), \ie the system is TM unstable only for perturbations with
wavevectors $a\, k < 1$.  On the other hand, for $a\, k \rrr 0$ the
growth rate given by Eq.\,(\ref{eq:gr_ideal}) diverges, but in this
regime the \emph{constant $\psi$ approximation} becomes invalid and
Eq.\,(\ref{eq:gr_ideal}) no longer holds.

We note that the width of the resistive layer is somewhat
arbitrary. Some authors \citep[cf.][]{gkp,Schnack} define the layer
extend to $|y| = \epsilon_{\mathrm{R}}$, whereas we define its
boundary to be located at
\begin{equation}
|y| = L_{\mathrm{R}} = 1.48 \epsilon_{\mathrm{R}},
\label{eq:yLR}
\end{equation}
which corresponds to the maximum of $\Phi(s)$ (\figref{fig:tm_phi}).
This is a convenient definition, because $\Phi(s_{\rm max})$
corresponds to the peaks of $\velo_y$, which can be determined from
simulations rather easily (compare \figref{fig:tm_phi} with the bottom
panels of \figref{fig:tm_example_4_in_1}).

Eqs.\,(\ref{eq:v1y_out}) and (\ref{eq:Phi_integral}) for the velocity
perturbation $\velo_y$, and Eq.\,(\ref{eq:b1y_out}) for the magnetic
field perturbation $b_{1y}$ (supplemented by Eq.\,(\ref{eq:gr_ideal})
and the transformations (\ref{eq:s_def}) and (\ref{eq:Phi}) constitute
a complete solution of the TM problem in resistive MHD.  In the inner
layer, the magnetic field perturbation $b_{1y}$ is approximately
constant, \ie
$b_{1y} (y \le y_{\mathrm{m}} ) \approx b_{1y}(y_{\mathrm{m}})$, and
the perturbations $\velo_x$ and $b_{1x}$ can be determined from
conditions (\ref{eq:div_v}) and (\ref{eq:div_b}), respectively.

As a next step, we will generalise the above results to
resistive-viscous MHD, because Eq.\,(\ref{eq:gr_ideal}) does not take
into account (numerical) viscosity which might be comparable to (or
even larger than) numerical resistivity.

\subsubsection{Viscous Case} \label{subsec:viscus_case}
The derivation of the TM solution in resistive-viscous MHD by FKR63 is
very similar to that of the non-viscous case. Therefore, we will only
sketch their procedure (for more details see FKR63).  One integrates
Eqs.\,(\ref{eq:flux_freezing}) and (\ref{eq:force_balance}) again
separately in the outer and inner region.  In the outer region, one
can neglect dissipative terms and plasma inertia as in the inviscid
case, \ie the magnetic field perturbation $b_{1y}$ and the velocity
perturbation $\velo_y$ are still given by Eqs.\,(\ref{eq:b1y_out}) and
(\ref{eq:v1y_out}), respectively.  In the inner region, FKR63
simplified the sixth-order ODE (\ref{eq:sixth}) for $\velo_y$ using
again the \emph{constant $\psi$ approximation} to a fourth-order ODE,
but now including viscous terms.  They further simplified the ODE by
neglecting terms with lower-order derivatives, which is an acceptable
approximation for magnetic Prandtl numbers
$\Pm \equiv \nu / \eta \gtrsim 1$.  Next, FKR63 introduced the
dimensionless variables
\begin{align}
  \tilde{s}    &=       \frac{y}{\epsilon_{\mathrm{RV}}},  \\
  \tilde{\Phi} &\propto \velo_{1y}, \label{eq:tile_phi}
\end{align}
(where we used the tilde symbol to explicitly stress that functions
\ref{eq:Phi_integral} and \ref{eq:tile_phi} differ) with the length
scale
\begin{equation}
\epsilon_{\mathrm{RV}} = (\eta \nu)^{1/6} 
                        \left(\frac{ a \sqrt{\rho_0}}{b_0 k} \right)^{1/3}.
\label{eq:epsilon_rv_org}
\end{equation}
The latter expression shows that both resistivity and viscosity affect
the size of the region, where dissipative effects are important. We
call this region the \emph{resistive-viscous layer}.  The width of
this layer $L_{{\mathrm{RV}}}$ is proportional to
$\epsilon_{\mathrm{RV}}$, and hence differs from the width
$L_{{\mathrm{R}}}$ of the resistive (inviscid) layer, which is
proportional to $\epsilon_{\mathrm{R}}$ (Eq.\,\ref{eq:epsilonr_def}).

In the resistive-viscous case, matching condition (\ref{eq:scales})
has to be replaced by
\begin{equation}
  \epsilon_{\mathrm{RV}} \ll y_\mathrm{m} \ll a,
\label{eq:scales_rv}
\end{equation}
and from matching of $\tilde{\Phi}(\tilde{s})$ with
Eq.\,(\ref{eq:v1y_out}), we obtain the TM growth rate in
resistive-viscous MHD:
\begin{equation}
 \gamma \approx \frac{ 2^{4/3 }}{3} \eta^{5/6} \nu^{-1/6} 
                \left(\frac{b_0 k}{\sqrt{\rho_0}} \right)^{1/3}
                 a^{-4/3} \left(\frac{1 }{a\, k} - a\, k \right).
\label{eq:gr_visc}
\end{equation}
The above expression differs from the result of FKR63 (see their
Eq.\,(H.8)), because these authors derived their equation in the
$a\, k \ll 1$ limit (in our units), whereas we did not neglect terms
proportional to $a\, k $.  For the background magnetic field
$b_{0x} = b_0 \tanh( y / a)$, we calculated the growth rate more
accurately.  Note that Eq.\,(\ref{eq:gr_visc}) is only an
approximation, because FKR63 only approximately solved ODE
(\ref{eq:sixth}). Therefore, the above equation could be off by a
small constant numerical factor.

Our numerical results differ from the analytic growth-rates derived
above.  Restrictions of grid resolution and computing time prevented
us from reaching the parameter regime where the derivation of the
analytic growth rates holds.  Therefore, in the following subsection,
we present an ``empirical'' ansatz for the TM growth rate which gives
much better predictions for our simulation results presented in
Sec.\,\ref{subsec:tm}.

\subsection{Empirical Approach}
\label{app:tm_semi}
In the analytically derived Eqs.\,(\ref{eq:gr_ideal}) and
(\ref{eq:gr_visc}), the TM growth rate is proportional to a product of
different powers of resistivity, viscosity, \alf speed, $a, k$, and
$(1 /(a\, k) - a\, k )$.  Based on this observation, we postulate an
ansatz:
\begin{equation}
  \gamma = N_0  \eta^{n_1} \nu^{n_2} 
           \left(\frac{b_0}{\sqrt{\rho_0}} \right)^{n_3} 
           k^{n_4} a ^{n_5} \left(\frac{1}{a\, k }- a\, k \right)^{n_6}, 
\label{eq:gr_ansatz}
\end{equation}
where $N_0$ is a (real) constant and $n_1,\dots,n_6$ are fractional
constants, which shall be determined by numerical simulations.  The
dimension of the growth rate is $[s^{-1}]$, which we abbreviate as
$ \mathrm{dim}( \gamma ) = [\s^{-1}]$. It has to be ``constructed''
from the other physical quantities.  Since
$\mathrm{dim}(\eta) = \mathrm{dim}(\nu) = [\cm^2 \ \s^{-1}]$,
$\mathrm{dim}(\ca) = [\cm \ \s^{-1}]$, $\mathrm{dim}(k) = [\cm^{-1}]$,
and $\mathrm{dim}(a) = [\cm]$, dimensional analysis provides the
following conditions:
\begin{align}
   n_1 + n_2 + n_3 &=1,  
\label{eq:nnn1} \\ 
 2( n_1 + n_2 ) + n_3 - n_4 + n_5 &=0. 
\label{eq:nnn2}
\end{align}
Similarly, the width $L_{\mathrm{RV}}$ of the resistive-viscous layer
should be equal to
\begin{equation}
  L_{{\mathrm{RV}}} = M_0 \epsilon_{\mathrm{RV}} = M_0 \eta^{m_1} \nu^{m_2} 
   \left(\frac{b_0}{\sqrt{\rho_0}} \right)^{m_3} k^{m_4} a^{m_5} 
   \left(\frac{1}{a\, k}- a\, k\right)^{m_6} ,  
\label{eq:vl_ansatz}
\end{equation}
where $M_0$ is a (real positive) constant and $m_1,\dots,m_6$ are
fractional numbers to be determined by simulations.  From the
dimensional analysis follows
\begin{align}
   m_1 + m_2 + m_3 &=0,  
\label{eq:mmm1} \\ 
 2 ( m_1 + m_2 ) + m_3 - m_4 + m_5 &=1. 
\label{eq:mmm2}
\end{align}

%%%%%%%%%%%%%%%%%%%%%%%%%%%%%%%%%%%%%%%%%%%%%%%%%%%%%%%%%%%%%%%%%%%%%%
\begin{table*}
  \caption{
    2D simulations performed to test and calibrate ansatz
    \bref{eq:gr_ansatz}  and \bref{eq:vl_ansatz}. 
    The columns give the series identifier, shear parameter $a$, 
    initial magnetic field strength $b_0$, viscosity $\nu$, and resistivity
    $\eta$. In all simulations the background density is set to 
    $\rho_0 = 1$, and an equidistant grid spacing of
    $\Delta x = \Delta y = (\pi/3) / 1024$ is used. The box length 
    is $2 L = 20\, \pi a / 3$ and the TM wavevector $k = 3/ (10 a)$.
    The estimators $n_i$, $N_i$, $m_i$, and $M_i$ are given by fits 
    according to Eqs.\,\bref{eq:ln_n1}, \bref{eq:ln_m1},  
    \bref{eq:ln_n3}, and \bref{eq:ln_m3}, respectively.
    In \tr{\#TMa} the estimators $n_1$, $N_1^\mathrm{a}$, $m_1$, and
    $M_1^\mathrm{a}$ are determined, whereas in \tr{\#TMar} the estimators 
    $n_1$ and $m_1$ are set to fractional values and $N_1^\mathrm{ar}$ 
    and  $N_1^\mathrm{ar}$ are determined. For series \tr{\#TMb, \#TMc,}
    and \tr{\#TMd} we proceeded analogously. }
\begin{center}
\hspace*{-2.3cm} 
\begin{tabular}{lccccllll}
\tableline
series & $ a $ &  $b_0$ &$\nu$ &  $\eta$ &  $n_i$ & $N_i$ & $m_i$ & $M_i$ 
\\
\tableline
% \\
\#TMa &  $ 0.1 $ & $1$ & $10^{-4}$ & $10^{-7} \dots 10^{-5}$ & $0.7994 \pm 0.0012$ & $5.377 \pm 0.015$ & $0.160 \pm 0.003$ & $ -2.08 \pm 0.04$ 
\\
\#TMar &   $ 0.1  $ & $1$ & $10^{-4}$  &  $10^{-7}  \dots  10^{-5}$   & $ 4/5$  & $ 5.385 \pm 0.001  $ & $1/6$ & $ -1.992\pm  0.004 $ 
 \\
\tableline
\#TMb &  $ 0.1  $ & $1$ & $10^{-5}$  &  $10^{-6} \dots  10^{-5}$   & $0.801 \pm 0.004$  & $5.84 \pm 0.05$ & $0.159 \pm 0.007$ & $ -2.393 \pm 0.006$ 
 \\
\#TMbr &  $ 0.1  $ & $1$ & $10^{-5}$  &  $10^{-6} \dots  10^{-5}$   & $4/5 $  & $ 5.826  \pm 0.003  $ & $1/6 $ & $ -2.393 \pm 0.006 $ 
 \\
\tableline
\#TMc &  $ 0.1  $ &  $0.5  \dots  10$ & $10^{-4}$  &  $ 5 \times 10^{-5} $   & $0.391 \pm 0.004 $  & $ -4.377  \pm 0.006  $ & $-0.364 \pm 0.017$ & $ -4.021   \pm 0.015  $ 
 \\
\#TMcr &  $ 0.1  $ & $0.5  \dots  10$ & $10^{-4}$  &  $ 5 \times 10^{-5} $   & $2/5  $  & $-4.385  \pm 0.006  $ & $ -1/3 $ & $-4.03  \pm 0.02 $ 
 \\
\tableline
\#TMd &  \hspace{-0.5cm}$0.05$\hspace{-0.5cm} & $0.5  \dots  4$ & $10^{-5}$  &  $10^{-6}$   & $0.411 \pm 0.008 $  & $ -4.058 \pm 0.005  $ & $ -0.329 \pm 0.017 $ & $ -5.17   \pm 0.01  $ 
 \\
\#TMdr & \hspace{-0.5cm}$0.05$\hspace{-0.5cm}  & $0.5  \dots  4$ & $10^{-5}$  &  $10^{-6}$   & $2/5 $  & $ -4.055 \pm 0.005  $ & $-1/3 $ & $  -5.17 \pm 0.01 $ 
 \\
%  \\
\tableline
\end{tabular}
\label{tab:semi_ana}
\end{center}
\end{table*}
%%%%%%%%%%%%%%%%%%%%%%%%%%%%%%%%%%%%%%%%%%%%%%%%%%%%%%%%%%%%%%%%%%%%%%

To determine $n_1$ and $m_1$, we performed the simulations series
\tr{\#TMa} (the setup is described in \tabref{tab:semi_ana} and the results
are given in \figref{fig:tm_calib_theory}, where the solid lines are
the theoretical predictions according to
Eqs.\,\ref{eq:appendix_gr_calib} and \ref{eq:appendix_vl_calib})
keeping all parameters constant but the resistivity.  To the measured
TM growth rates and widths of the resistive-viscous layer (as
described in \secref{sec:phys_tm}), we fit functions
\begin{align}
\label{eq:ln_n1}
 \ln(\gamma)  &=  n_1 \ln(\eta) + N_1, \\
\label{eq:ln_m1}
\ln(    L_{{\mathrm{RV}}})   &=  m_1 \ln(\eta) + M_1,
\end{align}
where according to ansatzes (\ref{eq:gr_ansatz}) and
(\ref{eq:vl_ansatz}) $N_1$ and $M_1$ should be constant for the models
of series \tr{\#TMa}, \ie
\begin{align}
N_1 &=  \ln \left[ N_0  \nu^{n_2} 
        \left(\frac{b_0}{\sqrt{\rho_0}} \right)^{n_3} k^{n_4} a^{n_5} 
        \left(\frac{\delta}{k }- \frac{k}{\delta}\right)^{n_6} \right], 
\\
M_1  &= \ln \left[M_0 \nu^{m_2} 
        \left(\frac{b_0}{\sqrt{\rho_0}} \right)^{m_3} k^{m_4} a^{m_5}  
        \left(\frac{1}{a\, k }- \frac{a\, k}{1} \right)^{m_6} \right].
\end{align}
From the obtained estimators (and their small errors; see
\tabref{tab:semi_ana}) $N_1^{\mathrm{a}}$ and $M_1^{\mathrm{a}}$ (the
upper index ``a'' denotes the simulation series, \ie \tr{\#TMa)}, we
conclude that our ansatzes hold (at least for resistivity) and that
$n_1 = 4/5$ and $m_1 = 1/6$.

In the simulation series \tr{\#TMb} (\figref{fig:tm_calib_theory}), we set
the value of the viscosity to $\nu = 10^{-5}$, keeping the other
parameters as in \tr{\#TMa}, and also vary the value of the
resistivity. The fits done according to Eqs.\,\eqref{eq:ln_n1} and
\eqref{eq:ln_m1} (see \tabref{tab:semi_ana}) confirm that $n_1 = 4/5$
and $m_1 = 1/6$.

We determine the dependence of the TM growth rate and of the width of
the resistive-viscous layer on viscosity in the following somewhat
indirect way.  To the results of simulations \tr{\#TMa}, we refit functions
\bref{eq:ln_n1} and \bref{eq:ln_m1}, but this time with
$n_1 \equiv 4/5$ and $m_1 \equiv 1/6$, to obtain $N_1^\mathrm{ar}$ and
$M_1^\mathrm{ar}$, respectively (\tabref{tab:semi_ana}; note that we
denote this series as \tr{\#TMar}, where ``r'' stands for ``refitted'').
In an analogous way, we obtain $N_1^\mathrm{br}$ and
$M_1^\mathrm{br}$.  According to ansatz (\ref{eq:gr_ansatz}), the
difference between $N_1^\mathrm{ar}$ and $ N_1^\mathrm{br}$ should be
\begin{equation}
   N_1^\mathrm{ar} - N_1^\mathrm{br} = 
   \ln(10^{-4 n_2}) - \ln(10^{ -5 n_2} )  =  n_2 \ln(10). 
\end{equation}
From the obtained estimators (\tabref{tab:semi_ana}), we have
$ N_1^\mathrm{ar} - N_1^\mathrm{br} = -0.441 \pm 0.004$, from which we
can infer $n_2 = -1/5$.  Analogously, from
$M_1^\mathrm{ar} - M_1^\mathrm{br} = 0.40 \pm 0.01$, we infer that
$m_2 = 1/6$ (theoretically this value should be
$ M_1^\mathrm{ar}- M_1^\mathrm{br} = (1/6) \ln(10) \approx 0.384$).

\begin{figure}
\centering
  \includegraphics[width=0.45\textwidth]{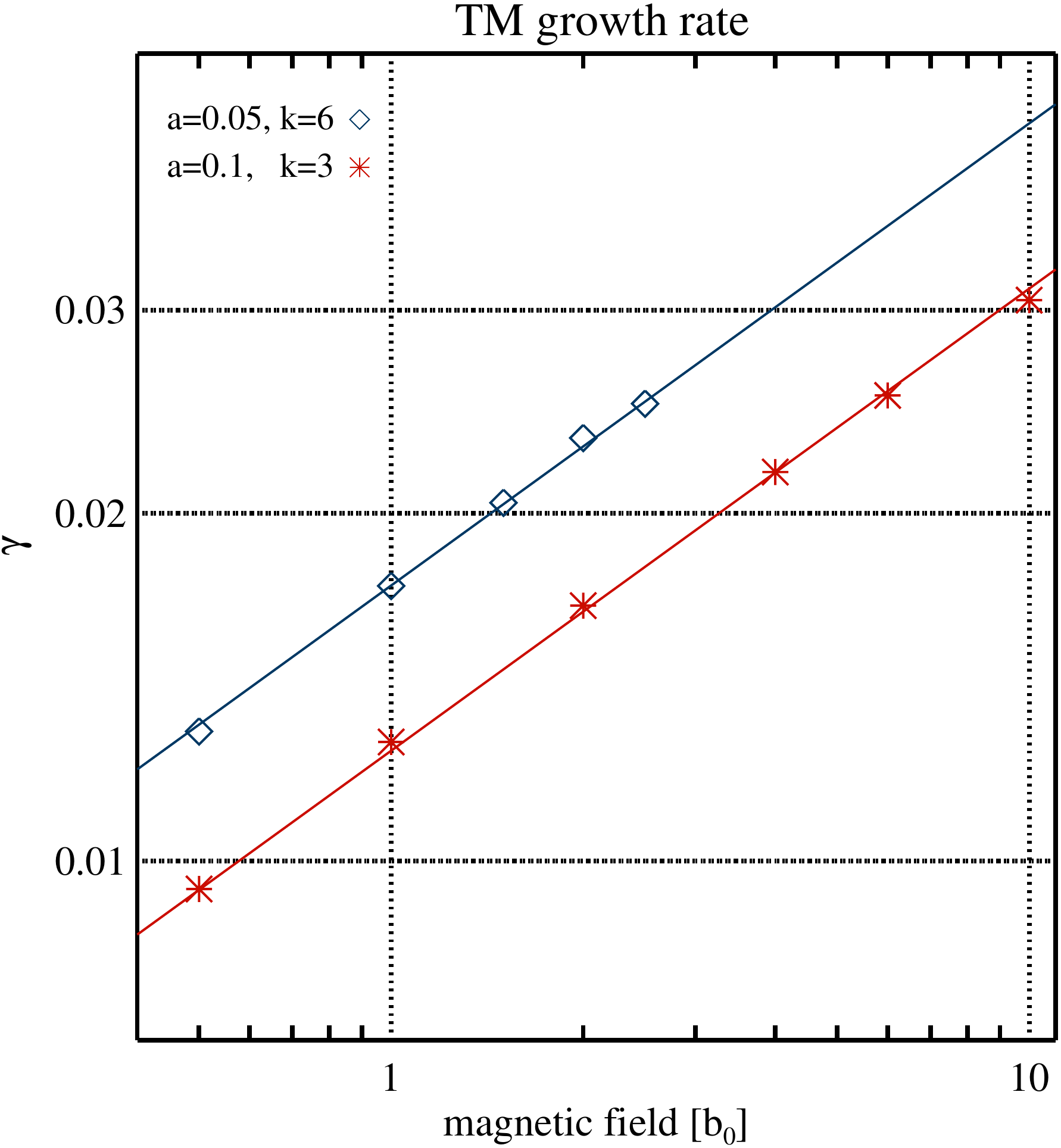} \phantom{MM}
  \includegraphics[width=0.45\textwidth]{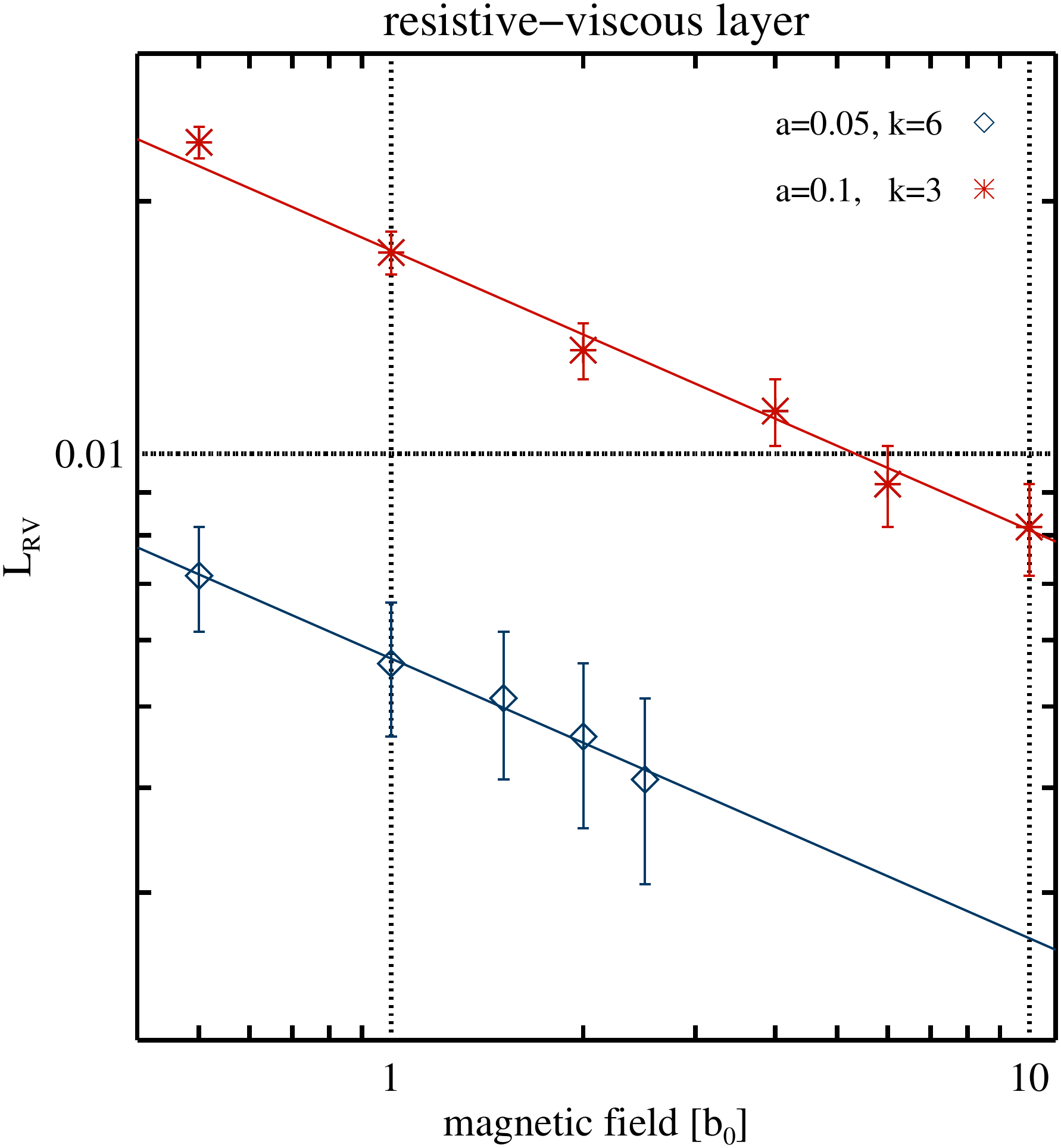}
  \caption{TM growth rate (\emph{left}) and resistive-viscous layer
    width (\emph{right}) as a function of background magnetic field
    strength.  Simulations \tr{\#TMd} and \tr{\#TMc} are depicted with
    blue diamonds and red asterisks, respectively.  Straight lines of
    the corresponding colours are linear fits to the logarithms of the
    simulation data.}
\label{fig:tm_calib_b}
\end{figure}

In two further sets of simulations (\tr{\#TMc} and \tr{\#TMd}), we determined
the dependence of the TM growth rate and of the width of the
resistive-viscous layer on the strength of the background magnetic
field.  To the simulation results (\figref{fig:tm_calib_b}), we fit
functions
\begin{align}
\label{eq:ln_n3}
 \ln(\gamma)  &=  n_3 \ln(b_0) + N_3, \\
\label{eq:ln_m3}
\ln(    L_{{\mathrm{RV}}})   &=  m_3 \ln(b_0) + M_3,
\end{align}
where $N_3$ and $M_3$ should be constant for a given simulation series.

%%%%%%%%%%%%%%%%%%%%%%%%%%%%%%%%%%%%%%%%%%%%%%%%%%%%%%%%%%%%%%%%%%%5
%%%%%%%%%%%%%%%%%%%%%%%%%%%%%%%%%%%%%%%%%%%%%%%%%%%%%%%%%%%%%%%%%%%5

From the obtained estimators (\tabref{tab:semi_ana}), we infer that
$ n_3 = 2/5$ and $m_3 = - 1/3$.  Note that these results are
consistent with our ansatzes, as from condition (\ref{eq:nnn1}), by
putting $n_1 = 4/5$ and $n_2 = -1/5$, we find
$n_3 = 1 - (n_1 + n_2) = 2/5$, and analogously from condition
\bref{eq:mmm1} we have $ m_3 = - m_1 - m_2 = -1/3$, as
$m_1 = m_2 = 1/6$.

With the results of simulation series \tr{\#TMc} and \tr{\#TMd}, one more aspect
of Eqs.\,(\ref{eq:gr_ansatz}) and (\ref{eq:vl_ansatz}) can be tested.
Even though we have not determined $n_4, n_5, n_6$ and
$m_4, m_5, n_6$, we expect from dimensional analysis that
$n_4 - n_5 = 8/5$ and $m_4 - m_5 = - 2/3$ (Eqs.\,\ref{eq:nnn2} and
\ref{eq:mmm2}, respectively).  Therefore, doubling $a^{-1}$ and $k$
(from $a = 0.1$, $k=3$ to $a = 0.05$, $k=6$) should increase the
instability growth rate by a factor of $2^{8/5}$ and decrease the
width of the resistive-viscous layer by a factor of $2^{-2/3}$
(because for a constant $a$ to $k$ ratio, the term
$(1 / (a\, k ) - a\, k )^{n_6}$ in Eq.\,\ref{eq:gr_ansatz} does not
change).  To the results of simulations \tr{\#TMc} and \tr{\#TMd}, we refitted
functions \bref{eq:ln_n3} and \bref{eq:ln_m3}, but this time with
$n_3 \equiv 2/5$ and $m_3 \equiv - 1/3$, obtaining
(\tabref{tab:semi_ana})
$\Delta_{\mathrm{N3}}^{ \mathrm{sim} } \equiv N_3^{\mathrm{cr}} -
N_3^{\mathrm{dr}} = 0.330 \pm 0.011$
and
$\Delta_{\mathrm{M3}}^{ \mathrm{sim} } \equiv M_3^{\mathrm{cr}} -
M_3^{\mathrm{dr}} = 1.14 \pm 0.03$.
Taking into account the different values of resistivity and viscosity
in these two series of simulations, we theoretically expect
$\Delta_{\mathrm{N3}}^{ \mathrm{th} } \equiv N_3^{\mathrm{cr}} -
N_3^{\mathrm{dr}} = 0.282$
and
$\Delta_{\mathrm{M3}}^{\mathrm{th} } \equiv M_3^{\mathrm{cr}} -
M_3^{\mathrm{dr}} = 1.11$.
Hence, the difference between theory and simulation is
\begin{align}
\Delta_{\mathrm{N3}}^{\mathrm{th} } -\Delta_{\mathrm{N3}}^{\mathrm{sim}} &=
  -0.048 \pm 0.011      ,\\
\Delta_{\mathrm{M3}}^{\mathrm{th} } - \Delta_{\mathrm{M3}}^{\mathrm{sim}} &= 
  -0.03 \pm 0.03.
\end{align}
Hence, the predictions for the resistive-viscous layer agree within
the measurement error.  Moreover, we tested that, as theoretically
expected, in the parameter range explored by us (incompressible
limit), the growth rate of the TM does neither depend on the
background pressure $p_0$ nor on the bulk viscosity \citep[see][for
details]{Rembiasz}.

So far, we have confirmed that our ansatzes for the TM growth rate
(Eq.\,\ref{eq:gr_ansatz}) and the width of the resistive-viscous layer
(Eq.\,\ref{eq:vl_ansatz}) hold and are given by
\begin{align}
  \gamma &=  N_0 \eta^{4/5} \nu^{-1/5} \left(
  \frac{b_0}{\sqrt{\rho_0}} \right)^{2/5} k^{n_4} a^{n_5} 
  \left( \frac{1}{a\, k }- \frac{a\, k}{1}\right)^{n_6},
\label{eq:appendix_gr_final} 
\\
  L_{{\mathrm{RV}}} &=   M_0 (\eta \nu)^{1/6} \left(
  \frac{b_0}{\sqrt{\rho_0}} \right)^{-1/3} k^{m_4} a^{m_5} \left(
  \frac{1}{a\, k }- \frac{a\, k}{1}\right)^{m_6} ,
\label{eq:appendix_vl_final}
\end{align}
where $N_0$ and $M_0$ are (real) constants, and $n_4, n_5, n_6$ and
$m_4,m_5$ are fractions.  Moreover, from conditions \bref{eq:nnn2} and
\bref{eq:mmm2} we have $n_4 - n_5 = 8/5$ and $m_4 - m_5 = - 2/3$,
respectively.  This allows us to calibrate these equations with the
help of estimators $N_1^\mathrm{ar}$ and $M_1^\mathrm{ar}$
(\tabref{tab:semi_ana}) for $k = 3$ and $a = 0.1$ obtaining%
\footnote{In these two equations we decided not to include the
  measurement errors, because for the applications discussed in
  Sec.\,\ref{sec:num_tm} they would be negligible anyway.}
\begin{align}
\label{eq:appendix_gr_calib}
\gamma(k=3, a = 0.1) &=  34.56  \eta^{4/5} \nu^{-1/5} \left(
  \frac{b_0}{\sqrt{\rho_0}} \right)^{2/5},  \\
\label{eq:appendix_vl_calib}
  L_{{\mathrm{RV}}}(k=3,a = 0.1) &=   0.634  (\eta \nu)^{1/6} \left(
  \frac{b_0}{\sqrt{\rho_0}} \right)^{-1/3}.
\end{align}

%%%%%%%%%%%%%%%%%%%%%%%%%%%%%%%%%%%%%%%%%%%%%%%%%%%%%%%%%%%%%
%% Bibliography
%%%%%%%%%%%%%%%%%%%%%%%%%%%%%%%%%%%%%%%%%%%%%%%%%%%%%%%%%%%%%

\label{lastpage}

\end{document}